\documentclass{aa}  

\usepackage{graphicx}
\usepackage{longtable}
\usepackage{txfonts}
\def\nHim1{n_{\langle\rm H\rangle,i-1}}
\def\nHip1{n_{\langle\rm H\rangle,i+1}}

    \def\half{{\scriptstyle{}^1\!\!/\!{}_2}}

\def\aap{A\&A}

\def\apjl{ApJL}
\def\apjs{ApJS}

\def\nr0{n_{r}^{\hspace{-0.9ex}^{\circ}}}

\begin{document}

\title{
   Sparkling nights and very hot days on WASP-18b:\\ the
     formation of clouds and the emergence of an ionosphere}
\titlerunning{ Sparkling nights and very hot days on WASP-18b}

   \author{Ch. Helling
          \inst{1,2, 3}
          \and
          P. Gourbin\inst{2,4}
          \and
          P. Woitke \inst{1,2}
          \and
          V. Parmentier\inst{5}
   }

   \institute{Centre for Exoplanet Science, University of St Andrews, St Andrews, UK\\
              \email{ch80@st-andrews.ac.uk}
         \and
         SUPA, School of Physics \& Astronomy, University of St Andrews, St Andrews, KY16 9SS, UK
         \and
         SRON Netherlands Institute for Space Research, Sorbonnelaan 2, 3584 CA Utrecht, NL
         \and
         Département de Physique,
Université Paris-Sud, Université Paris-Saclay,
91405 Orsay, France
         \and
         Atmospheric, Oceanic \& Planetary Physics, Department of Physics, University of Oxford, Oxford OX1 3PU, UK
                          }

   \date{Received XXX; accepted YYY}

 
  \abstract
      {WASP-18b is an utra-hot Jupiter
  with a temperature difference of upto 2500K between day and
  night. Such giant planets begins to emerge as planetary laboratory
  for understanding cloud formation and gas chemistry in well-tested
  parameter regimes in order to better understand planetary mass loss and for
  linking observed element ratios to planet formation and evolution.}
      {We aim to understand where clouds form,
      their interaction with the gas phase chemistry through depletion
      and enrichment, the ionisation of the atmospheric gas and the possible
      emergence of an ionosphere on ultra-hot Jupiters.}
   {We utilize 1D profiles from
     a 3D atmosphere simulations for WASP-18b as input for kinetic
     cloud formation and gas-phase chemical equilibrium calculations. We solve
     our kinetic cloud formation model for these 1D profiles that sample
     the atmosphere of WASP-18b at 16 different locations along the
     equator and in the mid-latitudes and derive consistently the gas-phase composition.}
   {The dayside of WASP-18b emerges as completely cloud-free due to
     the very high atmospheric temperatures. In contrast, the
     nightside is covered in geometrically extended and chemically
     heterogeneous clouds with disperse particle size distributions. The atmospheric C/O ratio increases to  $>0.7$ and the  enrichment of the atmospheric gas with cloud particles is  $\rho_{\rm d}/\rho_{\rm gas}>10^{-3}$.  The clouds that form at the limbs appear
     located farther inside the atmosphere and they are the least
     extended. Not all day-night terminator regions form clouds. The
     gas-phase is dominated by H$_2$, CO, SiO, H$_2$O, H$_2$S, CH$_4$,
     SiS. In addition, the dayside has a substantial degree of ionisation due
     to ions like Na$^+$, K$^+$, Ca$^+$, Fe$^+$. Al$^+$ and Ti$^+$ are
     the most abundant of their element classes. We find that
     WASP-18b, as one example for ultra-hot Jupiters, develops an
     ionosphere on the dayside. }
   {}


   \maketitle
%

\section{Introduction}
WASP-18b is a hot (T$_{\rm equ}\approx\,$2400K) Jupiter of 10\,M$_{\rm J}$ and 1.1 R$_{\rm J}$
(\citealt{2009Natur.460.1098H,2017ApJ...850L..32S}) orbiting a inactive late
F6-type host star (\citealt{2014Ap&SS.354...21F}) in 0.94 days on an orbit with a low eccentricity (e = 0.0085) and in almost perfect alignment with its host
star (\citealt{2010A&A...524A..25T}). 
\begin{figure*}
   \centering
   \includegraphics[width=0.8\textwidth]{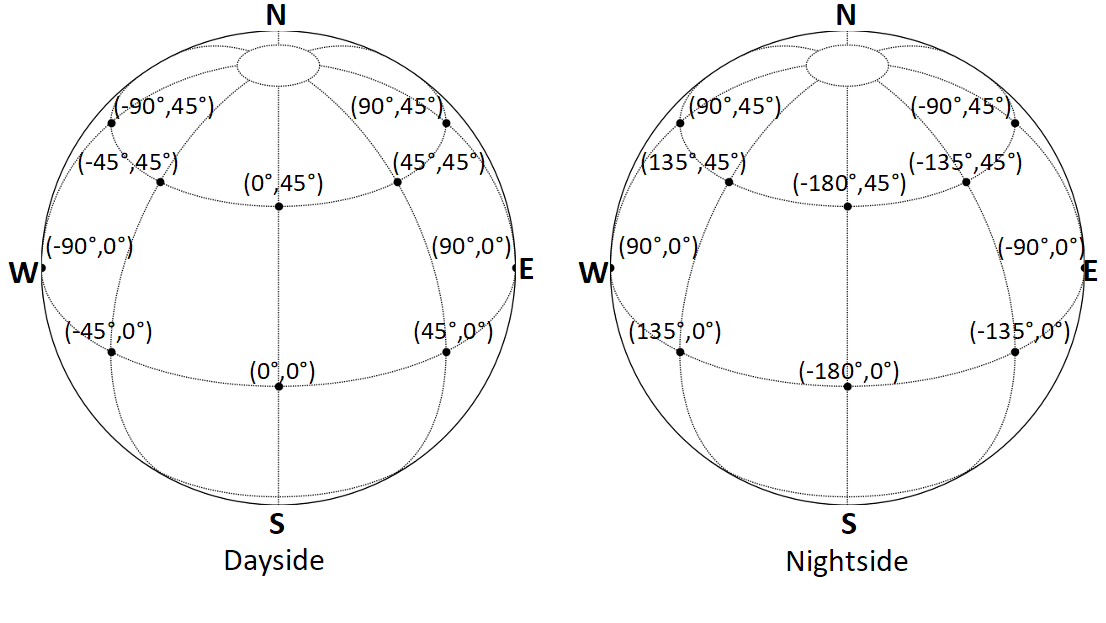}
   \caption{The 3D visualisation of the location of the 16 1D
     profiles (black dots) that are used to sample the 3D
     atmosphere for cloud formation, gas-phase chemistry and thermal
     ionisation on WASP-18b on the day (left) and the nightside
     (right). The sampled longitudes are $\phi=0^o, 45^o, 90^o, 135^o,
     -180^o, -135^o, -90^o$, the latitudes are at the equator and $\theta=0^o, 45^o$ in
     the northern hemisphere. The 3D simulations
     assume the southern hemisphere being similar to the northern
     hemisphere. The substellar point is $(\theta, \phi) = (0^o,
     0^o)$, the terminators are at $\phi= 90^o, -90^o$. The sketches indicate East and West which, however, need to be quoted {\it with respect to the day or the nightside}. }
   \label{globe}%
 \end{figure*}
WASP-18b is an ultra-hot Jupiter, being among the hottest close-in gas
giants known so far (\citealt{2018arXiv180500096P}).  Its ultra-short
period cause very high irradiation from its F-type star and,
hence leads to a extreme temperature difference between day- and
nightside (\citealt{2016ApJ...821...16K}).  \cite{2011ApJ...742...35N} suggest that
WASP-18b has an extremely low Bond albedo and very inefficient day/nightside
energy redistribution, a conclusion supported by
\cite{2013Icar..226.1719I,2015MNRAS.449.4192S}. The near Ks-band
secondary eclipse observations in combination with the previously
obtained Spitzer data suggest that a better thermal mixing is to be
expected at higher pressures deeper in the atmospheres
(\citealt{2015MNRAS.454.3002Z}). \cite{2017ApJ...850L..32S} find that
WASP-18b might have a non-solar C/O\,$\sim\,1$ based on a radiative
transfer retrieval method that includes a selected number of absorbing
species (H$_2$, H$_2$O, CH$_4$, NH$_3$, CO, CO$_2$, HCN, C$_2$H$_2$,
CIA of H$_2$-H$_2$, H$_2$-He; \citealt{2018MNRAS.474..271G}).  H$_2$O, TiO
and VO were expected in their dayside emission spectroscopy from HST
secondary eclipse observations but were
not found. \cite{2018ApJ...855L..30A} found that neglecting H$^-$ as
opacity source and the thermal dissociation of H$_2$O would push their
retrieval results to high C/O ratios for the evaluation of secondary
eclipse observations with HST WFC3 (1.1$\,\ldots\,1.7\mu$m) and combined
with Spitzer IRAC (3.5, 5.8 and 8.0 $\mu$m) both providing dayside
emission spectra. By including both effects, the
metallicity was constrained to approximately solar and an atmospheric C/O<0.85.
\cite{2018ApJ...855L..30A} confirm the minimal day-night energy
redistribution found by previous authors and the presence of a thermal
inversion on the dayside.

This paper begins a  consistent analysis of cloud
formation and gas-phase chemistry on ultra-hot Jupiters and analysis WASP-18b by applying numerical
simulations. We post-process 1D-profiles from a 3D simulation with
our kinetic cloud formation model (nucleation, growth/evaporation,
gravitational settling, element conservation) that includes a detailed
gas-phase calculation and evaluates the thermal ionisation of the
gas-phase. We find that WASP-18b forms clouds on the nightside and
that thermal ionisation causes the emergence of an ionosphere on the
dayside with Mg, Fe, Al, Ca, Na, and K being the most important electron donors in the collisional dominated parts of the atmospheres. Ti$^+$ is the most abundant Ti-species on the dayside, not TiO. The enrichment of the atmospheric gas with cloud particles (dust-to-gas ratio, $\rho_{\rm d}/\rho_{\rm gas}$) is rather
homogeneously of the order of $\geq10^{-3}$. The C/O ratio has increased
to $>0.7$ in the cloud forming regions of WASP-18b's atmosphere.

Our approach is outlines in Sect.~\ref{appr} which includes a summary
of the cloud formation model. Section~\ref{globcl} present our
results for the global cloud properties of WASP-18b, and
Sect.~\ref{gasc} the global day-night changes in gas-phase chemistry
on WASP-18b. Section~\ref{s:discu} offers a discussion on element
replenishment representations and a comparisons of WASP-18b to
HD\,189733b and HD\,209458b.  Section~\ref{con} concludes the paper.

 \begin{figure*}
   \centering
   \includegraphics[width=\textwidth]{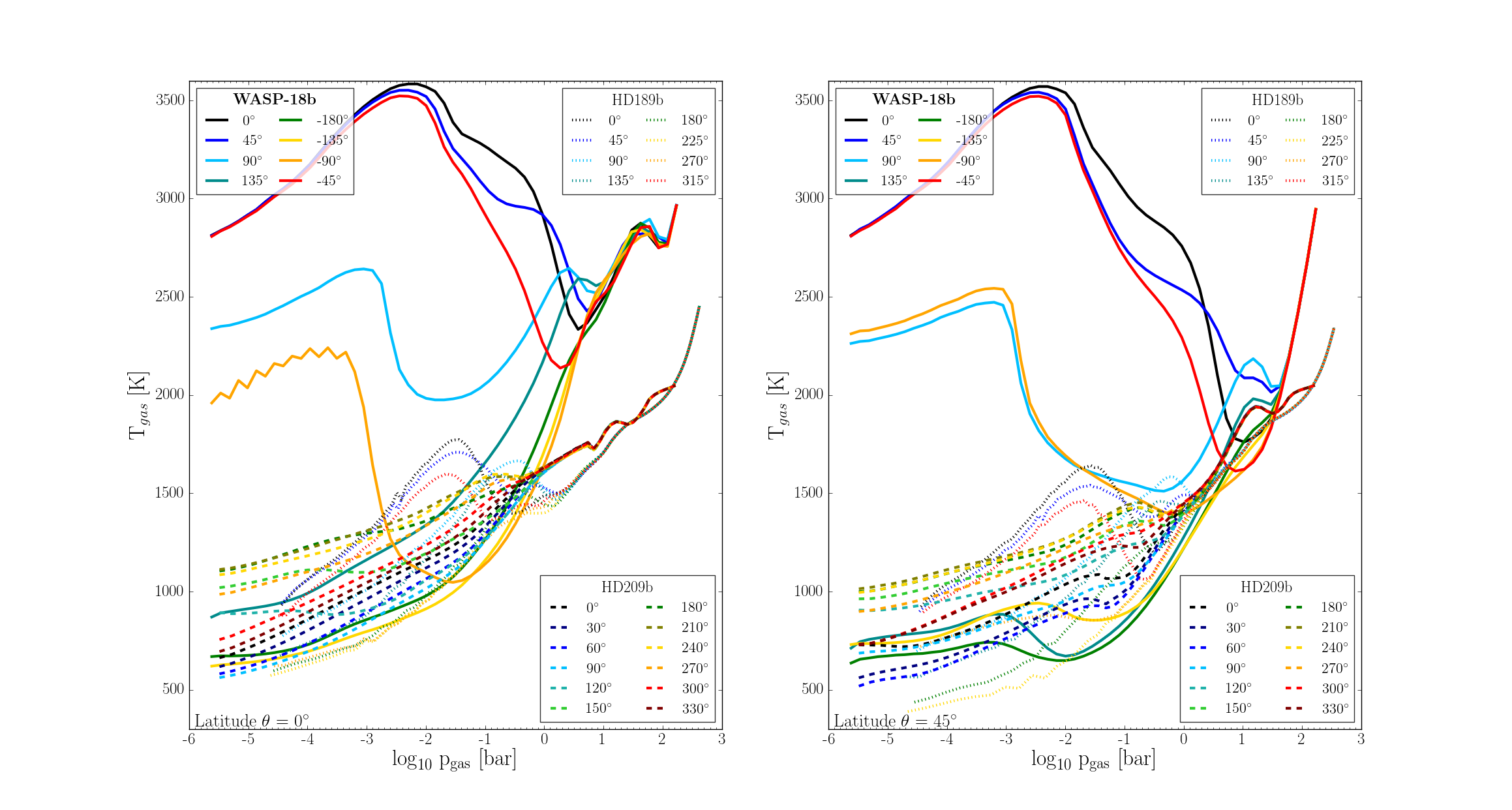}
   \caption{Input gas temperature, T$_{\rm gas}$ [K], and gas
     pressure, p$_{\rm gas}$ [bar], for WASP-18b
     (\citealt{2018ApJ...855L..30A}), and in comparison to HD\,189733b
     (\citealt{2013MNRAS.435.3159D}) and HD\,209458b
     (\citealt{2014A&A...561A...1M}).  WASP-18b is considerably
     different to HD\,189733b and HD\,209458b at the days-side but
     similar temperature regimes occur on the night-side.}
   \label{tp}
 \end{figure*}

 \begin{figure*}
   \centering
   \includegraphics[width=\textwidth]{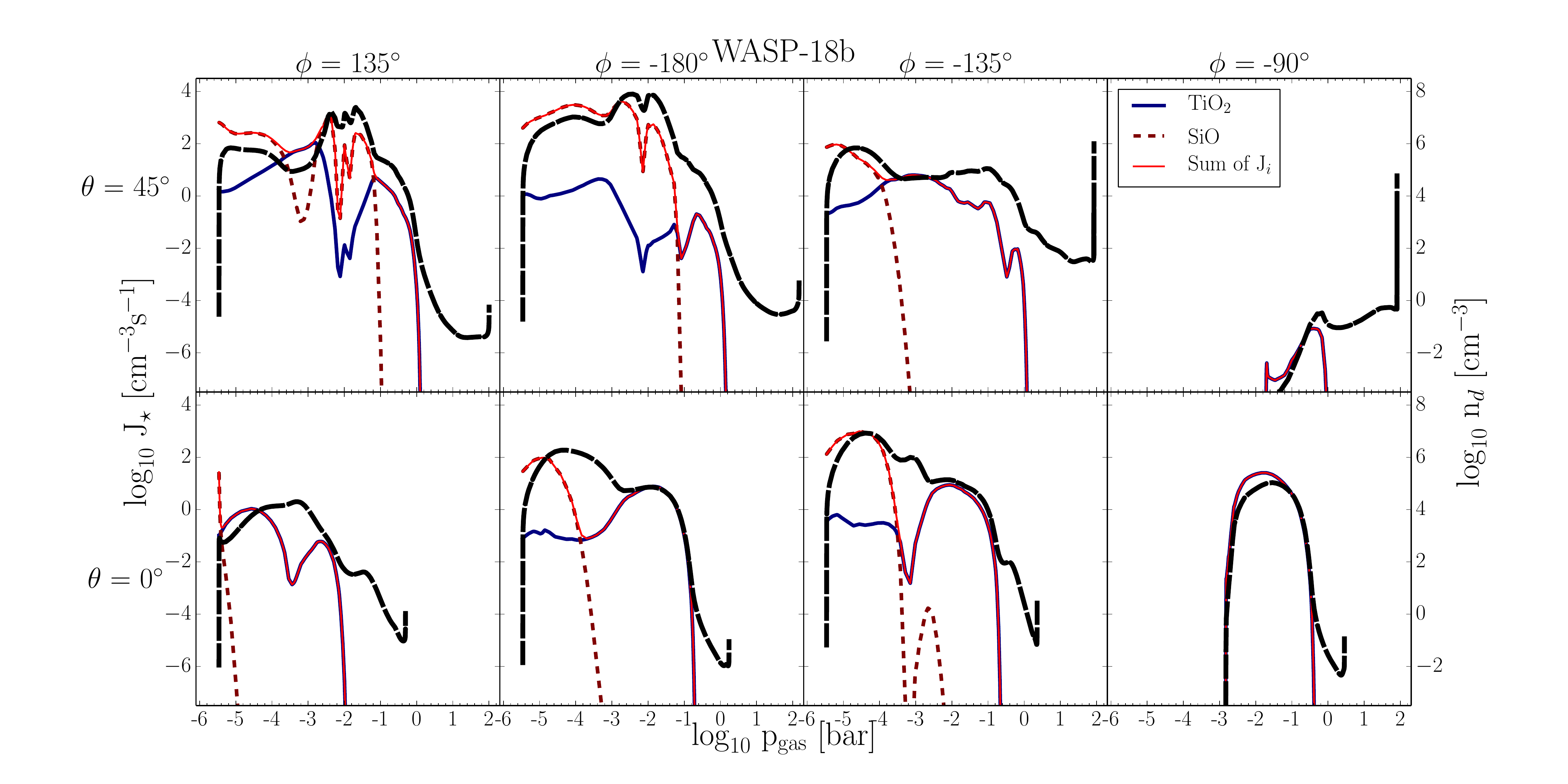}
   \includegraphics[width=\textwidth]{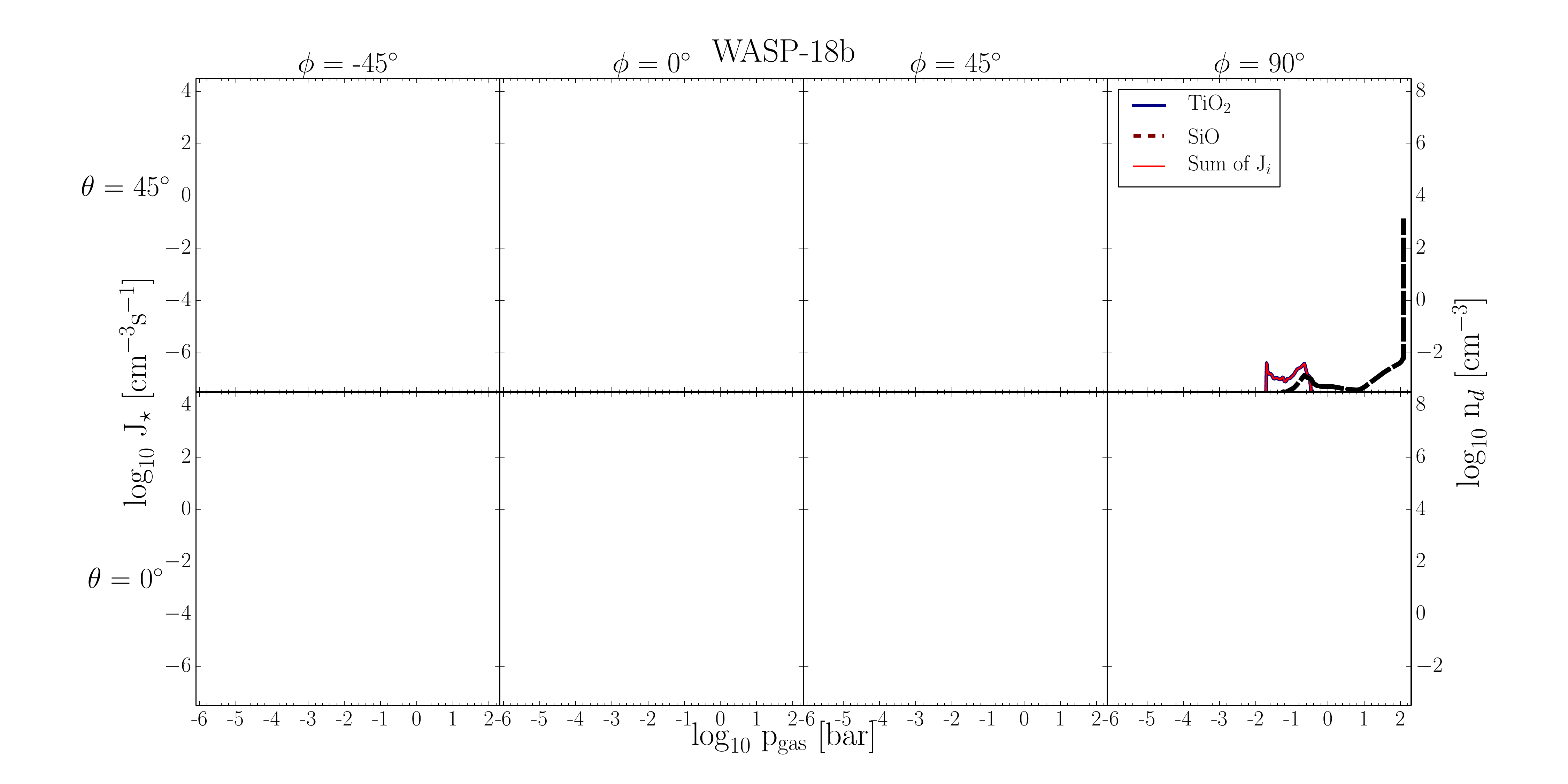}
   \caption{The seed formation rate,  J$_*$ [cm$^{-3}$ s$^{-1}$], that triggers the cloud formation
     (left axis) and the resulting cloud particle number densities, n$_{\rm d}$ [cm$^{-3}$], 
     (right axis, thick dashed line). The total seed formation rate
     (solid red line) is the sum of the nucleation rate for TiO$_2$
     (solid blue), SiO (dashed brown) and C (dashed gray). Only
     TiO$_2$ and SiO form condensation seeds, carbon does not condense
     in the oxygen-rich environment of WASP-18b. No seed
     formation occurs on the dayside, hence, no clouds can form, except in the
     mid-latitudes near the east day/night terminator.}
   \label{Jstarglob}%
    \end{figure*}

  \begin{figure*}
   \hspace{-1.2cm}
 \includegraphics[width=1.15\textwidth]{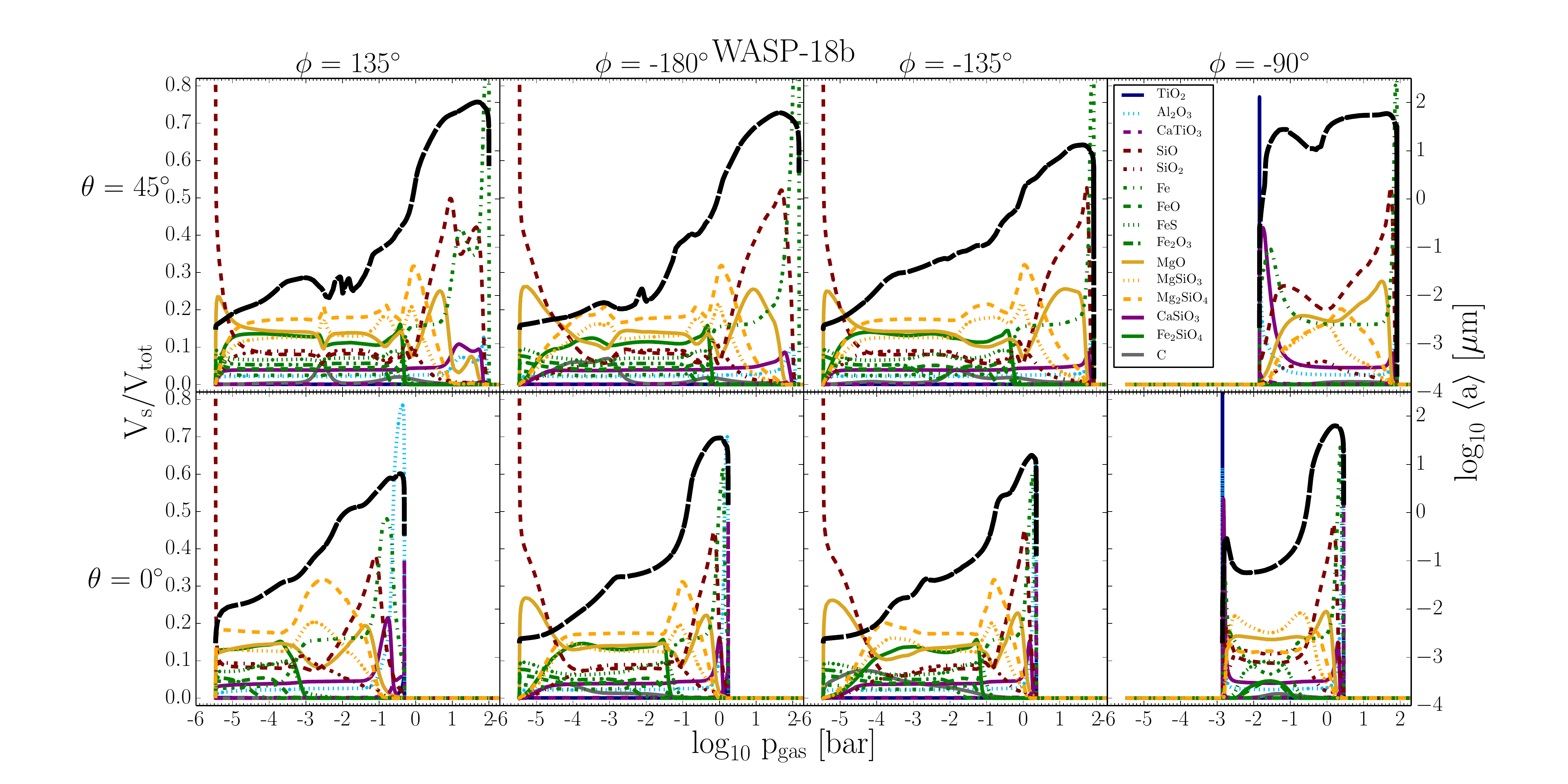}
   \caption{The cloud structure in terms of material composition  ($V_{\rm s}/V_{\rm tot}$, $s$ - 15 solid species, left axis, colour coded) and the mean cloud particles size ($\langle a\rangle$
     [$\mu$m], right axis, black dashed) on the nightside of WASP-18b. No clouds form on the dayside of WASP-18b. (top: 45$^o$ north, bottom: equator).  Some materials reach 100\% at the top (SiO[s] (brown dashed), TiO$_2$[s] (blue solid)) or at the bottom (Fe[s] at $\phi=135^o, -185^o$)  of the cloud. Carbon (grey solid) appears with a volume fraction of  $\approx 5-10\%$  at some locations.}
   \label{Vsnight}%
    \end{figure*}

\section{Approach}\label{appr}

We apply the two-model approach that we adopted to study the global
and the local cloud formation of HD\,189733b and HD\,209458b in
comparison (\citealt{2016MNRAS.460..855H}).  We extract 1D (T$_{\rm
  gas}$(z), p$_{\rm gas}$(z), v$_{\rm z}(x,y,z)$)-profiles
(T$_{\rm gas}$(z) -- local gas temperature [K], p$_{\rm gas}$(z)
    -- local gas pressure [bar], v$_{\rm z}$(x,y,z) -- local vertical
    velocity component [cm s$^{-1}$]) from 3D hydrodynamic atmosphere
  simulations across the globe and use these structures as input for
  our kinetic, non-equilibrium cloud-formation code {\sc Drift}
  (\citealt{Woitke2003,Woitke2004,Helling2006,Helling2008}). The 3D
  thermal and wind structure of WASP-18b were calculated using the
  SPARC/MITgcm (\citealt{Showman2009}) .  We refer to the individual
  papers regarding more details on the 3D simulations and also the
  cloud formation modelling (see below).  SPARC/MITgcm used here did not implement radiative feedback by clouds. This approach has the
  limitation of not taking into account the potential effect of
  horizontal winds on the cloud formation. We note that the cloud
  formation processes are determined by the local thermodynamic
  properties which are the result of the 3D dynamic atmosphere
  simulations. Horizontal winds would affect the cloud formation
  profoundly if the horizontal wind time-scale would be of the order
  of the time-scales of the microscopic cloud formation processes.
  Another aspect is the transport of existing cloud particles through
  horizontal advection which can not be considered in the approach
  that we follow in this paper. Horizontal advection as transport
  mechanisms for cloud particles will play a role if the frictional
  coupling between gas and cloud particles is sufficient.  A
  frictional decoupling emerges for large enough cloud particles or
  for low enough gas densities as explored in
  \cite{Woitke2003}. Horizontal transport will only affect our results
  if the wind blows in the right direction (night $\rightarrow$ day),
  and if the advected cloud particles remain thermally
  stable. Vertical decoupling is included in the approach used here as
  part of the cloud formation formalism
  (Sect.~\ref{ss:cm}). \cite{Woitke2003} have further shown that
  latent heat release is negligible for the condensation of the
  materials considered if forming from a solar element abundances gas.

 \subsection{Kinetic formation of cloud particles from oxygen-rich gases}\label{ss:cm}

Cloud formation in extrasolar atmospheres requires the formation of
seed particles because giant gas planets have no crust like rocky
planets.  Rocky planets, like Earth, sweep up cloud condensation nuclei (CCNs) through sand storms, volcano outbreaks, wild fires, and ocean spray. Condensation seeds provide a surface onto which other materials can condense more easily as surface reactions are
considerably more efficient than the sum of chemical gas phase
reactions leading to the formation of the seed. The formation of the
first surface out of the gas phase proceeds by a number of subsequent
chemical reactions that eventually result in small seed
particles. Such a chain of chemical reactions can proceed by adding e.g. a
molecular unit during each reaction step (e.g.,
\citealt{jeong2000,plane2013}). \cite{gou2012}, for example, show that
condensation occurs from small gas-phase constituent like MgO and SiO,
which will lead to the formation of bigger units like Mg$_2$SiO$_4$
during the condensation process. It is important to realize that big
molecules like Mg$_2$SiO$_4$ or Ca$_4$Ti$_3$O$_{10}$ do not exist in
the gas phase (see \citealt{2017arXiv171201010W}). In this paper, we
follow our kinetic cloud formation approach and refer the reader for details on the theoretical background to the references provided below.

\noindent
    {\it Nucleation (seed formation):} We apply the concept of homogeneous nucleation
    to model the formation of TiO$_2$, SiO and carbon seed particles
    (\citealt{helling2013RSPTA,2015A&A...575A..11L,2018A&A...614A.126L}).  We consider the
    simultaneous formation of these three different nucleation species
    in contrast to our previous works. In order to do so, we combine
    our previous works \cite{2015A&A...575A..11L} (for TiO$_2$ and
    SiO) and \cite{2017A&A...603A.123H} (for carbon).  The effective
    nucleation rate, $J_* = \sum_i J_{\rm i=TiO2, SiO, C}$ [cm$^{-3}$
      s$^{-1}$], determines the number of cloud particles, n$_{\rm d}$
    [cm$^{-3}$], and hence, the total cloud surface (as sum of the
    surface of the cloud particles). 

\noindent
    {\it Bulk growth/evaporation:} It is essential that the seed
    forming species are also considered as surface growth material as
    both process (nucleation and growth) compete for the participating
    elements (here: Ti, Si, O, C). We consider the formation of 15
    bulk materials (TiO$_2$[s], Mg$_2$SiO$_4$[s], MgSiO$_3$[s],
    MgO[s], SiO[s], SiO$_2$[s], Fe[s], FeO[s], FeS[s], Fe$_2$O$_3$[s],
    Fe$_2$SiO$_4$[s], Al$_2$O$_3$[s], CaTiO$_3$[s], CaSiO$_3$[s],
    C[s]) that form from 9 elements (Mg, Si, Ti, O, Fe, Al, Ca, S, C)
    by 126 surface reactions (Table~\ref{tab:chemreak}). We solve
    moment equations for the cloud particle size distribution function
    that consider nucleation, growth/evaporation, gravitational
    settling and mixing
    (\citealt{Woitke2003,Helling2006,Helling2008,2008A&A...485..547H,
      helling2013RSPTA}). The growth speed is $\chi$ [cm s$^{-1}$]
    the sign of which is determined by the effective supersaturation
    ration, $S_{\rm eff}$. If $S_{\rm eff}<1$, $\chi<0$ and the cloud particles
    evaporate. The approach presented in \cite{Woitke2003}, and
      utilized in this paper, applies force balance between friction
      and gravity to derive a size-dependent drift velocity which is
      required to determine a drift dependent growth term in the
      moment equations (gravitational settling).

In addition to the local element abundances,  $\varepsilon(z)$,
also the local thermodynamic properties $T_{\rm gas}(z)$ and
$\rho_{\rm gas}(z)$ (local gas density, [g cm$^{-3}$]) determine if
atmospheric clouds can form, to which sizes, ($\langle a \rangle$
  [$\mu$m] -- mean cloud particle size), the cloud particles grow and
of which materials, s (e.g. TiO$_2$[s], Mg$_2$SiO$_4$[s], MgSiO$_3$[s], $\ldots$), they will be composed. The gravitational
settling  velocity ($v_{\rm drift}$ [cm s$^{-1}$]) is determined
by the local gas density, $\rho_{\rm gas}(z)$, and the cloud particle
size.

\noindent
{\it Element conservation:} An additional set of equations for all
involved elements is solved with source/sink terms for
nucleation, surface growth/evaporation, gravitational settling.

\noindent
{\it Element replenishment:} Cloud particle formation depletes the
local gas phase, and gravitational settling causes these elements to
be deposited for example in the inner (low pressure) atmosphere where the cloud
particles evaporate. For a stationary cloud to form, element
replenishment needs to be modelled. We apply the approach outline in
\cite{Lee2015} (Sect.~2.4) using the local vertical velocity to
calculate a mixing time scale, $\tau_{\rm mix}\sim v_{\rm
  z}(r)^{-1}$. We  discuss the use of constant
vertical diffusion coefficient, $\tau_{\rm mix} \sim K_{\rm zz}^{-1}$
in Sect.~\ref{s:Kzz}.

\subsection{Chemical gas composition}

We apply chemical equilibrium (thermochemical equilibrium) to
calculate the chemical gas composition  (represented in terms of
  number densities n$_{\rm x}$ [cm$^{-3}$]) of the atmospheres as part
of our cloud formation approach.  We use the 1D (T$_{\rm gas}(z)$,
p$_{\rm gas}(z)$) profiles and element abundances $\varepsilon_{\rm
  i}(z)$ (i=O, Ca, S, Al, Fe, Si, Mg, Ti, C) depleted by the cloud
formation processes.  All other elements are assumed to be of solar
abundance. We use the {\sc GGChem} routines that recently were made
publicly available through \cite{2017arXiv171201010W}. A combination
of $156$ gas-phase molecules, $16$ atoms, and various ionic species
were used under the assumption of LTE.  The respective material data
are benchmarked (\citealt{2017arXiv171201010W}). High velocities
and/or strong radiation may cause departure from
LTE. \cite{2006ApJ...648.1181V,2010ApJ...716.1060V,2009ApJ...701L..20Z,2010ApJ...717..496L,2012ApJ...745...77K,2011ApJ...737...15M,2012A&A...546A..43V}
have shown that in warm exoplanet atmospheres (T>1200 K), the chemical
timescales are in fact short, and hence thermo-chemical equilibrium
prevails in particular in the cloud-forming regions that we are
interested.

No condensates are part of our chemical equilibrium calculations
in contrast to equilibrium condensation models. The influence of cloud
formation on the gas phase composition results from the reduced or
enriched element abundances due to cloud formation and the cloud
opacity impact on the radiation field, and hence, on the local gas
temperature and gas pressure.  The element depletion or enrichment due
to cloud formation is therefore directly coupled with the gas-phase
chemistry calculation.

\subsection{Input and boundary conditions}\label{1dtraj}

{\it Element abundances:} We assume that WASP-18b has an oxygen-rich
atmosphere of approximately solar element composition. We use the
solar element abundances from \citet{Grevesse2007}
(Table~\ref{tab:mr}) as initial values for the cloud formation
simulation and outside the cloud forming domains.

\noindent
{\it Input profiles from a 3D atmosphere simulation for WASP-18b:}
We use 1D profiles from a 3D atmosphere code which are globally
distributed as shown in Fig.~\ref{globe}. The geometry is north-south
symmetric.  The 3D thermal and wind structures were
calculated using the SPARC/MITgcm (\citealt{Showman2009}). The
hydrodynamic model solves the primitive equations on a cube sphere
grid. It has been successfully applied to a wide range of hot Jupiters
(\citealt{Showman2009,2013A&A...558A..91P,2015ApJ...801...86K,2016ApJ...821....9K,2017arXiv170600466L})
including a few ultra hot Jupiters
(\citealt{2018arXiv180309149Z,2018AJ....156...17K}).

 Molecular and atomic abundances in the 3D code are calculated using a modified
 version of the NASA CEA Gibbs minimization code
 as part of a 
grid previously tabulated and used to explore gas
 and condensate equilibrium chemistry in substellar objects
 \citep[][]{Moses2013,Skemer2016,2015ApJ...801...86K,Wakeford2017,Burningham2017,Marley2017DPS}
 over a wide range of atmospheric conditions. We consider $\approx 500$ gas-phase species and condensates
 containing the elements H, He, C, N, O, Ne, Na, Mg,  Al, Si, P, S, Cl,
 Ar, K, Ca, Ti, Cr, Mn, Fe, and Ni.
We assume solar
 elemental abundances and local chemical equilibrium with rainout of
 condensate material taken into account, but no interaction between the gas phase
 and the solid phase. 

 The opacities of major molecules (CO, H$_2$O, CH$_4$, NH$_3$, TiO, VO, CrH, FeH,
 CO$_2$, PH$_3$, H$_2$S), alkali atoms (Na, K, Cs, Rb, and Li), and continuum
 (collision induced absorption due to H2-H2, H2-He and H2-H,
 bound-free absorption by H and H- and free-free absorption 
 are taken into account following~\citet{Freedman2008}
 and~\citet{Freedman2014}. The radiative transfer is performed within
 8 k-coefficients inside each of the 11 wavelengths
 bins~\citep{Kataria2015}.
 We used a timestep of
 25s, ran the simulations for 300 days and averaged all quantities
 over the last 100 days. The above modelling process is the same as
 that described in~\citet{Parmentier2016}, using the WASP-18 system
 parameters from~\citet{Southworth2009}

The model, as all global circulation models studying these hot planets
suffers from some important limitations. The model does not include
magnetic interactions between the planetary magnetic field and the
ionised gas~\citep{Rogers2014,Rogers2017}, latent heat transport
through the recombination of $\rm H_2$~\citep[e.g.][]{Bell2018} nor
cloud opacities~\citep{Parmentier2016,Roman2017}. We therefore expect that
large scale structure such as the day/night contrast or the east/west
terminator difference to be accurate within an order of magnitude, but exact
temperatures and winds speeds might change ~\citep[e.g.]{Koll2018}.

{\it WASP-18b's planetary parameter:} We use $T_{\rm equ}=2411$K
(\citealt{2017ApJ...850L..32S}), assume a constant value $g=19043$ cm
$s^{-2}$ ($R_{\rm P}=8.328818\cdot 10^9$ cm, $M_{\rm P}=1.979614 \cdot
10^{31}$g, \citealt{2009Natur.460.1098H}) and a constant mean
molecular weight for a H$_{\rm 2}$ dominated gas ($\rm\mu$ = 2.3
m$_{\rm\mu}$). The local gas density $\rho_{\rm gas}(z)$ is calculated
from the given gas pressure by applying the ideal gas law, $\rho_{\rm
  gas}(z) = P_{\rm gas}(z)\mu m_{\mu}/(k_{B}T_{\rm gas}(z))$. In order to find
the height of each of the atmospheric layer inside the 53 vertical
computational domains, hydrostatic equilibrium is assumed in vertical
direction. The hydrostatic equilibrium equation is integrated  to
convert the given pressure into a height coordinate. The inner
integration boundaries are the planet radius where $p_{\rm gas}=1$
bar.

We introduce a small uncertainty by using $\rm\mu$ = 2.3 m$_{\rm\mu}$
if H$_2$ would dissociate. We show in Fig.~\ref{H} that thermal H$_2$
dissociation only occurs on the dayside of WASP-18b where no clouds
can form. Hence, this assumption does not affect our cloud formation
results for which the above procedure is required in order to derive
the geometric extension of the numerical grid.

\section{The global cloud properties of WASP-18b}\label{globcl}

We study if and which kind of clouds could form on WASP-18b by
sampling 8 different profiles at the equator and 8 in the
mid-latitude region. We endeavour to provide a first insight into the
global and the local cloud structure for a planet that is expected to
show vastly different day and nightsides.

Figures~\ref{tp} summarise the input profiles that
we use to study cloud formation at the day- (longitude $\phi=0^o$, $45^o$,
$-45^o$) and the nightside (longitude $\phi=-180^o$, $135^o$, $-135^o$) of
WASP-18b as well as at the day/nightside terminators (longitude
$\phi=90^o$ and $\phi-90^o$). The differences in the thermodynamic structures
are very large: Day and nightside temperature have a difference of
upto 2500K. The dayside profiles can be as hot as 3500K at a
relatively low pressure of p$_{\rm gas} = 10^{-3}$ bar. The terminator
regions show strong temperature inversions causing a steep local
(inward) drop in temperature of up to 1000K. Such thermodynamic
differences suggest that cloud and gas-phase chemistry will
differ strongly between the day and the nightside of the planet.
This temperature inversion may cause the appearance of emission
features due the emergence of an outward increasing temperature
gradient.

{\it WASP-18b in comparison to HD\,189733b and HD\,209458b:} In
reference to our previous studies (\citealt{2016MNRAS.460..855H}),
Fig.~\ref{tp} shows the WASP-18b 1D profiles in comparison to the
1D profiles from HD\,189733b and HD\,209458b. Both, HD\,189733b
and HD\,209458b have atmospheres that are filled with clouds on the
day and the nightside, though their detailed characteristics (like
cloud particle size, material composition) differ. WASP-18b has a much
hotter inner atmosphere compared to HD\,189733b and HD\,209458b at the
equator and in the hemispheres, but reaches comparable low
temperatures at the nightside. The dayside is hotter by about 2500K
compared to HD\,189733b and HD\,209458b in the equator region. The
day/nightside temperature difference for HD\,189733b and HD\,209458b
are $\approx 500K$ compared to the 2000K on WASP-18b at local gas
pressure of $10^{-5}\,...\,10^{-2}$bar  (see Fig.~\ref{tp}).

\begin{figure*}
  \vspace*{-0.3cm}
   \includegraphics[width=0.54\textwidth]{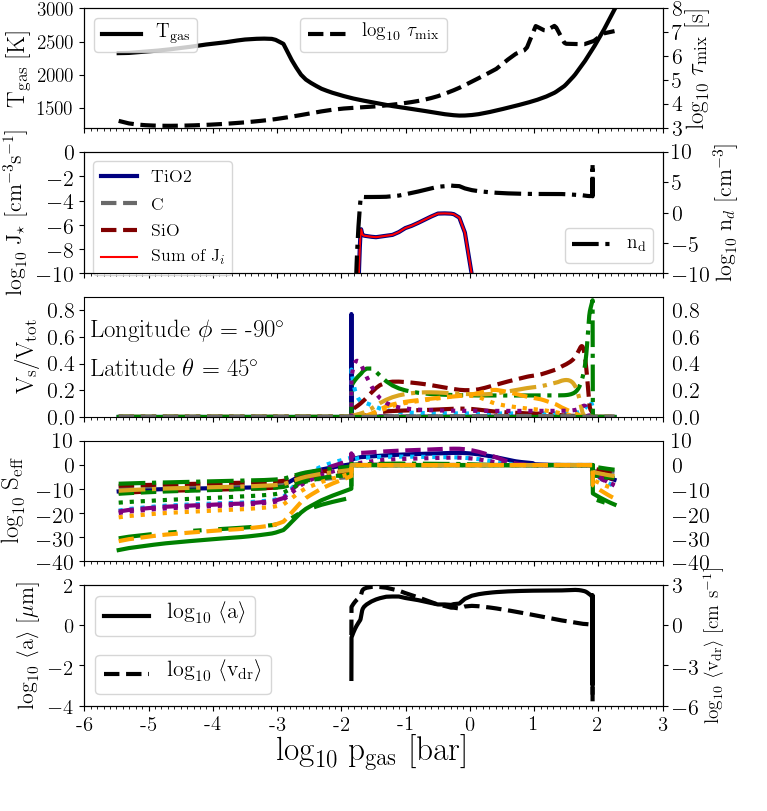}
   \hspace{-0.8cm}\includegraphics[width=0.54\textwidth]{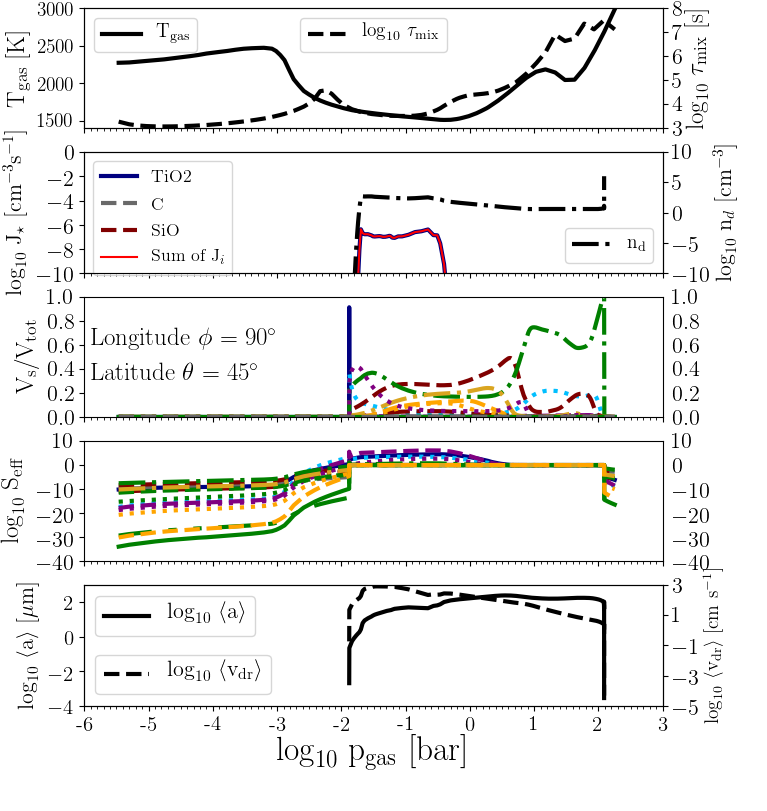}\\
   \includegraphics[width=0.54\textwidth]{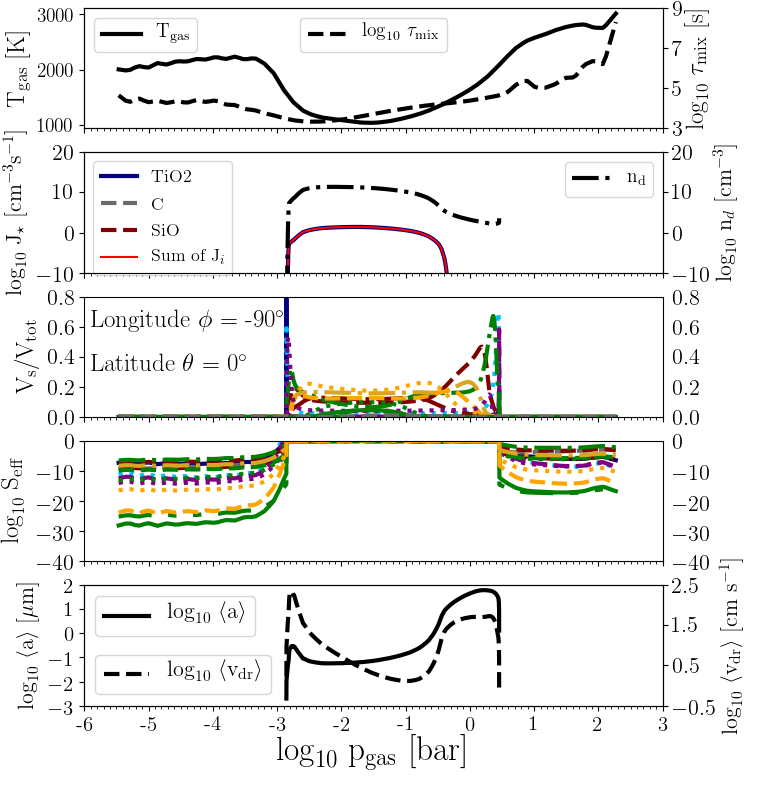}
    \hspace{-0.8cm}\includegraphics[width=0.55\textwidth]{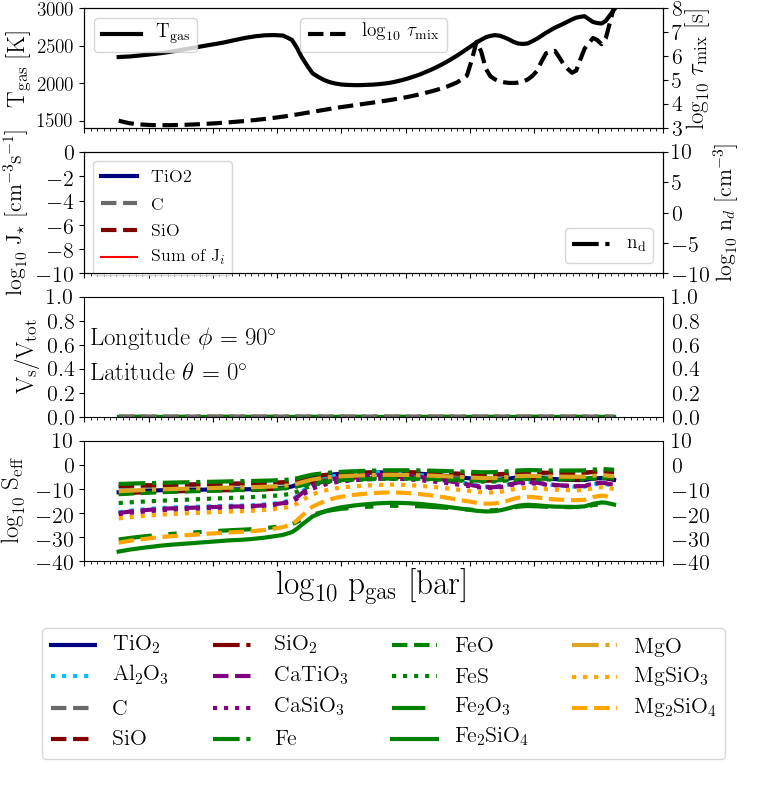}\\[-0.7cm]
   \caption{The combined view on the cloud structure of the profiles
     probing each of day-night-terminators of WASP-18b (top: 45$^o$
     north, bottom: equator; left: west, right: east  seen from the dayside). Following
     earlier works (e.g. Fig. 9 in \citealt{2014A&ARv..22...80H}), we
     summarise the fundamental cloud properties for each of the limb
     profiles to demonstrate the differences in cloud details and
     their causes. For each profile the plots show the following
     properties as function of the local gas pressure p$_{\rm gas}$
     [bar] : {\it 1st panel:} local gas temperature, T$_{\rm gas}$ [K]
     (solid, left), mixing time scale $\tau_{\rm mix}$ [s] (dashed,
     right). {\it 2nd panel:} total nucleation rate, J$_{*}$ [cm$^{-3}$
       s$^{-1}$] (red), and the individual rates for TiO$_2$ (blue),
     SiO (brown dashed), carbon (green dashed); number density of
     cloud particle, n$_{\rm d}$ [cm$^{-3}$].  {\it 3rd panel:}
     effective growth velocity, $\chi$ [10$^{-7}$ cm s$^{-1}$]. {\it
       4th panel:} material volume fraction, $V_{\rm s}/V_{\rm tot}$
     with $V_{\rm s}$ the volume of the material s, $V_{\rm tot}$ the
     volume of all cloud particles. {\it 5th panel:} effective
     supersaturation ratio, $S_{\rm eff}$, per material. {\it 6th
       panel:} mean cloud particle size, $\langle a \rangle$ [$\mu$m]
     (solid, left), drift velocity $v_{\rm drift}$ [cm s$^{-1}$]. The
     materials are line and colour coded as follows: TiO$_2$[s] -
     solid, dark blue; Al$_2$O$_3$[s] - dotted, light blue; C[s] -
     dashed, green; SiO[s] - dashed, brown; SiO$_2$[s] - dash-dot,
     crown; CaTiO$_3$[s] - dashed, purple; CaSiO$_3$[s] - dotted,
     purple; Fe[s] - dash-dot, green; FeO[s] - dahsed, green; FeS[s] -
     dotted, green; Fe$_2$O$_3$[s] long-dash-dot, green;
     Fe$_2$SiO$_3$[s] - solid, green; MgO[s] - dash-dot, dark organg;
     MgSiO$_3$[s] - dotted, organg; Mg$_2$SiO$_4$[s] - dashed,
     organg.  Superrotation prohibits cloud formation on the east
     equatorial  terminator ($\phi=90^o$, $\theta=0^o$; lower left panels).}
   \label{termin}
   \vspace*{-1cm}
      \end{figure*}

{\it Our first result} is that no clouds form on the dayside (Fig.~\ref{Jstarglob}), hence,
a non-depleted warm gas-phase chemistry should be observed. However, a
depletion of the gas-phase could occur if cloud particles can be
transported horizontally from the nightside to the dayside and if,
at the same time, the local supersaturation is large enough to allow
for surface growth processes.  A depletion of the dayside gas-phase could also occur if the depleted gas from the nightside is advected on the dayside faster than the vertical mixing occurs (see \citealt{2013A&A...558A..91P}). In-situ cloud formation only takes
place in the day/nightside terminator regions and on the nightside
equator and in the hemispheric regions on WASP-18b. The northern
dayside terminator profile ($\theta=45^o$, $\phi=90^o$) does form
seed particles, but the equator dayside terminator profile
($\theta=0^o$, $\phi=90^o$) remains cloud-free as superrotation
affects the equatorial temperature such that the mid-latitudes are
colder than the equator (compare also Fig.~\ref{tp}). The
supersaturation of the gas is very low ($<10^{-5}$) at the $(\theta, \phi)=(0^o, 90^o)$ terminator (see Fig.~\ref{termin}). Cloud formation on the dayside remains therefore
very unlikely even if condensations seeds would be swept along with
the winds given that the temperatures on the dayside are even
higher. On the nightside, a vast amount of cloud particles form
leading to a dust-to-gas-ratio of $\geq10^{-3}$ and an increase in the
atmospheric C/O to $>0.7$.  \cite{2018ApJ...855L..30A} derive a disk-averaged C/O $<0.85$ from their emission spectre, hence for the day-side of WASP-18b. 
We predict that the dayside C/O ratio should be unaffected by cloud formation. If this primordial C/O were close to the upper bound of 0.85, then cloud formation could drive the C/O on the nightside to values large enough to change significantly the atmospheric chemistry and lead to an observable signature (\citealt{2014Life....4..142H}). Molecules like HCN and CN might be detectable. The increase of C/O to $>0.7$ (starting from solar values) does only occur on the nightside in our simulation.

\subsection{The nightside of WASP-18b}

Figure~\ref{Jstarglob} demonstrawhere in the atmosphere WASP-18b cloud
formation is triggered through the nucleation of condensation seeds. We
consider the simultaneous formation of TiO$_2$ (solid blue), SiO
(brown dashed) and C (solid gray) seed particles (left axis). The sum
of these seeds then provides the total number of cloud particles (long
dashed black lines, right axis) locally. The comparison between the
total nucleation rate, $J_*$ [cm$^{-3}$s$^{-1}$], the cloud particle
number density, $n_{\rm d}$ [cm$^{-3}$] and Fig.~\ref{Vsnight} reveals
that the vertical cloud extension encompasses a larger volume (plotted in terms
of pressure) than the nucleation regions would suggest. This
difference demonstrates that cloud particles are transported
vertically through gravitational settling into the deeper atmosphere.

Figure~\ref{Jstarglob} further demonstrates that SiO and TiO$_2$ do
efficiently nucleate in the uppermost atmospheric regions until
$\approx$ 2200K in the model structures used here. Carbon does not
nucleate for all probed profiles. TiO$_2$ remains efficient also
at higher temperatures due to the combined effect of element
consumption by material growth and temperature. TiO$_2$ is the sole
nucleation species for the hotter nightside east-terminator ($\phi=-90^o$) at
the equator and in the mid-latitudes.

The dayside of WASP-18b does have no seed formation except at the
dayside west terminator ($\phi=90^o$). This is a clear indication for
that WASP-18b will not form any clouds in situ at its dayside because
the local temperatures are simply too high. This conclusion is further
emphasized by our study of the gas-phase composition in
Sect.~\ref{gasc} and Fig.~\ref{Ti} which shows that Ti$^+$ is the
dominating Ti-carrier at the days-side of WASP-18b, and not TiO$_2$ as
is the case for most of the nightside. Small cloud particles can move along with
the flow and could therefore be transported form the night to the
dayside. They could, hence, serve as condensation seeds in the
absence of in-situ seed formation. Our investigation of the local
gas-phase, however, shows that the dayside of WASP-18b will be too
hot to allow for a sufficient supersaturation of the gas phase for
condensation to occur.

Figure~\ref{Vsnight} present the  vertical cloud extension, their detailed
material composition in units of relative volumes ($V_{\rm s}/V_{\rm
  tot}$, $s$ - solid species; colour codes lines on left axis) and the
vertical distribution of the mean cloud particles ($\langle a\rangle$
[$\mu$m], black solid line on right axis) for the equator (bottom row)
and the 45$^o$-latitude on the northern hemisphere. These 1D cloud
maps visualize that the clouds extend over a larger pressure range in
the (norther) hemisphere than at the equator region. At the (northern)
hemisphere nightside profiles ($\phi=135^o, -180^o, -135^o$), the
entire computational domain is filled with cloud particles. The cloud
extension becomes more confined the more westward ($\phi=-135^o
\longrightarrow -180^o \longrightarrow 135^o$) we probe the
atmosphere on the nightside in the equatorial regions. This confinement does not occur in the northern/southern hemisphere. The mean cloud particles sizes follow the 
profile of the atmospheric gas that generally has an outward
decreasing gas density: The cloud particles size increases inwards and
reaches it's maximum in the densest regions just before it decreases
and drops to zero because the local temperature becomes to hot and the
cloud particles evaporate. It is interesting to note that the
hemispheric ($\theta=45^o, -45^o$) day-night terminator at
$\phi=-90^o$ forms the largest cloud particles as result of a
considerably lower nucleation rate. 

The material composition of the cloud particles changes as the local
thermodynamic conditions change. The top-most layer is determined by
the seed particle forming materials and MgO[s] for all but the
$\phi=-135^o$-equator profile. As soon as Mg-Si-O and Fe-Si-O
materials become thermally stable, a combinations of them makes up the
matrix (the bulk) of the cloud particles. Mg$_2$SiO$_4$[s],
MgSiO$_3$[s], MgO[s],Fe$_2$SiO$_4$[s] make up 60\% of the volume with
Mg$_2$SiO$_4$[s] and MgSiO$_3$[s] contributing most. This is followed
by 20-30\% made of SiO[s], SiO$_2$[s], and FeS[s]. All other materials
remain below the 10\% level providing a colorful mix of minerals in
the Mg$_2$SiO$_4$[s]/MgSiO$_3$[s]-dominated part of the cloud. We note
that the cloud particles change in sizes from $\langle a\rangle
\approx 10^{-3}\,\ldots\,1\mu$m in this cloud region. The temperature
increases when going deeper into the atmospheres at the nightside
which causes the Mg-Si-O/Fe-Si-O materials to become thermally
unstable and, hence, to evaporate. This results in a substantial
material peak of SiO[s] before it evaporates and the high-temperature
condensates determine the material composition of the cloud at its
hottest rim  at low altitudes. The SiO[s] dominates substantially in the warmer
regions. 75\% of the cloud volume in this layer made of $\approx
1\,\ldots\,2 \mu$m-sized particles is made of SiO[s] with MgO[s]
($\approx 20$\%) and Fe[s] ($\approx 15$\%) with inclusions from
CaSiO$_3$[s] and Al$_2$O$_3$[s]. The equatorial region has again the
smaller particles also in this cloud region. Once SiO[s] has
evaporated, a thin cloud layer made of Fe[s] with CaSiO$_3$[s] and
Al$_2$O$_3$[s] inclusions follows. At the equator, one to two
additional cloud regions made of almost pure Al$_2$O$_3$[s] and below
that CaTiO$_3$[s] form.  Al$_2$O$_3$[s] and CaTiO$_3$[s] are the most
stable materials in our setup. In the mid-latitudes, the largest cloud
particles reach sizes of 100$\mu$m and are made of Fe[s] with
CaSiO$_3$[s] and Al$_2$O$_3$[s] inclusions. In the equator region, the
largest particles reach 10$\mu$m and are made of - possibly sparkling
- Al$_2$O$_3$[s] and CaTiO$_3$[s].  If the cloud particle undergo a
heating or are exposed to large pressures, amorphous materials can
turn into their crystalline counterparts (e.g. \citealt{2009IJAsB...8....3H})  possibly causing clouds to  sparkle.

The implication of the cloud formation on the gas-phase chemistry will
be discussed in detail in Sect.~\ref{gasc}.

 \subsection{The day/nightside terminator clouds on WASP-18b}

WASP-18b is interesting with respect to cloud and gas-chemistry as its
thermodynamic set-up varies greatly over the globe
(Sect.~\ref{1dtraj}). The change-over between 'no clouds' and 'plenty
of clouds' occurs at the day/nightside terminators which we discuss
here separately. Both terminator-regions at the equator and in the
northern hemisphere show strong temperature inversions causing an
extended, local minimum of the temperature inside the atmosphere. 
      
Figure~\ref{termin} summarized the cloud details for the four 1D
terminator profiles investigated in this paper (top: 45$^o$ north,
bottom: equator; left: west, right: east  seen from the dayside). The top panel of this
figures also show the input properties (T$_{\rm gas}$, $p_{\rm gas}$)
(solid line, left axis) and the mixing time scale, $\tau_{\rm mix}$
(dashed line, right axis). Cloud formation takes place at three out of
the four terminators. The east (seen from the dayside) terminator
($\phi=90^o$) at the equator is too hot for clouds to form. But winds
will not transport cloud particles from the night side here because
the wind moves from the day into the nightside at $\phi=90^o$. The
terminator exposed to the wind from the nightside ($\phi=-90^o$) is
cold enough for in situ cloud formation.

The vertical cloud extension varies between the equatorial ($p_{\rm
  gas}\approx 10^{-3}\,...\,10^{0.5}$ bar) and the norther/southern
hemisphere ($p_{\rm gas}\approx 10^{-2}\,...\,10^{2}$ bar) terminator
as result of the local temperature. This will have implications for
transit spectroscopy. The mean cloud particle sizes vary widely at the
equator and less in the norther/southern hemisphere
terminator. Differences emerge also for the material compositions. The
northern/southern hemisphere terminators (east \& west) have cloud
particles mainly composed of SiO[s], Fe[s], MgO[s] being less
dominated by Mg$_2$SiO$_4$[s]/MgSiO$_3$[s] with inclusions from other
materials. The top and the bottom parts of the terminator clouds are
dominated by high-temperature condensates: The top is made of thin
layers of pure TiO$_2$[s], followed by a thin layer of a
Al$_2$O$_2$[s]/CaTiO$_3$[s] mix, and then a thin
Fe[s]/CaTiO$_3$[s]-dominated layer with inclusions from
Al$_2$O$_2$[s], SiO[s] and others to a lesser extend. The bottom layer
is made of big Fe[s]-dominated particles with some Al$_2$O$_2$[s]. The
$\phi=90^o$ terminator region has a substantial Fe[s]-dominated  low-altitude
layer in contrast to the $\phi=-90^o$ terminator that has a
rather thin Fe[s]-dominated  inner layer  at low-altitudes. The whole  low-altitude portion for
$p<1$bar is made of big particles of $\approx 32\mu$m which are
decelerated by the inward increasing gas density (dashed line, right
axis).

The equatorial terminator that forms clouds on WASP-18b is in the west  terminator seen from the dayside
($\phi=-90^o$).  Its cloud top is made of subsequent thin layers of
TiO$_2$[s], Al$_2$O$_2$[s], CaTiO$_3$[s] and Fe[s]-dominated and
become more and more a mix of many materials. The major vertical portion of the cloud is
made of Mg$_2$SiO$_4$[s]/MgSiO$_3$[s]/MgO[s]/SiO[s]/Fe[s] with
inclusions from the other materials. The bottom of the cloud changes
from SiO[s] dominated, to Fe[s], then Al$_2$O$_2$[s], and then
CaTiO$_3$[s] dominated. These cloud bottom layers are composed of the
biggest particles of $\approx 60 \mu$m falling with a fall speed of
1.5 cm s$^{-1}$ (dashed line, right axis).
      
            \begin{figure*}
   \centering
   \includegraphics[width=\textwidth]{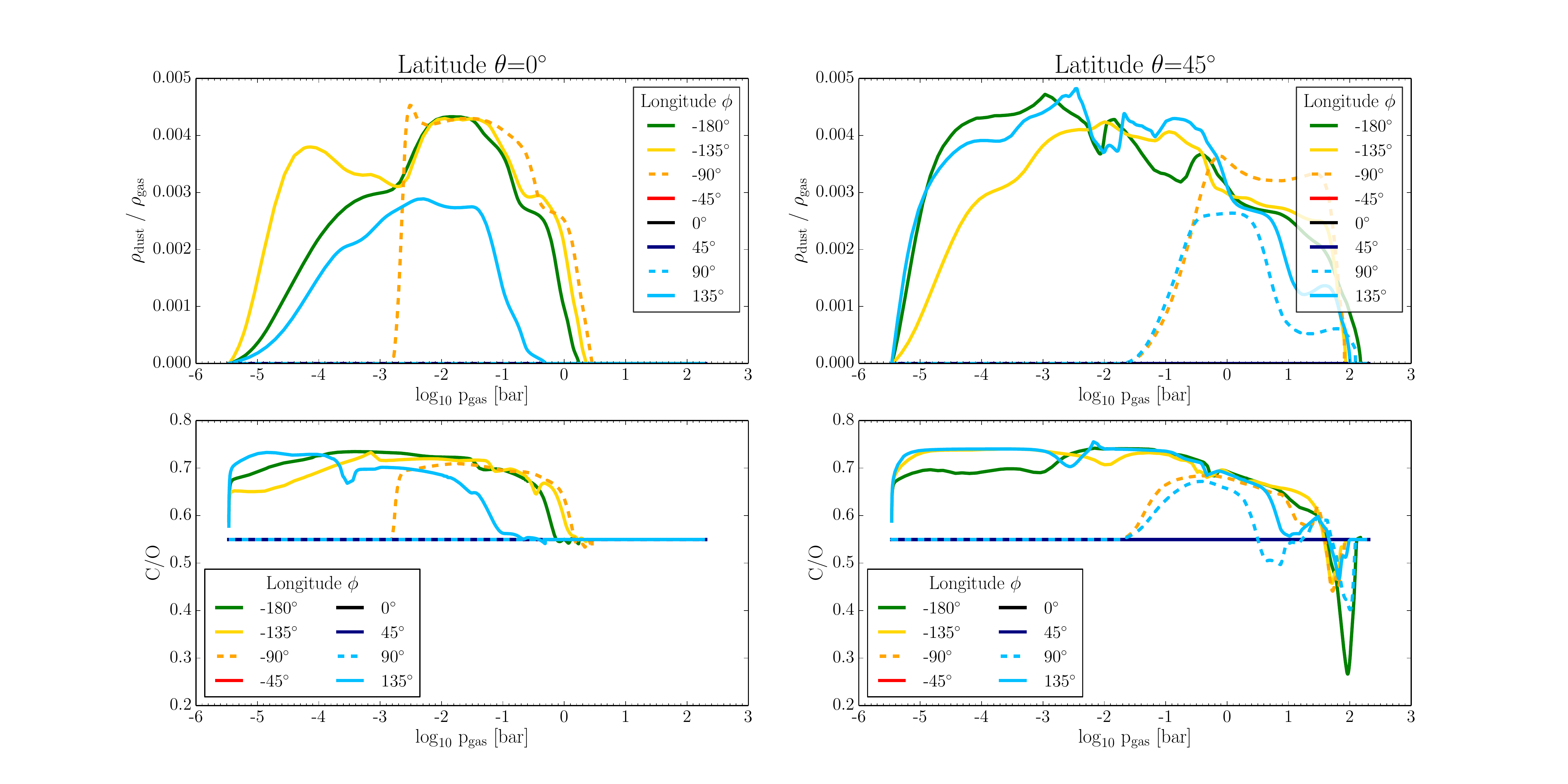}
   \caption{The cloud particle load   in terms of mass density ratios (top; $\rho_{\rm dust}/\rho_{\rm gas}$) and the C/O element ratio
     (bottom) at the equator (left) and in the norther hemisphere
     (right) of the WASP-18b atmosphere profiles studied.  C/O remains at the solar value of 0.53 for profiles without clouds forming ($\phi=-45^o, 135^o$). For
     other mineral rations like Fe/Si, Mg/Si etc., please refer
     to Appendix~\ref{a:chem} (Figs.~\ref{O}\,--~\ref{Ti}).}
   \label{C/O}
      \end{figure*}

   \begin{figure*}
      \includegraphics[angle=0, width=1.1\textwidth]{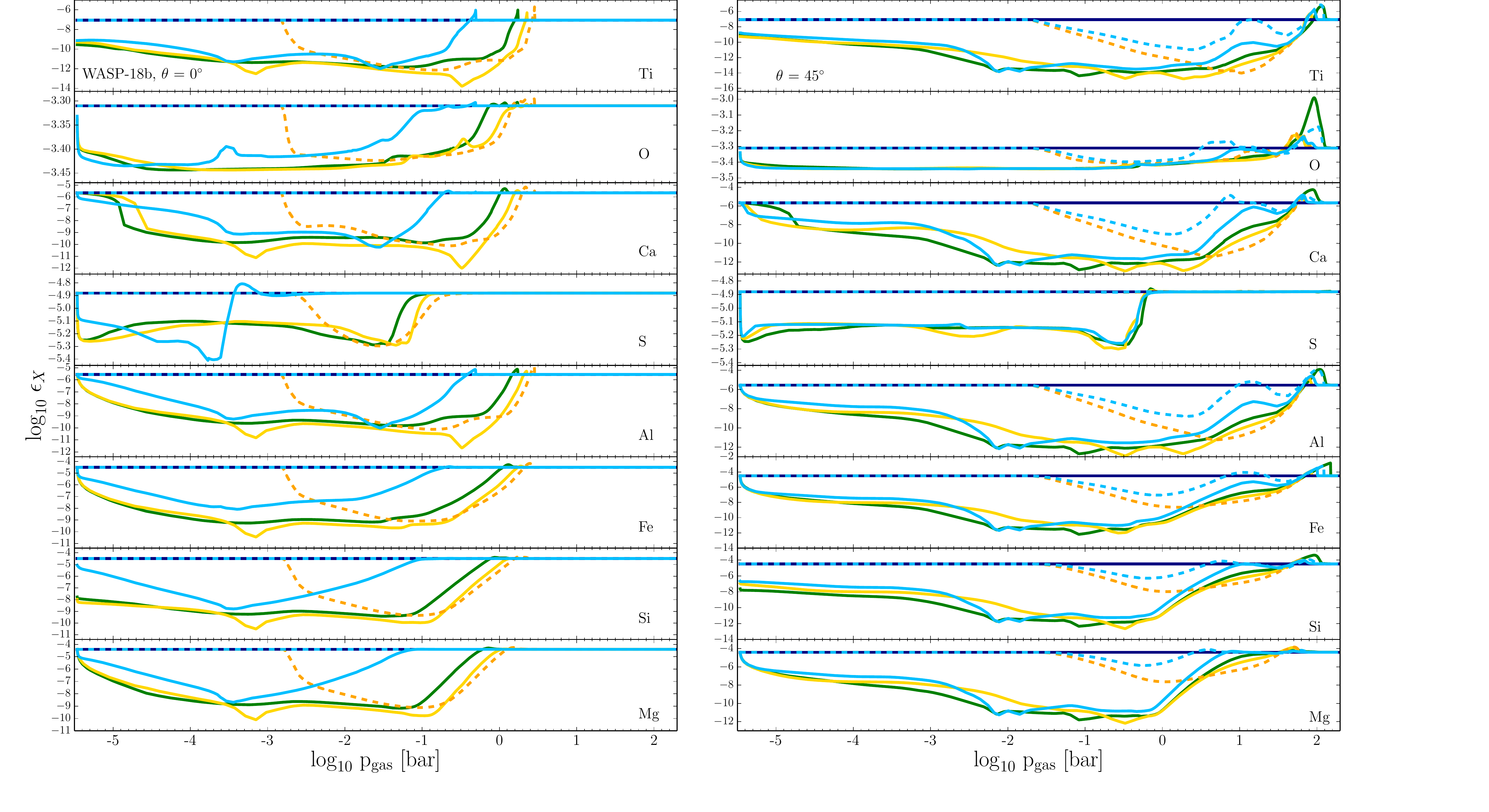}
   \caption{The abundances, $\epsilon_{\rm x}$, of those element
     involved in cloud formation for the equatorial (left, $\theta=0^o$) and the
     northern hemisphere (right, $\theta=45^o$) profiles of WASP-18b. The colour code for the atmospheric profiles is that same as in Fig.~\ref{C/O} and all figures throughout the paper. The day-night terminators are depicted as dashed lines. Straight lines indicate no element depletion for these longitudes. Cloud formation mostly causes element depletion. Element enrichment occurs where cloud particles evaporate, most noticeable for $\epsilon_{\rm O}$. The element
     abundances change due to cloud formation throughout the
     atmosphere and across the globe.  The clouds affect the element
     abundances over a larger pressure range at the norther (southern)
     hemisphere compared to the equatorial region. }
   \label{elm}
   \end{figure*}

 \begin{figure*}
\hspace*{-0.8cm} \includegraphics[width=1.03\textwidth]{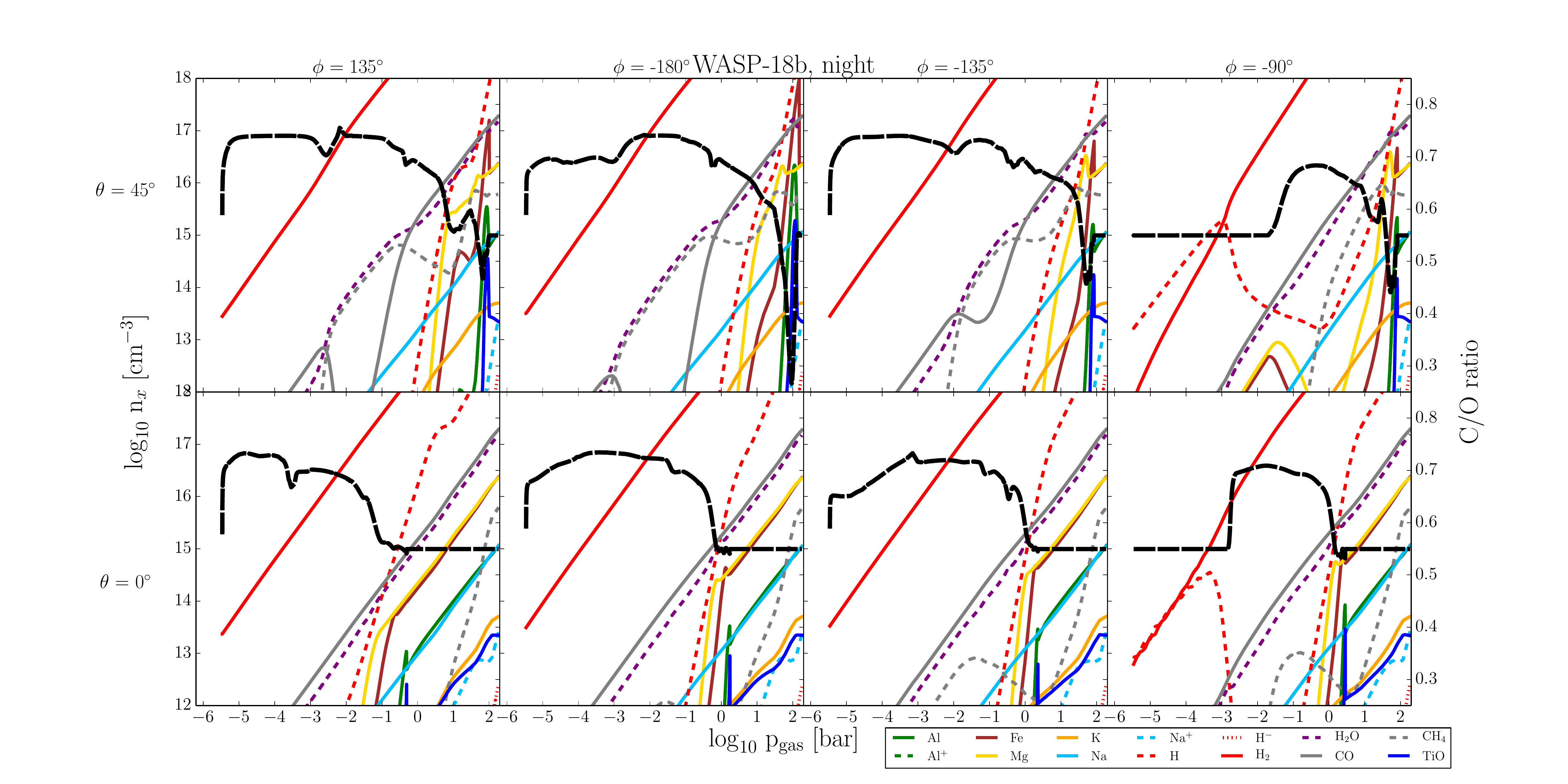}\\*[-0.2cm]
\hspace*{-0.8cm}    \includegraphics[width=1.03\textwidth]{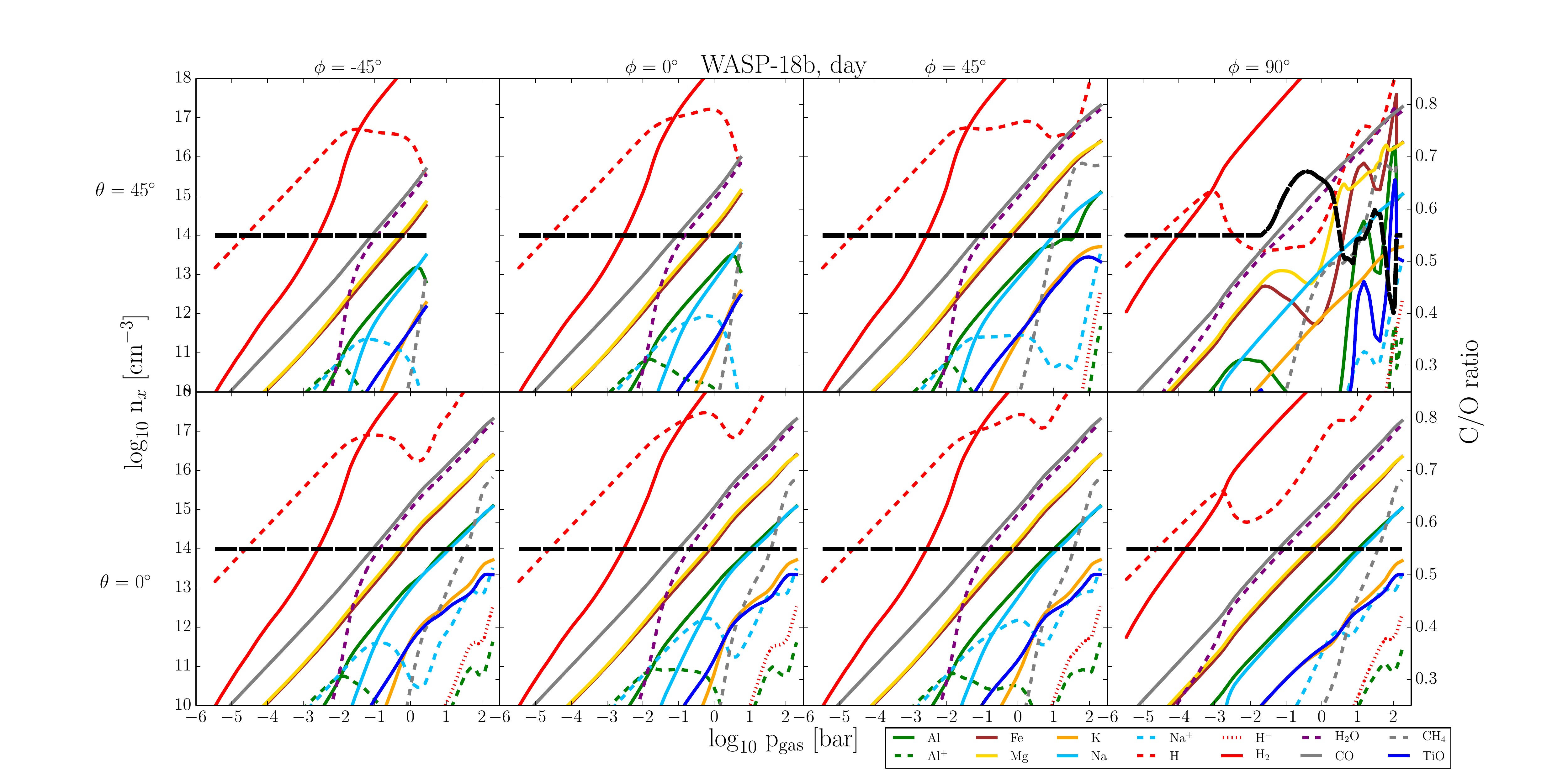}
   \caption{Composition of the atmospheric gas in chemical equilibrium with respect to the
     most abundant molecules, atoms and ions (number density n$_{\rm x}$ [cm$^{-3}$]) from the night (top) to the dayside  (bottom) on WASP-18b. The day- and the nightside have very different chemical atmosphere gas composition. The dayside is dominated by atomic H, the nightside by H$_2$. The next most abundant gas species are CO, H or H$_2$ depending ($\theta, \phi$). C/O is over plotted to visualise where clouds form. C/O remains at its initial (solar) value on the dayside where no clouds are forming, but increases on the nightside.  A complete account of all species per element is
     provided in Appendix~\ref{a:chem}.}
   \label{atomsions}
 \end{figure*}

      \subsection{Cloud particle load of the atmosphere of WASP-18b}

An interesting measure regarding the cloud particle load of the
atmosphere is the so-called dust-to-gas mass ratio, $\rho_{\rm
  dust}/\rho_{\rm gas}$ (Fig.~\ref{C/O}, top). This measure for the
enrichment of a gaseous medium with condensates (often solid) is
widely used in disk modelling where it is set to a constant value
(e.g. \citealt{2018MNRAS.tmp.1107A}), or to study the dust enrichment
of cometary tails
(e.g. \citealt{2010A&A...522A..63F,2011Icar..213..280L}) or AGB star
envelopes
(e.g. \citealt{2000A&A...358..651H,2011A&A...531A.148R}). \cite{2017arXiv171201010W}
have demonstrated that the dust-to-gas mass ratio changes with
temperature for a system in thermal equilibrium (i.e. everything that
can condense has condensed). The main contributors are materials
involving Al, Ca, Fe, Si and Mg. Condensates composed of titanium,
nickel, vanadium, chromium, manganese, sodium and potassium contribute
only very little as their initial element abundances are very low. The
maximum value reached in thermal equilibrium is $\rho_{\rm
  dust}/\rho_{\rm gas} \approx 0.004\,\ldots\,0.0052$ for gases
$>500$K and (Fig. 5 in \citealt{2017arXiv171201010W}).

The cloud particle load of the atmospheric gas in WASP-18b varies
vertically with the maximum of $\rho_{\rm dust}/\rho_{\rm gas} \approx
0.0042$ being reached at $p\approx 10^{-2}$bar in the equatorial
region and $\rho_{\rm dust}/\rho_{\rm gas} \approx 0.0049$ at lower
pressure of $p\approx 10^{-3}$bar in the mid-latitudes on the
nightside.  $\rho_{\rm dust}/\rho_{\rm gas}$ clearly shows that the
cloud layers are located much farther inside the atmosphere and are
considerably less extended at the day-night-terminators. This
indicates the transitional character of the planet's limbs where a hot
dayside transits into a cold nightside. No one $\rho_{\rm
  dust}/\rho_{\rm gas}$ can describe the profiles studied here.
However, the $\rho_{\rm dust}/\rho_{\rm gas}$ is of the order of
$10^{-3}$ for all probed profiles suggesting a globally rather
homogeneous cloud particle load in regions where clouds form.

\section{The global changes of the atmospheric gas composition on WASP-18b}\label{gasc}

The large temperature (and pressure) difference between the day- and
the nightside on WASP-18b and the transitional character of the
terminator regions lead to a distinct cloud distribution on
WASP-18b. We therefore expect the atmospheric gas composition on
WASP-18b to be vastly different on the day- and the nightside, too.
We have considered 15 materials which affect the element abundances of
the 9 (Mg, Si, Ti, O, Fe, Al, Ca, S, C) elements by cloud particle
growth and evaporation by 126 surface reactions. The growth process
reduces the element abundances, the evaporation process enriches the
element abundances. The nucleation process has only very little affect
on the element involved (Si, Ti, O, C), but it is strongly affected by element depletion.

 \subsection{Element depletion and enrichment}

One of the most important outcomes of a cloud formation model is the
feedback on the gas phase through element depletion or enrichment. The
essential quantities to start with are the element abundances as the
number of elements needs to be conserved for each element
individually. Figures~\ref{elm} demonstrates the consistency of our
cloud formation model because cloud formation only affects the
involved elements at the nightside and at the day/night-terminators.

Each of the elements is individually depleted and the depletion varies
widely through the cloud structure (along the pressure axis) and
across the globe. The largest variations in element depletion for any
one element occurs along the equator for WASP-18b (Fig.~\ref{elm}, left plots). The
day/nightside terminators stand out by depleting a less extended
atmosphere of the vertical atmosphere at the equator and in the
hemispheres.

A closer look at Fig.~\ref{elm} shows that the individual depletion of
the elements varies by order of magnitudes. Ti is strongest depleted
but it also has the smallest initial element abundance of all involved
cloud forming elements. Some elements show a substantial enrichment at
the inner cloud boundary located at low-altitudes due to the evaporation of cloud particles
that have been falling until such high temperatures. Most noticeable
is the effect in the mid-latitudes (right of Fig.~\ref{elm}) for
Ti, O, Ca, Al and Fe. Si is enriched, too, compared to the initial solar
values but far less than the other elements. Therefore, element
enrichment occurs in deep, unobservable layers of the
atmosphere. Observable layers are expected to be depleted in all elements
that are involved in cloud formation.

Oxygen is one of the most abundant elements in a solar set of
element abundances and it is involved in almost every growth material
that easily forms. We therefore portrait the oxygen-depletion also as
C/O ratio in Fig.~\ref{C/O}. 

\subsection{C/O and [X/Si] element ratio}\label{ss:XSi}

The most prominent element ratio studied in the literature is the
carbon-to-oxygen ratio, C/O. C/O is of interest with respect to planet
formation scenarios and the link to the chemistry in planet forming
disk (e.g. \citealt{2014Life....4..142H,2018A&A...613A..14E}).  C/O,
Mg/Si and Fe/Si are discussed in the literature as control parameter for the
amount of carbides and silicates formed in planets
(\citealt{2017Ap.....60..325A,2018arXiv180109474S}).
Our results show that the thermodynamic properties of the
local gas determine the condensation processes from which element
ratios like C/O (Fig.~\ref{C/O}) and X/Si (Figs.~\ref{O}\,--~\ref{Ti})
result.

The largest change in the C/O throughout the cloud-affected part of
the atmosphere occurs in the northern/southern hemisphere. Here the
larges and the smallest C/O is reached. C/O is bound between 0.28 and
0.73. Only a C/O ration above 0.93 will allow first carbon-binding
molecules to emerge (TiC; \citealt{2017arXiv171201010W}).
\cite{2018arXiv180202576B} show that C/O varies between 0.4 and 0.6
for the solar twins in the solar neighborhood.  No one atmospheric C/O will
suffice to describe the atmosphere of WASP-18b. First, it varies
vertically throughout the atmosphere and it varies globally. C/O
remains solar at the dayside where no clouds form and becomes
enriched/depleted on the cloud-forming nightside. The C/O at the
night side shows the change from a thermally stable to a thermally
unstable cloud. C/O increases above the initial (solar) value of 0.53
in the thermally stable part of the atmosphere but decreases
substantially where the cloud particles evaporate. The increase in C/O
is mainly caused by oxygen consumption and the decrease in C/O is mainly caused by
oxygen being transported into deeper atmospheric regions due to
falling cloud particles. The width of the O-enriched zone shows that
the cloud particles do not evaporate instantaneously. If carbon-materials form, carbon depletion/enrichment will affect the C/O ratio, too. Our calculations show the condensation of only very little carbon on the nightside of WASP-18b  (5-10\%; Fig~\ref{Vsnight}). 

Similar behaviour occurs for all elements involved in the cloud
formation processes and the detailed results for [X/Si] are provided in
Appendix~\ref{a:chem} where each figure shows the gas-abundance and
the respective X/Si curve (thick, black dashed line).
\cite{2018arXiv180202576B} show that their Mg/Si for solar twins in
the solar neighborhood are greater than one. Our initial, undepleted
(Mg/Si)$^0$ = 1.23\footnote{\cite{2018arXiv180202576B,2010ApJ...715.1050B}
  cite (Mg/Si)$^0$=1.05 based on the element abundance data from
  \cite{2005ASPC..336...25A} with $\epsilon^0_{\rm Mg}=7.53$. We use
  \cite{2009ARA&A..47..481A} with $\epsilon^0_{\rm Mg}=7.6$. Both
  sources have used $\epsilon^0_{\rm Si}=7.51$.} and cloud formation
pushes Mg/Si as high as 8 at the inner cloud boundary at the  $\phi=-90^o$
terminator (Fig.~\ref{Mg}), and well above 4 at the inner cloud
boundary of all other cloud-affected profiles probed in our
study. Mg and Si are very correlated in the regions where both
participate in cloud formation, but the inner, low-altitude cloud regions are more
affected by SiO than any Mg-binding condensate. Hence, the large Mg/Si
occur where Si is still strongly depleted which is also reflect in the
Si/O ratio (Fig.~\ref{Si}).  Fe/Si shows similar features in regions
where Si is strongly depleted, like in the nucleation zone at
$\phi=-134^o$ (Fig.~\ref{Fe}). In principle, the same story is told by
all the other mineral ratios. Examples like K/Si and Na/Si demonstrate
very well where the Si depletion kicks in because K and
Na are not depleted (Figs.~\ref{K},~\ref{Na}) as they do not condense
in the temperature regimes of WASP-18b.  Consequently, the lowest
metalicity gas exhibits the highest Mg/Si. The detailed finding for
the mineral ratios are provided in Appendix~\ref{a:chem}.

\cite{2017arXiv171201010W} investigated the change of the gas-phase C/O
ratio in the limiting case of thermal stability which allows to
consider the effect of complex materials like phyllosilicates. Their
results demonstrate that the condensation of Mg/Si/O-binding minerals
increases C/O to $>0.7$, and that a further increase to $>0.8$ is
caused if phylosilicates are included. They also demonstrate that
deriving the C/O ratio from a gas only made of H/C/N/O as elements
leads to false results compared to the whole set of solar elements.

The changing element ratios demonstrate that an atmospheric ratio (as
for C/O, for example) will differ from the bulk value or in fact from
any value derived for the warmer, deeper atmospheric regions. The
pristine, unaltered atmospheric element ratios should be recovered. An
added complication may arise for an extended inner, low-altitude cloud which is
enriched by elements (C/O decreases) rather than depleted (C/O
increases).

\begin{table*}[h!]
  \begin{center}
    \caption{An account of the dominating gas-phase species per element for WASP-18b}
    \label{tab:table1}
    \begin{tabular}{p{1cm}|p{4.0cm}|p{4.3cm}|p{4cm}|p{4cm}} 
      & \textbf{Day} & \textbf{Night} & \multicolumn{2}{|c}{\textbf{Terminators}}\\
      \hline
      & & & \textbf{Longitude $\phi=-90^{\circ}$} & \textbf{Longitude $\phi=90^{\circ}$}\\
      \hline
      Al & p<10$^{-2}$bar: Al$^{+}$\newline p>10$^{-2}$bar: Al\newline at high $p_{\rm gas}$: AlH, AlOH
      & AlOH, AlO$_{2}$H\newline at high $p_{\rm gas}$: Al, Al$^{+}$
      & $p_{\rm gas}<10^{-2}$bar: Al\newline $p_{\rm gas}>10^{-2}$bar: AlOH\newline at high $p_{\rm gas}$: Al, AlH, AlOH
      & \underline{\textbf{Lat $\theta=45^{\circ}$:}}\newline $p_{\rm gas}<10^{-2}$bar: Al\newline $p_{\rm gas}>10^{-2}$bar: AlOH\newline  at high $p_{\rm gas}$: Al, AlH, AlOH.\newline
      \underline{\textbf{Lat $\theta=0^{\circ}$}}:\newline $p_{\rm gas}<10^{-3}$bar: Al$^{+}$\newline $p_{\rm gas}>10^{-3}$bar: AlH, AlOH\newline at high $p_{\rm gas}$: Al, AlH, AlOH \\*[0.2cm]
      Ca & $p_{\rm gas}<10^{-2}$bar: Ca$^{+}$\newline $p_{\rm gas}>10^{-2}$bar: Ca
      & $p_{\rm gas}$<10$^{-3}$bar:\,Ca(OH)$_{2}$,\,CaCl$_{2}$\newline $p_{\rm gas}$=10$^{-3}...<1$bar: Ca(OH)$_{2}$\newline $p_{\rm gas}>$1bar: Ca
      & Ca\newline if clouds: Ca(OH)$_{2}$, CaCl$_{2}$ & Ca \\*[0.2cm]
      C & CO & CO, CH$_{4}$ & CO ; CH$_{4}$ at high p & CO ; CH$_{4}$ at high p \\*[0.2cm]
      Fe & Fe, Fe$^{+}$ & Fe; low $p_{\rm gas}$: Fe(OH)$_{2}$\newline high  $p_{\rm gas}$: FeH & Fe & Fe \\*[0.2cm]
      H & $p_{\rm gas}<10^{-1}$bar: H\newline $p_{\rm gas}>10^{-1}$bar: H$_{2}$ & H$_{2}$; at high $p_{\rm gas}$: H & H$_{2}$ ;  $p_{\rm gas}<10^{-3}$bar : H & H$_{2}$ ; p<10$^{-3}$bar: H \\*[0.2cm]
      Mg  & Mg ;  $p_{\rm gas}<10^{-2}$bar: Mg$^{+}$\newline \textit{(many differences for Mg at each profile)}& \underline{\textbf{Lat $\theta=45^{\circ}$:}}\newline  $p_{\rm gas}$<10$^{-3}$ \&$p_{\rm gas}$>10$^{-1}$: Mg\newline  $p_{\rm gas}$=10$^{-3}...10^{-1}$bar: Mg(OH)$_{2}$\newline \underline{\textbf{Lat  $\theta=0^{\circ}$ :}}\newline Mg\newline $p_{\rm gas}<10^{-4}$bar: Mg(OH)$_{2}$\newline highest $p_{\rm gas}$: MgH & Mg\newline MgH at high $p_{\rm gas}$ & Mg\newline MgH at high $p_{\rm gas}$ \\*[0.2cm]
      O & CO, $p_{\rm gas}$<10$^{-2}$bar: O\newline $p_{\rm gas}>10^{-2}$bar: H$_{2}$O & CO, H$_{2}$O & CO\newline all $p_{\rm gas}$/ $p_{\rm gas}$>10$^{-3}$bar: H$_{2}$O & CO; $p_{\rm gas}>10^{-3}$: H$_{2}$O\\*[0.2cm]
      Si & $p_{\rm gas}<10^{-2}$: Si\newline $p_{\rm gas}>10^{-2}$: SiO & SiS, SiO & SiO & SiO \\*[0.2cm]
      S & low $p_{\rm gas}$: S\newline high $p_{\rm gas}$: H$_{2}$S & H$_{2}$S & $p_{\rm gas}<10^{-3}$: S\newline $p_{\rm gas}>10^{-3}$: H$_{2}$S & $p_{\rm gas}<10^{-3}$: S\newline $p_{\rm gas}>10^{-3}$: H$_{2}$S \\*[0.2cm]
      Ti & $p_{\rm gas}<10^{-2}$: Ti$^{+}$\newline $p_{\rm gas}>10^{-2}$: Ti, TiO & $p_{\rm gas}<1$bar: TiO$_{2}$\newline  $p_{\rm gas}>1$bar: TiO\newline highest $p_{\rm gas}$: Ti & TiO\newline TiO$_{2}$ when clouds form & TiO\newline  $p_{\rm gas}<10^{-3}$/ highest $p_{\rm gas}$: Ti
    \end{tabular}
  \end{center}
  \label{tab:gph}
\end{table*}

 \begin{figure*}
   \centering
   \includegraphics[width=\textwidth]{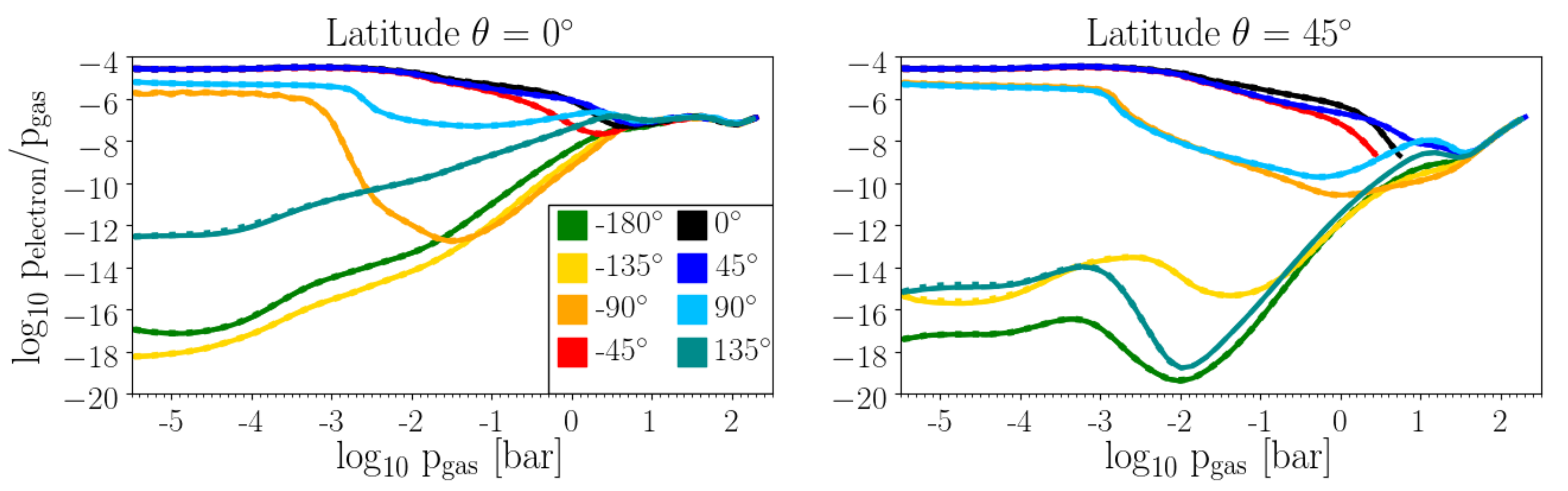}
   \caption{ WASP-18b's dayside ionosphere due to a partially ionised
     gas in the atmosphere ($\phi= -90^o, -45^o, 0^o, 45^o, 90^o$): The degree of thermal ionisation from the day to the night
     side at the equator (left) and in the northern hemisphere (right)
     of WASP-18b. The ionosphere more extended in the equatorial regions ($\theta=0^o)$ than in  northern hemisphere ($\theta=45^o$).}
   \label{fe}
 \end{figure*}

 \subsection{The composition of the neutral gas-phase on WASP-18b}

The most direct information about an atmosphere can be derived if gas
species can be observed spectroscopically because individual atoms and
molecules are finger prints for specific temperature and pressure
regimes. Such a direct access is hampered if clouds form, but WASP-18b
provides us with a cloud-free dayside. Figures~\ref{H}-~\ref{Ti}
provide a detailed account of the gas-phase composition of the
atmosphere of WASP-18b sorted by elements. Each of these plots
contains an element ratio (C/O, Si/O, Mg/Si, Na/Si etc.) which allows
to trace the cloud forming regions as discussed in
Sect.~\ref{ss:XSi}. Here, we confine ourselves to features of general
and also of specific interest in order to build our understanding for
the chemical composition in such thermodynamically diverse
planets.

The most abundant species of the collisional dominated parts of the atmospheric gas on WASP-18b are in
decreasing order H$_2$, CO, SiO, H$_2$O, SiS, and MgH on the
dayside. This hierarchy changes somewhat on the nightside to: H$_2$,
CO, H$_2$O, CH$_4$, H$_2$S. The most abundant gas-phase species are
neutral molecules despite the large temperature difference between day
and night (Fig.~\ref{atomsions}). A summary of
dominating gas species is provided in Table~\ref{tab:table1}.

Despite H$_2$ being the most abundant molecule, it is not everywhere
the most abundant H-binding species. This has recently been pointed
out by \cite{2018ApJ...855L..30A}.  The dayside of WASP-18b is
dominated by atomic hydrogen, H, up to $10^{-2}$bar, and at both day/night terminators up to $10^{-3}$bar (Fig.~\ref{H}). We note that
at the equatorial  $\phi=-90^o$-terminator, H and H$_2$
appear with very similar number densities.  The nightside is H$_2$
dominated as the gas temperature is too low for thermally dissociating
H$_2$.

For the O-complex, CO does dominate the gas phase on the dayside,
followed by atomic oxygen in the outer layers and by H$_2$O in regions
of high densities. The night side is affected by the thermodynamics of
the atmosphere and element depletion of oxygen through the formation of
silicates.  The drop in CO correlates well with the peak in C/O which
indicates efficient cloud formation and it is compensated by an
increase of CH$_4$ and H$_2$O (see $\phi=-135^o, -180^o$,
Fig.~\ref{O}).

Aluminium and titanium are elements that play a key role for cloud
formation as they form materials that are thermally stable up to rather
high temperatures compared to Mg/Si/Fe/O-silicates. Both, the Al and
the Ti complex (Figs.~\ref{Al},~\ref{Ti}) replicate what is commonly
observed for alkali metals (Na, K, Ca): Their positive ions become
more abundant than the neutral atom, in fact, than any of their
neutral counterparts: The Al-complex (Fig.~\ref{Al}) is dominated by
AlH in the high-altitude atmosphere on the nightside, Al$^+$/Al at the
dayside and Al/AlOH/AlH at the day/night terminator. The equatorial
area at $\phi=-90^o$ is dominated by atomic aluminium.  The Ti-complex
(Fig.~\ref{Ti}) is dominated by TiO$_2$ on the nightside ($\phi=135^o,
-180^o, -135^o$), Ti$^+$/Ti at the dayside and Ti/TiO at the limbs ($\phi=90^o,-90^o$)

Neither Fe not Si exhibit an ion that is more abundant than any of their neutral atoms or molecules

The C-complex is dominated by CO on almost all profiles except for
the northern profiles at the nightside  (Fig.~\ref{C}).

Figure~\ref{atomsions} demonstrates how profoundly the globally changing thermodynamic structure affects the local molecular number densities (day-night difference) and how strongly the cloud formation affects the actual values (the over plotted C/O serves as guid for the cloud location on the pressure axis) of atoms and molecules.

\subsection{The dayside ionosphere of the partially ionised atmosphere of WASP-18b}

Brown dwarfs have a long tradition of being studied as analogs for
giant gas planets because spectral observations are more feasible than
for extrasolar planets (e.g. \citealt{2018ApJ...854..172C}). Brown
dwarfs irradiated by white dwarfs have recently been discovered and
emission lines that origin from heated upper atmospheric regions at high altitudes have
been observed from the irradiated brown dwarf WD0137-349B
(\citealt{2017MNRAS.471.1728L}).  While the specific ionised species
(Fe$^+$ vs. Na$^+$) observed will be determined by the temperature
that can be reached in these hot, outer regions, 
WASP-18b has a hot enough dayside for atomic ions to emerge.

\smallskip
\noindent
{\it Ions dominating:} The temperature on the day side of WASP-18b is high enough that
various elements appear in their second ionisation state (singly
ionised) as shown in Fig.~\ref{atomsions} (top). The most abundant
ions that are also more numerous than the neutral atoms are Na$^+$,
K$^+$, and Ca$^+$. Also Al$^+$ and Ti$^+$ are more abundant than their
atomic form (Fig.~\ref{Al}, ~\ref{Ti}) but are far less numerous than
Na$^+$, K$^+$, or Ca$^+$. H$^-$ is far more abundant at the dayside
and in the terminator regions than at the nightside, but never the
dominating H-species. This supports the finding in
\cite{2018ApJ...855L..30A}.  Figure~\ref{atomsions} has over-plotted
the total electron number density (dark gray) demonstrating that the
most important electron donors on the dayside of WASP-18b are Mg, Fe,
Al, Ca, Na, and then K.  Mg and Fe had been identified as dominating
electron donors in the inner, warmest  low-altitude parts of non-irradiated giant
gas planet and brown dwarf atmospheres
(\citealt{2015MNRAS.454.3977R}).

\smallskip
\noindent
{\it Possible emission:} We note that the atomic hydrogen abundance,
as well as that of Al$^+$, Ti$^+$, and Fe$^+$ (and atomic O, C, Si)
increases outward where the local temperature increases outwards on
WASP-18b. This may suggest that WASP-18b shows emission from H$\alpha$
(and O, C, Si, Al$^+$, Ti$^+$, Fe$^+$) from its dayside and
terminator regions.  Possible emission lines could include $\lambda = 4243.47, \ldots, 5262.02, \ldots, 7494.75, \ldots, 8286.72 \AA$ for Fe$^+$ (Fe II),  $\lambda = 6231.06, \ldots, 6822.69, \ldots 8925.81\AA$ for Al$^+$ (Al II),  $\lambda = 8445.44 \AA$ for O (OI) (e.g. as seen in post-AGB stars by  \citealt{2018MNRAS.481.3935A}),  $\lambda = 4571. 98, 5129.15\AA$ for  Ti$^+$ (Ti II),  $\lambda = 8824.221\AA$ for Fe (Fe I) (e.g. as seen in accretion outburst of an M5-dwarf with a protoplanetary disk by \citealt{2017A&A...607A.127S}).  The occurrence of these lines will depend on the local temperature and density, and a proper radiative transfers calculation will be required for providing a more profound theoretical support for this suggestion. H$\alpha$ is used as measure for planetary mass
loss. The possibility of mass loss on ultra hot Jupiters was discussed
in \cite{2018arXiv180500038L} for the hottest example among the super-hot Jupiters,  KELT-9b.

\smallskip
\noindent
{\it Dayside ionosphere on ultra-hot Jupiters:} The
changing state of thermal ionisation of the individual elements causes
a difference of 15 order of magnitudes in the local degree of thermal
ionisation, $f_{\rm e}$, between the day and the nightside on
WASP-18b (Fig.~\ref{fe}). The nightside is cool enough that only little
thermal ionisation occurs in almost the entire vertical atmosphere.
The degree of ionisation on the dayside reaches an approximately
constant level at $p<10^{-2.5}$bar of $f_{\rm e}\approx
10^{-4.5}$. Most of the  high-altitude atmosphere on the dayside and at the
day/night terminators has $f_{\rm e}> 10^{-7}$. The value of $f_{\rm
  e}> 10^{-7}$ has been suggested to be a threshold above which a gas
may exhibit plasma behavior without being fully ionised
(\citealt{2015MNRAS.454.3977R}). Hence, the dayside of WASP-18b can
be expected to show plasma phenomena, incl. magnetic field coupling.
This conclusion is in line with works on other ultra-hot
Jupiters. For example does \cite{2018AJ....156...17K} conclude that a magnetic drag
force is required to explain the phase curve of WASP-103b.

The dayside of WASP-18b reaches temperatures as high as 3500K which
is too low to ionised the atomic hydrogen that dominates the daysides
high-altitude atmosphere thermally (Fig.~\ref{H}). The nightside is completely
neutral and all elements appear in their first (neutral) state of
ionisation. The dayside is composed of a partially ionised gas which
has a electron number density comparable to and above Earth's
ionosphere ($10^6$cm$^{-3}$). We therefore postulate that WASP-18b and
other ultra-hot Jupiters like WASP-121b, WASP-12b, WASP-103b,
WASP-33b, Kepler-13Ab, Kepler 16b, KOI-13b, MA-1b have an ionosphere
on their dayside that stretches into the terminator regions. Planets
right of the 20\% line in Figure 13 in \cite{2018arXiv180500096P} will
fall into this category. 

Would the nightside develop an ionosphere, too, for example due to
scattered XUV photons from  the host star (stellar XUV  has a huge effect on the
mass loss of small planets; e.g. \citealt{2018MNRAS.478.1193K}) or other high-energy irradiation from e.g. the interplanetary environment? It is
reasonable to expect that external radiation increases the ionisation
of the atmospheric gas in its upper, high-altitude regions similar to what has been
shown for brown dwarfs. A considerable increase of the degree of
ionisation results in a thin shell of an almost or fully ionised gas if the  irradiation were comparable to the ISM radiation field
(\citealt{2018arXiv180409054R}). Cosmic rays will not significantly
contribute to the formation of an ionosphere because they are rather efficiently attenuated
(\citealt{2013ApJ...774..108R}).

\begin{figure*}
   \centering
   \includegraphics[width=\textwidth]{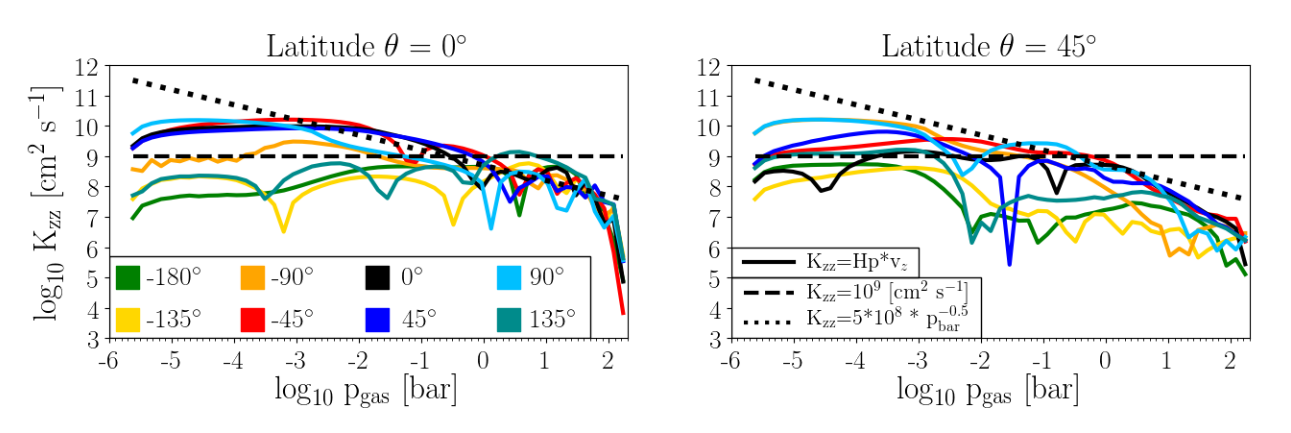}
   \caption{The vertical diffusion coefficient, $K_{\rm zz}$, calculated from various parametrisations:  $K_{\rm zz}=H_{\rm p}v_{\rm z}$, $K_{\rm
  zz}=10^9$ cm$^2$s$^{-1}$, $K_{\rm zz}=5\cdot 10^8 p^{-0.5}$ cm$^2$ s$^{-1}$.  The representation $K_{\rm zz}=f(p)$ is the same as derived in \cite{2013A&A...558A..91P}. }
   \label{Kzz}
\end{figure*}
 
\begin{figure*}
   \centering
   \includegraphics[width=\textwidth]{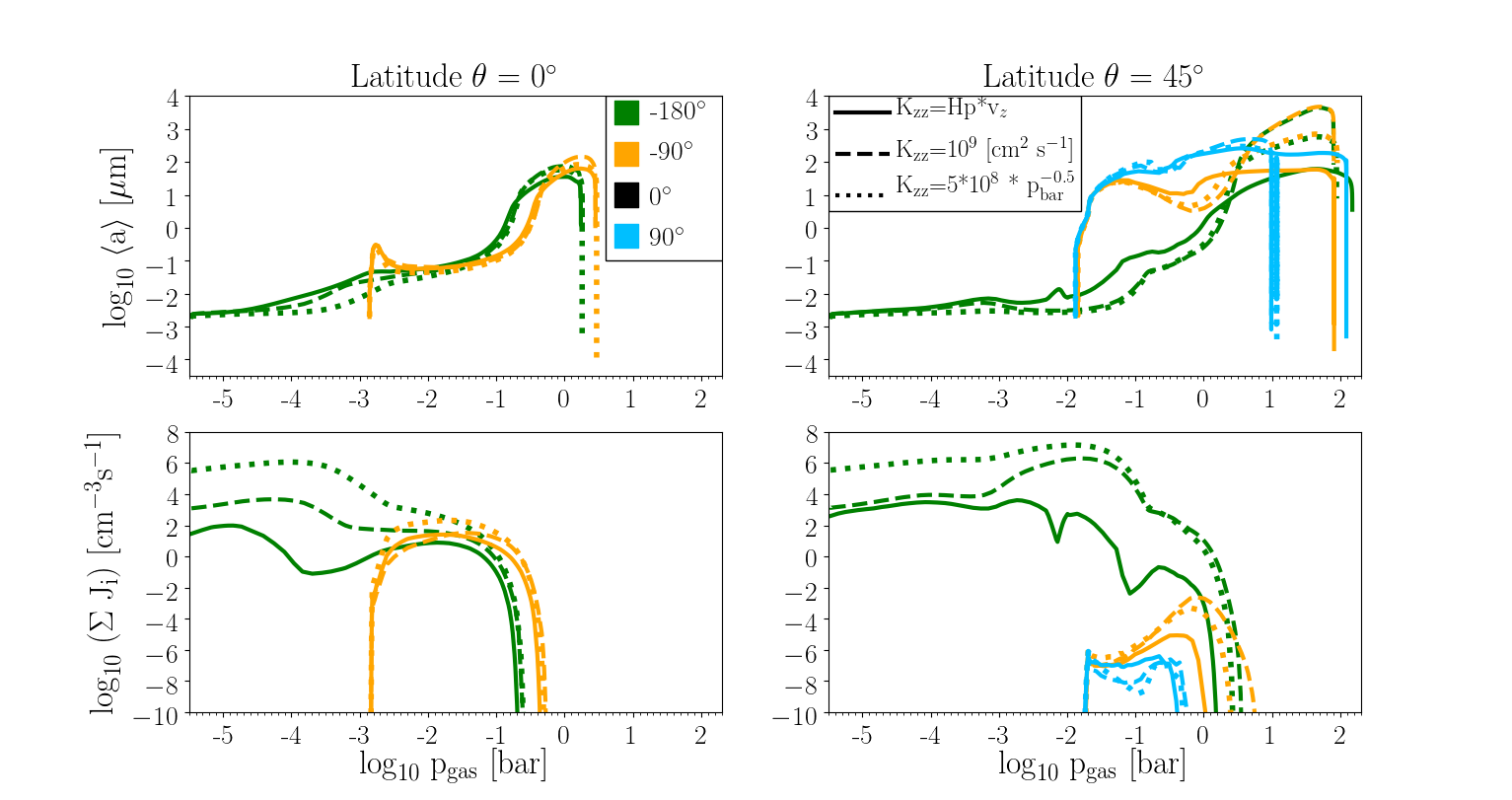}\\*[-0.0cm]
   \caption{Different element replenishment representations and the nucleation rate, $J_* = \Sigma J_i$, and the local mean particle size, $\langle a\rangle$, of WASP 18b.}
   \label{compKzz1}
 \end{figure*}

\begin{figure*}
   \centering
      \includegraphics[width=\textwidth]{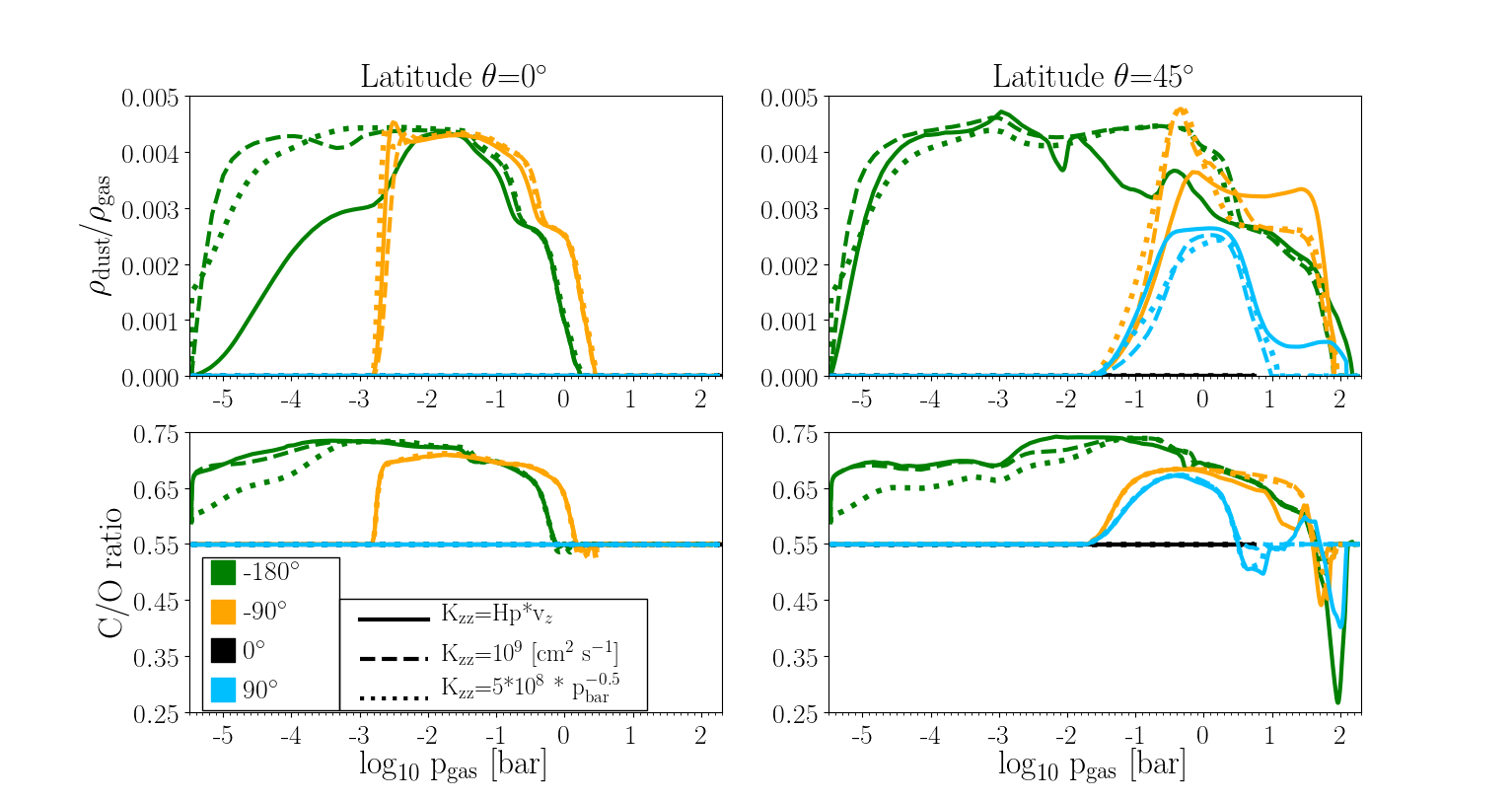}
   \caption{Different element replenishment representations and global cloud properties of WASP-18b: $\rho_{\rm d}/\rho_{\rm gas}$ (top) and C/O ratio (bottom) .}
   \label{compKzz2}
\end{figure*}

\section{Discussion}\label{s:discu}

\subsection{Element replenishment representations}\label{s:Kzz}

The solution of the Navier-Stokes equations in combination with a
radiative transfer, gas phase chemistry, element conservation and
cloud formation enables a parameter-free solution. Only material
properties and numerical parameters remain to be adjusted. Not so in
1D. 1D approaches are numerically fast, hence, can be run on a high
cadence as required, for example, in the retrieval approaches
(e.g. \citealt{2017ApJ...848..127B}). Our cloud formation model
provides us with the tool that we need to predict cloud properties
based on a fundamental understanding of microphysical processes like
cluster formation, frictional interaction, surface reactions. But we
have demonstrated that a replenishment mechanism is required to
describe cloud formation in a 1D quasi-static atmospheric environment
(\citealt{Woitke2004}, Appendix A). Therefore, element replenishment
is parameterised in 1D cloud simulations. \cite{2008MNRAS.391.1854H}
and \cite{2018ApJ...854..172C} summarized the approaches applied in
the brown dwarf / giant gas planet literature where $\tau_{\rm
  mix}\sim H_{\rm p}(z)^2/K_{\rm zz}$.  Here, we test our classical
approach of $K_{\rm zz}=H_{\rm p}v_{\rm z}$ against a constant $K_{\rm
  zz}=10^9$ cm$^2$s$^{-1}$ and a scaling with the local pressure
$K_{\rm zz}=5\cdot 10^8 p^{-0.5}$ with p in [bar]. The $K_{\rm
  zz}$-approach originates from modelling mixing as diffusion process,
and the $\tau_{\rm mix}$-approach models mixing as large-scale
convection.  The $K_{\rm zz}$-scaling with the local gas-pressure was
derived from a 3D cloud-free GCM for HD209758b which has a different
temperature structure as demonstrated in
Fig.~\ref{tp}. Figure~\ref{Kzz} shows how the different
parameterisations vary for the 1D profiles used here.

Figures~\ref{compKzz1} show the nucleation rate and the mean cloud
particle sizes for the three different $K_{\rm zz}$
parameterisations. Different profiles are shown in different
colors, and the different $K_{\rm zz}$ cases are visualised by
different line stiles. The cloud results for the day/nightside
terminators remain rather unchanged for the three approaches tested
because the nucleation rates are very similar.  For all other
profiles, the largest differences occur to the approach that
scales with the local gas pressure (dotted lines). The nucleation rate
differs here up to 4 orders of magnitude compared to the constant
value (dashed lines). The consequence is that the mean particles sizes
vary but just by maximally 1-1.5 order of magnitudes in the inner, low-altitude, hence optically tick 
cloud. The cloud particle load of the atmosphere reflects this, too.

While it is elusive to discuss which of the approaches could be
called 'correct', the comparison provides some idea about
uncertainties that are imposed by differences in $K_{\rm zz}$. We have
demonstrated this in terms of cloud particle load, $\rho_{\rm
  d}/\rho_{\rm gas}$, and in terms of the C/O ratio
(Fig.~\ref{compKzz2}). The differences in the mixing/diffusion don't
have much effect on the results at the equatorial, cloud-forming
terminator ($\phi=-90^o$) and on the northern terminator at
$\phi=90^o$ regarding $\rho_{\rm d}/\rho_{\rm gas}$ and C/O. The
largest uncertainties occur  at the high altitudes of the cloudy
atmosphere for both, the cloud itself and the feedback on the gas
phase.

We note that the three approach give roughly the same answer, but the
parametrized Kzz's give a much smoother variation of parameters like
the particle size. Hence, small scale variations in the cloud
properties are due to local variations of $K_{\rm zz}=H_{\rm p}v_{\rm
  z}$ and are possibly not too realistic. Furthermore, horizontal
mixing can not be considered in the approach followed in this paper,
but will homogenize the global element abundance and cloud particle
distribution to a certain extend in  the  higher altitude regions  where the 
gas pressure is low if the wind blows into the right direction (night $\rightarrow$ day) and thermally stability prevails. For example, the cloud particle size differences
between the equator and the mid-latitudes, and the limb and the
anti-stellar point would be lessened if the gas is supersaturated. For
WASP-18b, the dayside will remain cloud free as no material achieves
supersaturation for temperatures as high as $>2200$K (\citealt{2017arXiv171201010W}). This
condition worthens with decreasing pressure. {\it It will therefore
  be essential to measure the atmospheric C/O on the day and on the
  nightside in order to link the atmospheric C/O to the bulk C/O of
  extrasolar planets.}

\subsection{Comparing WASP-18b to HD\,189733b and HD\,209458b}

Despite the limits to our approach, we can offer some general
comparisons between the three giant gas planets that we have studied
so far with respect to cloud formation and chemistry feedback. This comparisons should
not go into details as the 3D models for HD\,189733b and HD\,209458b
take the clouds consistently into account for the gas-phase chemistry
and for the radiative transfer, while this is not yet the case for
WASP-18b.

WASP-18b with $T_{\rm eq}\approx 2400$K is far hotter than HD\,189733b
and HD\,209458b with $T_{\rm eq}\approx 1000$K and $T_{\rm eq}\approx
1500$K, respectively, and its surface gravity is one order of
magnitude larger. Figure~\ref{tp} shows, that despite these
substantial differences in global parameters, the night side of all
three giant gas planets exhibit very comparable temperatures. And all
three planets develop temperature inversions on their dayside (rather
shallow in HD\,209458b). In the low-pressure regions of the nightside
profiles, WASP-18b has higher pressures for any given local gas
temperature. Higher pressures enable condensation processes at higher
temperatures as thermal stability expands into higher temperature
regimes with increasing pressure.

The overall material composition of the cloud particles are very
similar in all three planets, though differences do emerge in
the details. The  high-altitude atmosphere is dominated by Mg/Si/O-materials with
$\approx 15$\% of Fe[s]. The nucleation species determine the
composition of the thin, uppermost cloud boundary. WASP-18b has a
distinct SiO[s] layer where Mg/Si/O-materials have evaporated. This
SiO[s] layer does not appear in HD\,189733b and HD\,209458b.

The overall mean particle sizes are comparable and span a similar
range in all three planets. The actual height-dependent size
distribution does vary between the planets. For example 
WASP-18b's $\phi=-90^o$-terminator show only very minimal changes of cloud
particles sizes with height compared to all other profiles
sampled. Here the material matrix is dominated by SiO[s] over most of
the cloud volume, except at the cloud top where distinct layers of
Fe[s] and CaTiO$_2$[s] appear.

The total  vertical cloud extension reaches farther into the low-pressure
atmosphere on WASP-18b despite having a surface gravity that is one
order of magnitude larger than that of HD\,189733b. HD\,209458b has
clouds that reach into the lowest pressure regimes: The upper
nightside cloud boundary is located at $p_{\rm gas}\approx
10^{-5.5}$bar on WASP-18b, at $p_{\rm gas}\approx 10^{-4.2}$bar on
HD\,189733b and at $p_{\rm gas}\approx 10^{-7}\,...\,10^{-5}$bar on
HD\,209458b (compare Fig.~\ref{Vsnight} and Fig.~8 in
\cite{2016MNRAS.460..855H}). Our test of mixing prescriptions in
Sect.~\ref{s:Kzz} suggests that the upper cloud boundary is not
affected by mixing here and is hence determined by the local
thermodynamic conditions.

 \section{Conclusion}\label{con}
WASP-18b provides us with a laboratory to study the atmosphere
chemistry of ultra-hot Jupiters that have very hot days and cold nights.

 Hot Jupiters do not have all their constituents in the gas phase, and
 cloud formation strongly affects the chemistry on the nightside and
 on the day/night terminators. The WASP-18b dayside is hot enough to
 thermally dissociate H$_2$, and the elements Na, K, Ca, Ti, Al as
 well as Fe, Mg, Si but to a lesser degree. VO, TiO and H$_2$O are not among the most abundant gas-phase species on the dayside which is in line with the non-detection in HST secondary eclipse observations.  TiO and H$_2$O are important for gas pressures above 10$^{-2}$ bar on the dayside. The low-density regime of the dayside has CO but also atomic species like Si and S, and ions like Na$^+$, Ca$^+$, K$^+$.  The low-pressure regime of the terminator gas-phase chemistry is made of bi-atomic molecules like CO, SiO, TiO and atoms like for example Fe and Mg. We further find that:

 \begin{itemize}
\item  WASP-18b has two very different sides: The nightside is cloudy and
 element depleted, the dayside is cloud-free and forms a thermal
 ionosphere that reached deep into the atmosphere.
   
  \item The largest cloud particles form at the $\phi=-90^o$ (west) day/night terminator, and superrotation causes a cloud-free $\phi=90^o$ (east) day/night
    terminator.

    \item Clouds become more extended towards the west of the
      nightside. Clouds are located farther inside the atmosphere and
      less extended at the limbs compared to the nightside.

\item The cloud particle load (dust-to-gas-ratio $\rho_{\rm
  d}/\rho_{\rm gas}$) is rather homogeneous of the order of $10^{-3}$
  despite large temperature difference in the cloud forming regions of
  WASP-18b.

 \item  Element enrichment occurs deep inside the
 atmosphere and the observable layers appear depleted in all elements
 that participate in cloud formation. 

\item At the dayside, where no cloud formation happens, the atmospheric C/O is
  constant and  remains at its undepleted value (here: solar). In the nightside, C/O is roughly constant but
  enhanced (C/O$\sim$0.7) within the photosphere. At the limbs the C/O
  ratio varies vertically from 0.7 to 0.5 within the layers probed by
  transmission spectroscopy.  Cloud formation does enhance the C/O in general.
  
  \item The cloud particles are made of a mix of materials that changes depending on the local temperature in the atmosphere. The mix is predominantly made of O-binding minerals, smaller oxides or iron, but 5-10\% of carbon can be mixed in occasionally. The cloud particles sizes change with height and location.
  
  \item The molecular number densities vary by orders of magnitudes between the day and the nightside, and the limbs.

 \end{itemize}

 \begin{acknowledgements}
     Jacob Arcangeli is thanked for hepful discussions of the manuscript.
\end{acknowledgements}

%
%

\bibliographystyle{aa}
\bibliography{bib}

\appendix

\section{Details on chemical composition}\label{a:chem}
Here we provide the detailed composition of the gas-phase in chemical
equilibrium for the eight profiles probed at the equator and in
the northern hemisphere. We note that GCM models are north/south
symmetric such that the southern hemisphere shows the same behavior
like the northern hemisphere.

We first provide an overview of how the abundances of the dominating
molecular species change (Fig.~\ref{mols}) along the equator on the
day- and the night side. Figures~\ref{O}\,--~\ref{Ti} detail the
chemical gas-phase composition with respect to the individual elements
H, O, Al, C, Ca, Fe, Mg, Si, S, Ti. Each of these plots
(Figs~\ref{O}--~\ref{Ti}) also shows an element ratio (mineralogical
ratios) at the right axis (Al/Si, C/O, Ca/Si, Fe/Si, Mg/Si, Si/O,
S/Si, Ti/Si).

\begin{table}[h!]
  \begin{center}
    \caption{Solar element abundances, $\epsilon^0$ (\citealt{2009ARA&A..47..481A}), and the solar mineral rations [X/Si]=$\epsilon^0_{\rm X}/\epsilon^0_{\rm Si}$, Si/O and C/O. \cite{2018arXiv180202576B} and \cite{2010ApJ...715.1050B} cite Mg/Si=1.05 based on the
  element abundance data from \cite{2005ASPC..336...25A} with
  $\epsilon^0_{\rm Mg}=7.53$.}
    \label{tab:mr}
    \begin{tabular}{l |l || l| l }
      &  $\epsilon^0$ & &  $\epsilon^0_{\rm X} /\epsilon^0_{\rm Si, O} $\\
      \hline
    H & 1.000 &   H/Si  & 3.0902$\cdot 10^{4}$\\
    He & 8.511$\cdot 10^{-2}$ & He/Si & 2.6301e3\\
    Li & 1.122$\cdot 10^{-11}$ & Li/Si & 3.4672$\cdot 10^{-7}$\\
    C & 2.692$\cdot 10^{-4}$ &C/Si & 8.3189\\
    N & 6.761$\cdot 10^{-5}$ &N/Si & 2.0893\\
    O & 4.898$\cdot 10^{-4}$ &O/Si & 1.5136e1\\
    Na & 1.738$\cdot 10^{-6}$ &Na/Si & 5.3708$\cdot 10^{-2}$\\
    Mg & 3.981$\cdot 10^{-5}$ &Mg/Si & 1.2302 \\
    Al & 2.818$\cdot 10^{-6}$ &Al/Si & 8.7082$\cdot 10^{-2}$\\
    S & 1.318$\cdot 10^{-5}$ &S/Si &  0.40729\\
    Cl & 3.162$\cdot 10^{-7}$ &Cl/Si &  9.7713$\cdot 10^{-3}$\\
    K & 1.072$\cdot 10^{-7}$ &K/Si & 3.3127$\cdot 10^{-3}$\\
    Ca & 2.188$\cdot 10^{-6}$ & Ca/Si & 6.7614$\cdot 10^{-2}$\\
    Ti & 8.913$\cdot 10^{-8}$ &Ti/Si & 2.7543$\cdot 10^{-3}$\\
    Fe & 3.162$\cdot 10^{-5}$ &Fe/Si & 9.7713$\cdot 10^{-1}$\\
    Si & 3.236$\cdot 10^{-5}$ &Si/Si & 1\\
  \hline
   & & C/O & 0.5495\\
   & & Si/O & 6.6067$\cdot 10^{-2}$
    \end{tabular}
  \end{center}
\end{table}

\begin{figure*}
   \centering
   \includegraphics[width=\textwidth]{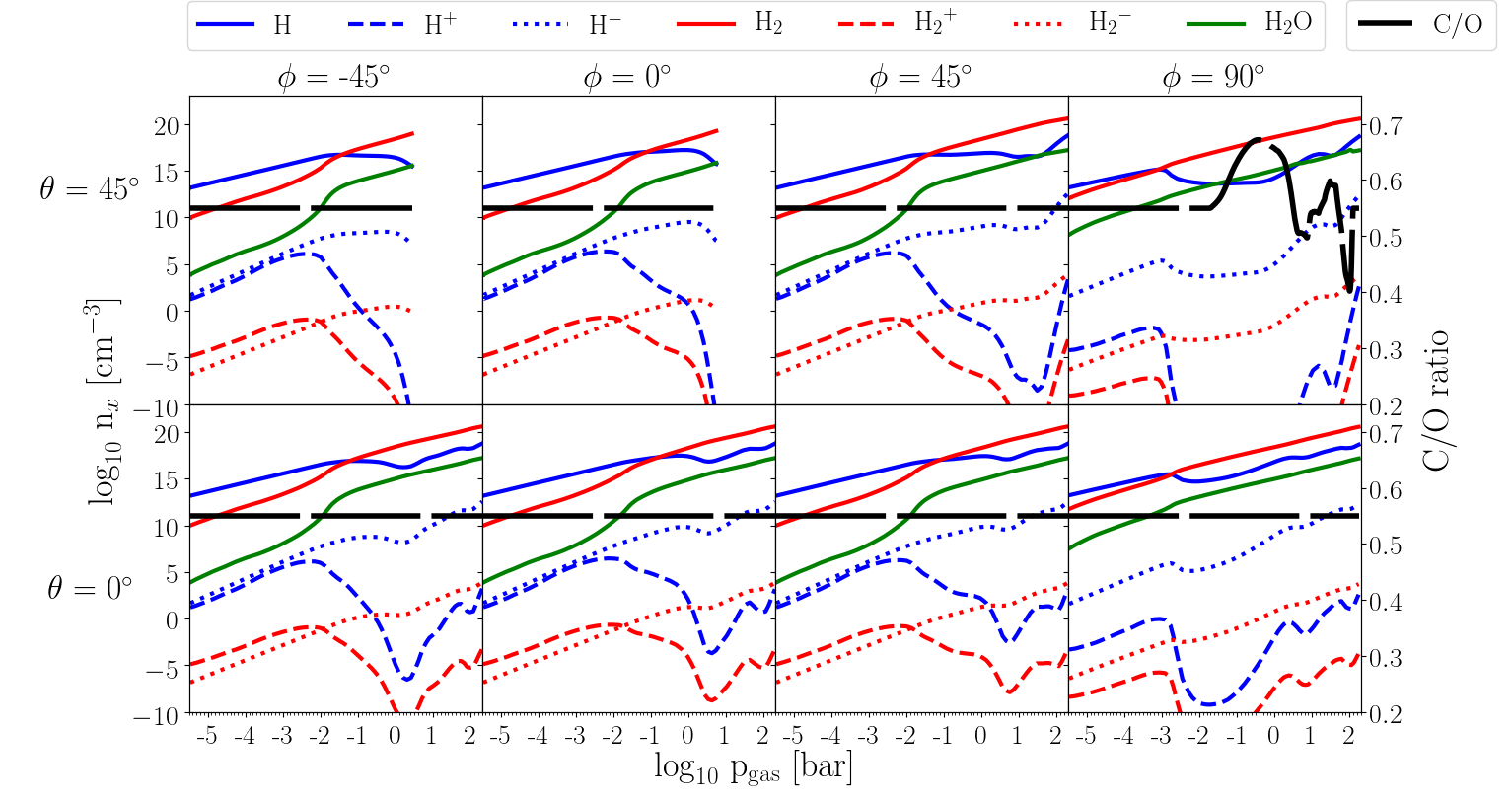}
   \includegraphics[width=\textwidth]{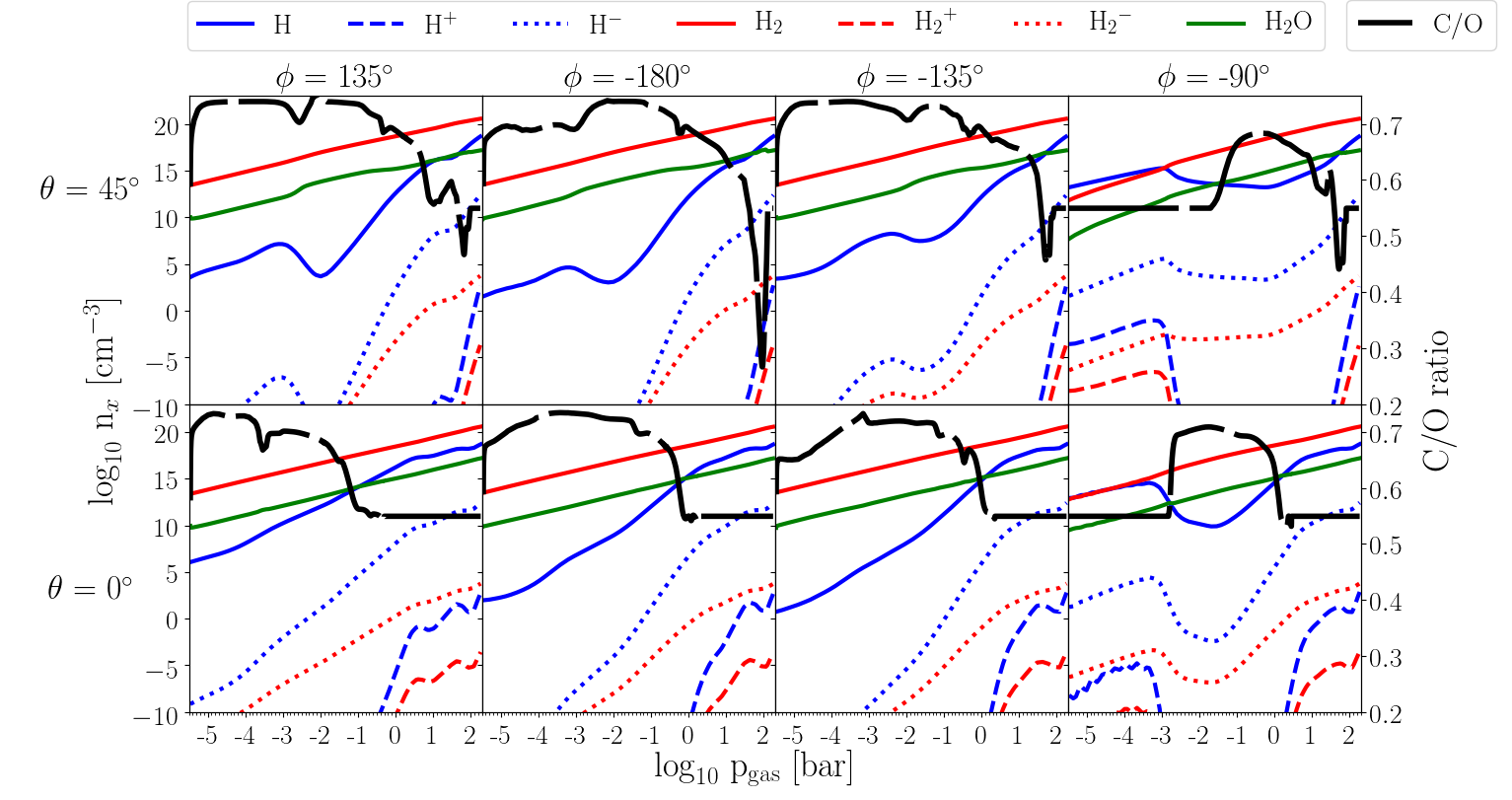}
   \caption{Number densities, log $n_{\rm x}$ [cm$^{-3}$],  of H-binding gas-species (color coded, left axis). The C/O is overplotted and shows where cloud affects the atmosphere (black long-dashed line, right axis).}
   \label{H}
\end{figure*}

\begin{figure*}
   \centering
   \includegraphics[width=\textwidth]{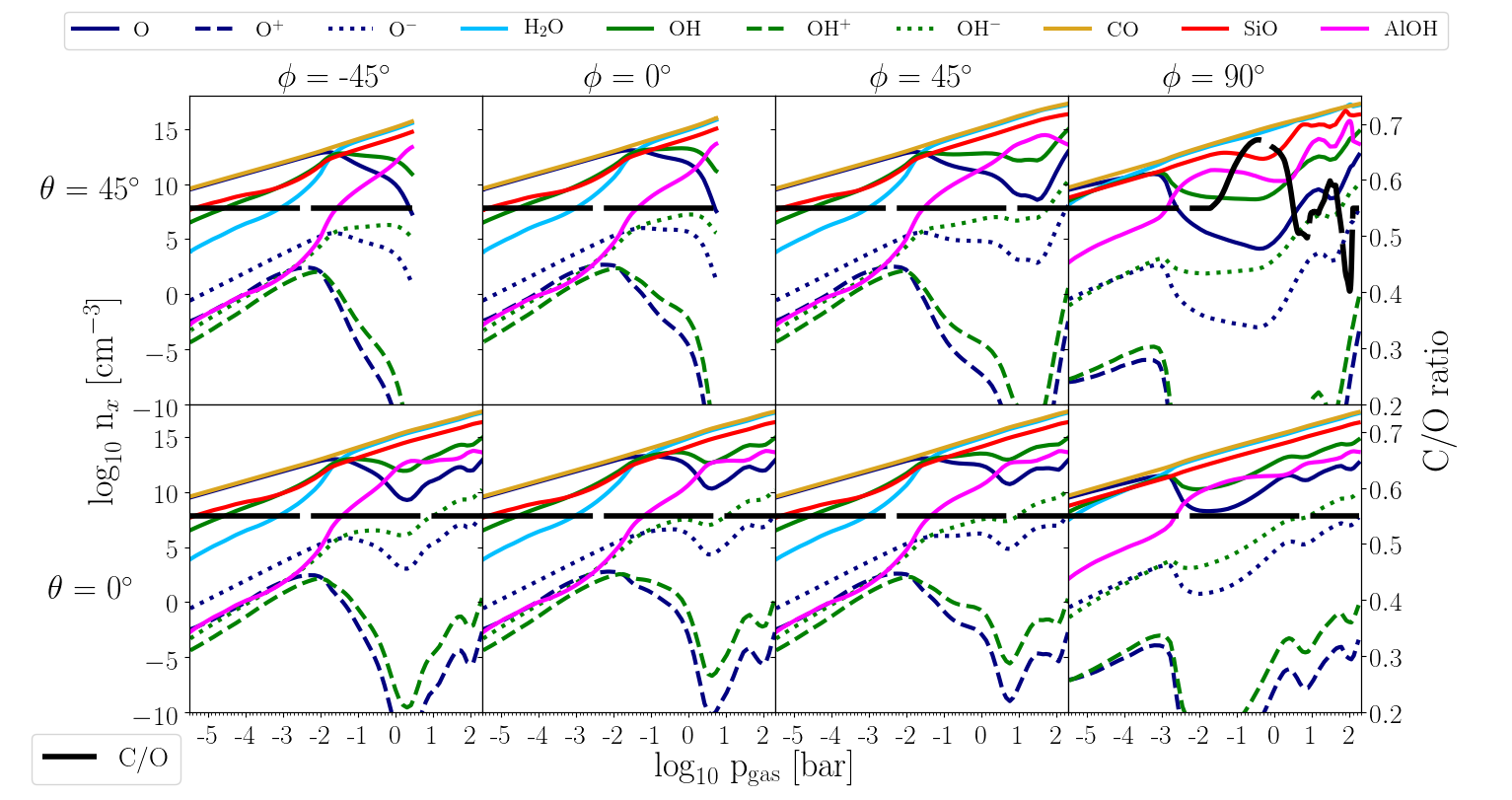}
   \includegraphics[width=\textwidth]{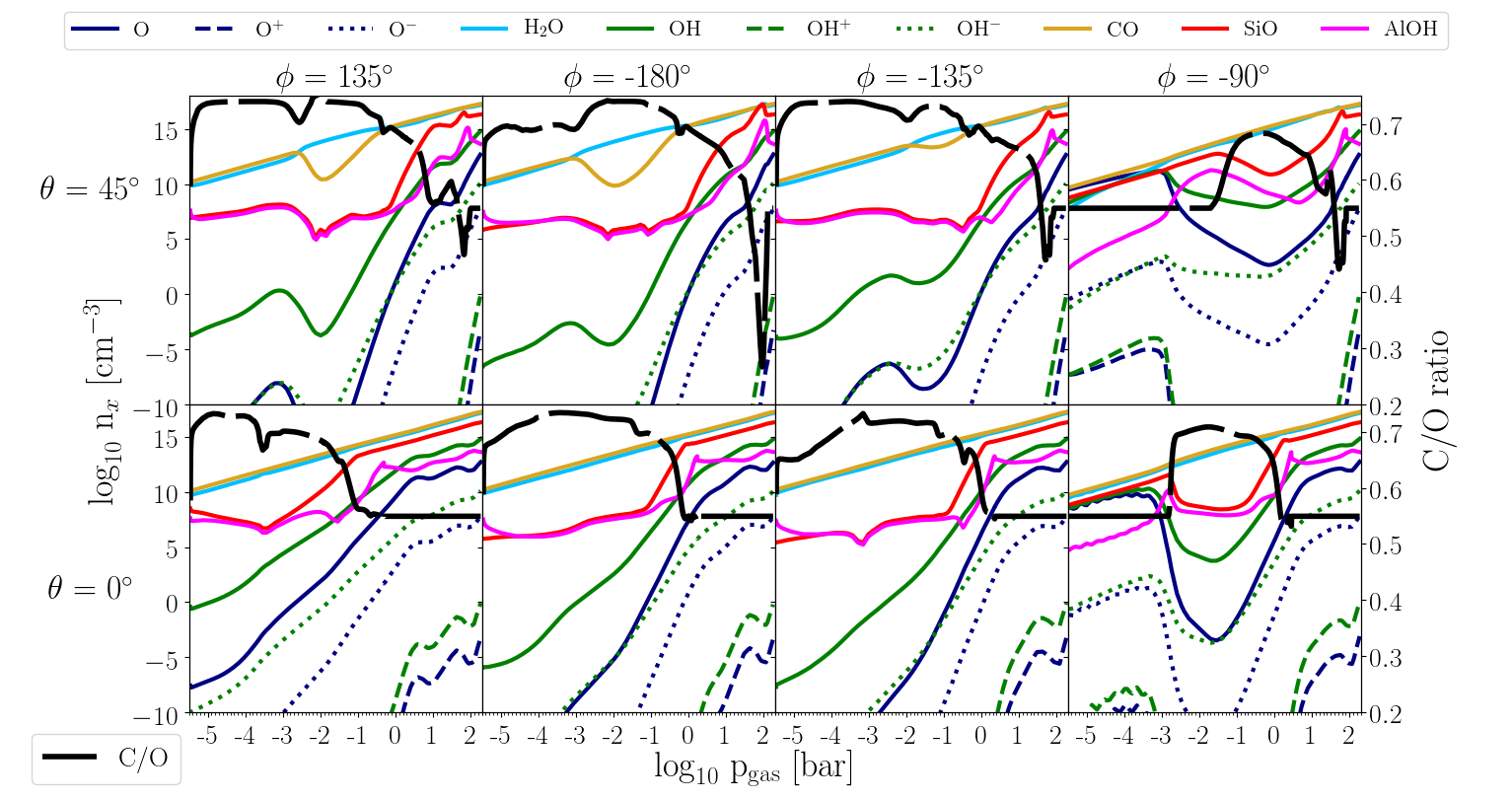}
   \caption{Number densities, log $n_{\rm x}$ [cm$^{-3}$],  of  O-binding gas-species (color coded, left axis). The C/O is overplotted and shows where cloud affects the atmosphere (black long-dashed line, right axis).}
   \label{O}
\end{figure*}

\begin{figure*}
   \centering
   \includegraphics[width=\textwidth]{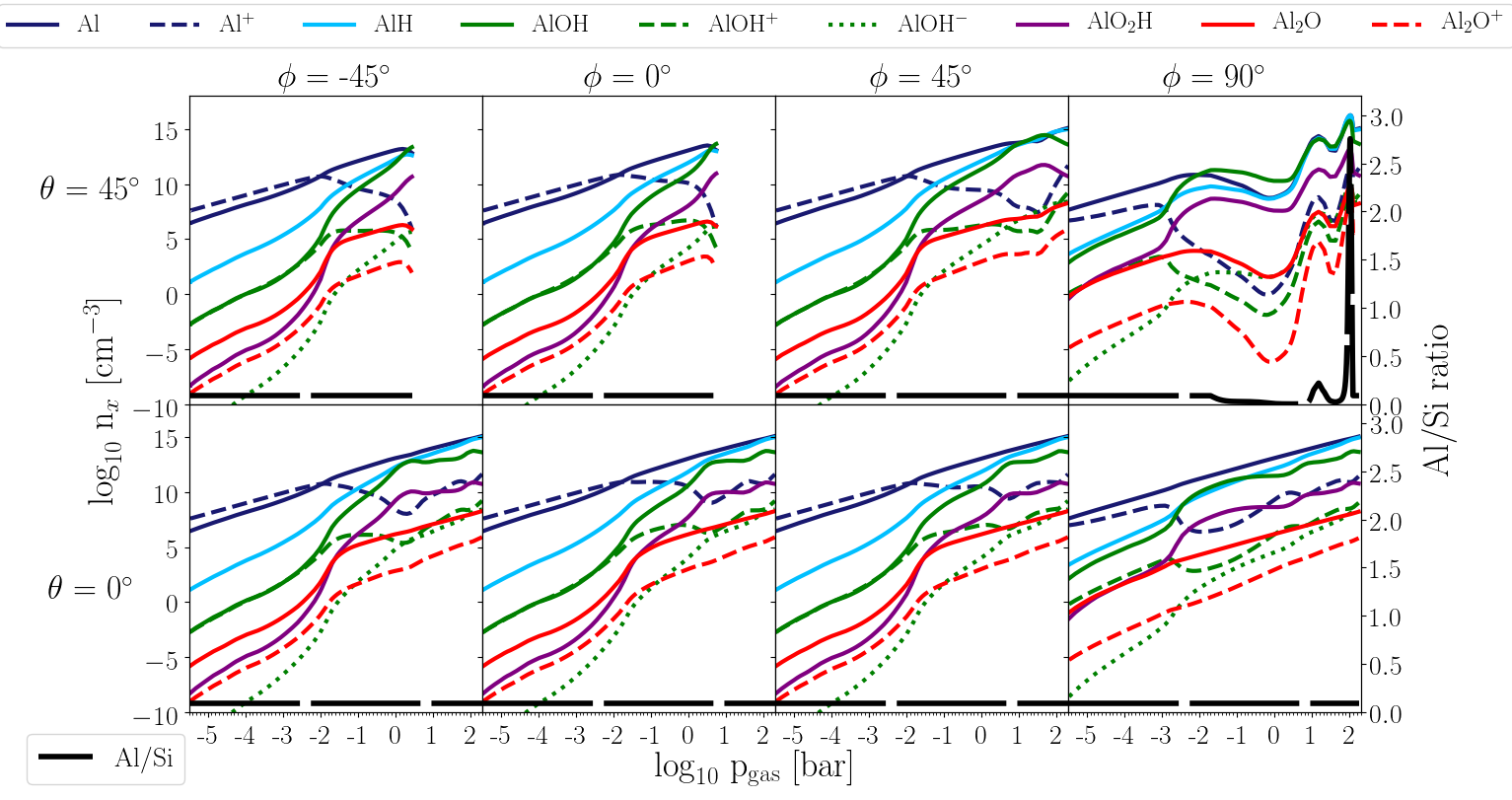}
   \includegraphics[width=\textwidth]{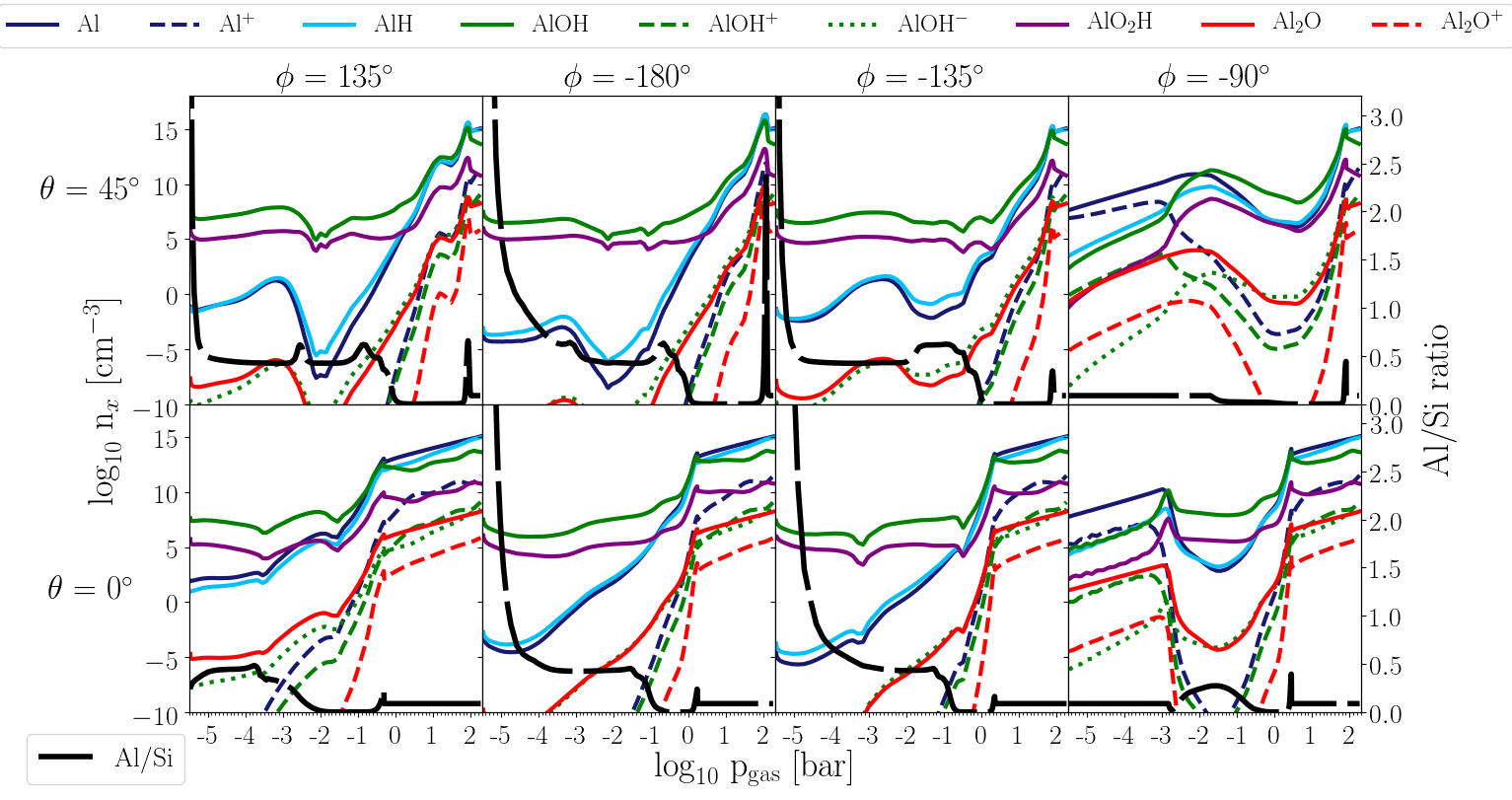}
   \caption{Number densities, log $n_{\rm x}$ [cm$^{-3}$],  of  aluminum binding gas-species (color coded, left axis). The Al/Si ratio is overplotted and shows where cloud affects the atmosphere (black long-dashed line, right axis).}
   \label{Al}
\end{figure*}

\begin{figure*}
   \centering
   \includegraphics[width=\textwidth]{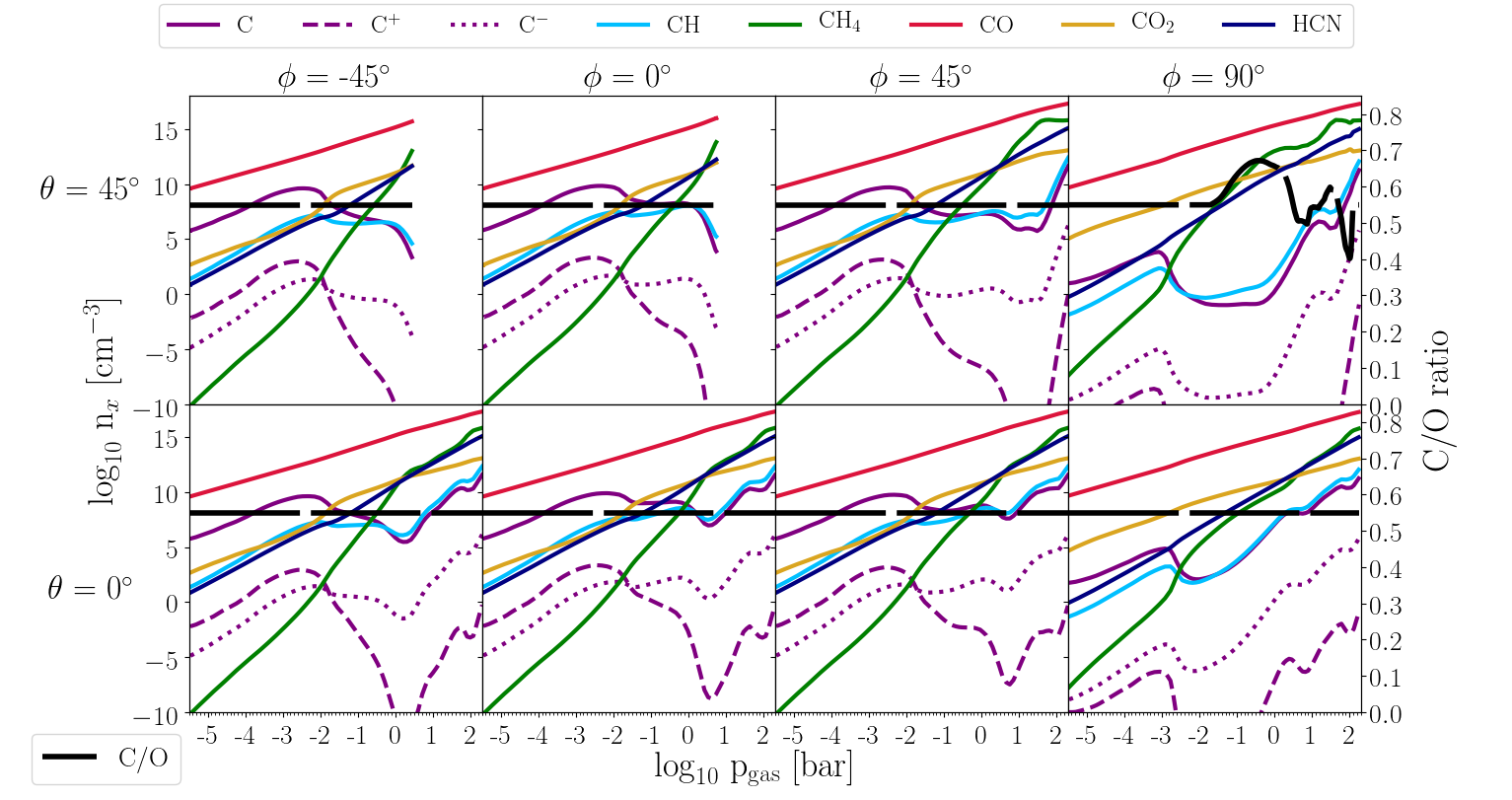}
   \includegraphics[width=\textwidth]{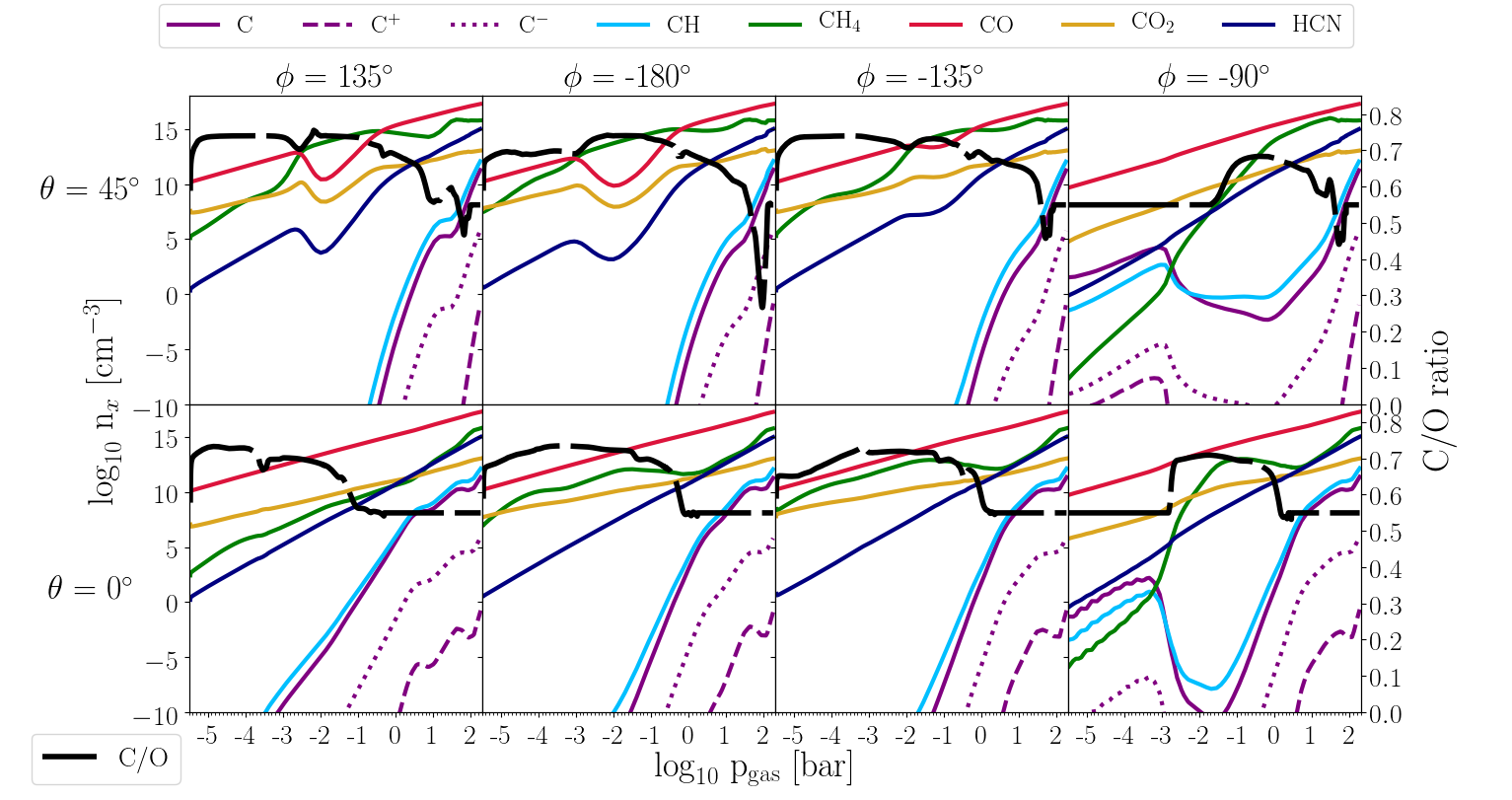}
   \caption{Number densities, log $n_{\rm x}$ [cm$^{-3}$],  of  carbon binding gas-species (color coded, left axis). The C/O is overplotted and shows where cloud affects the atmosphere (black long-dashed line, right axis).}
   \label{C}
\end{figure*}

\begin{figure*}
   \centering
   \includegraphics[width=\textwidth]{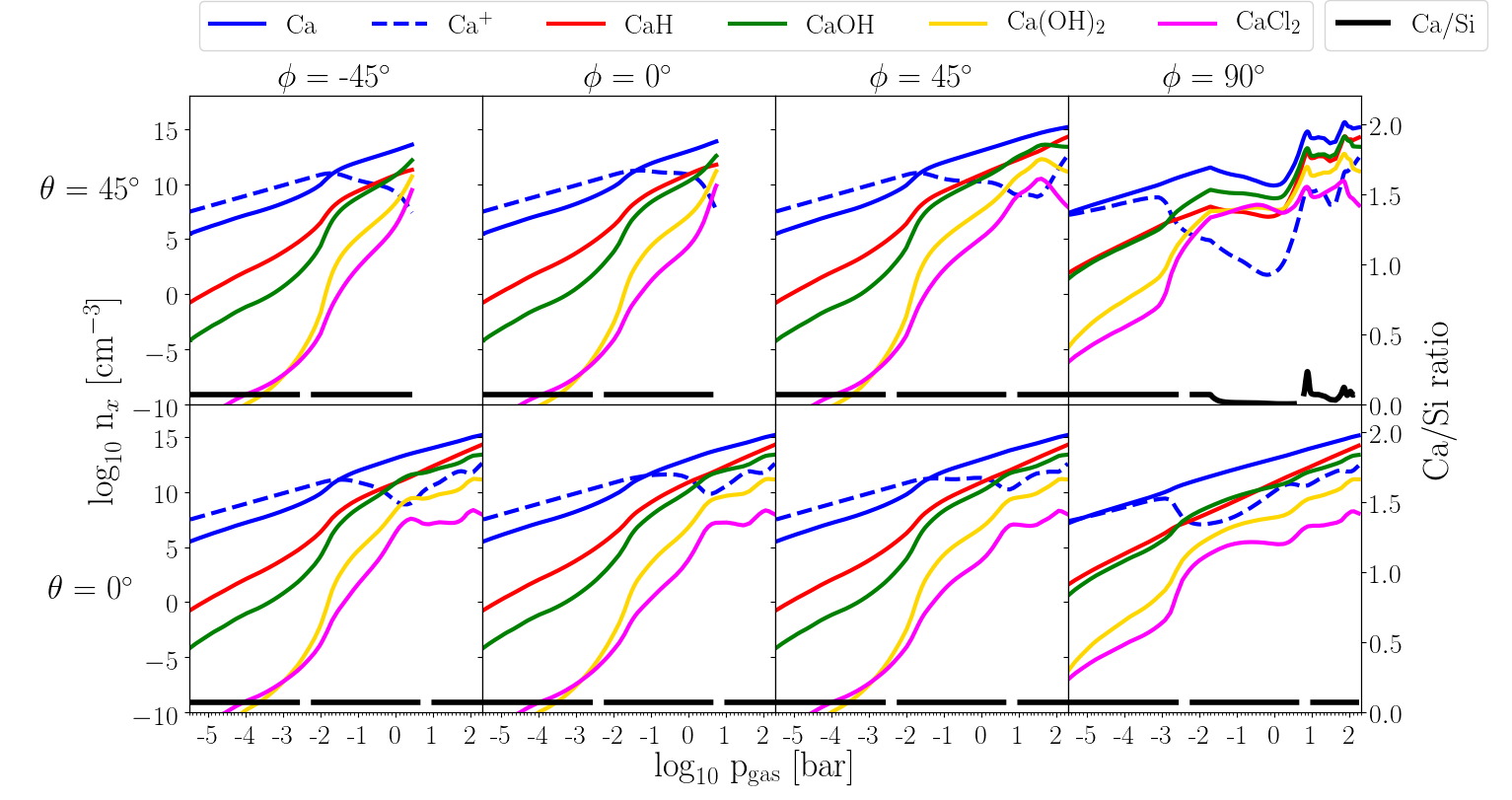}
   \includegraphics[width=\textwidth]{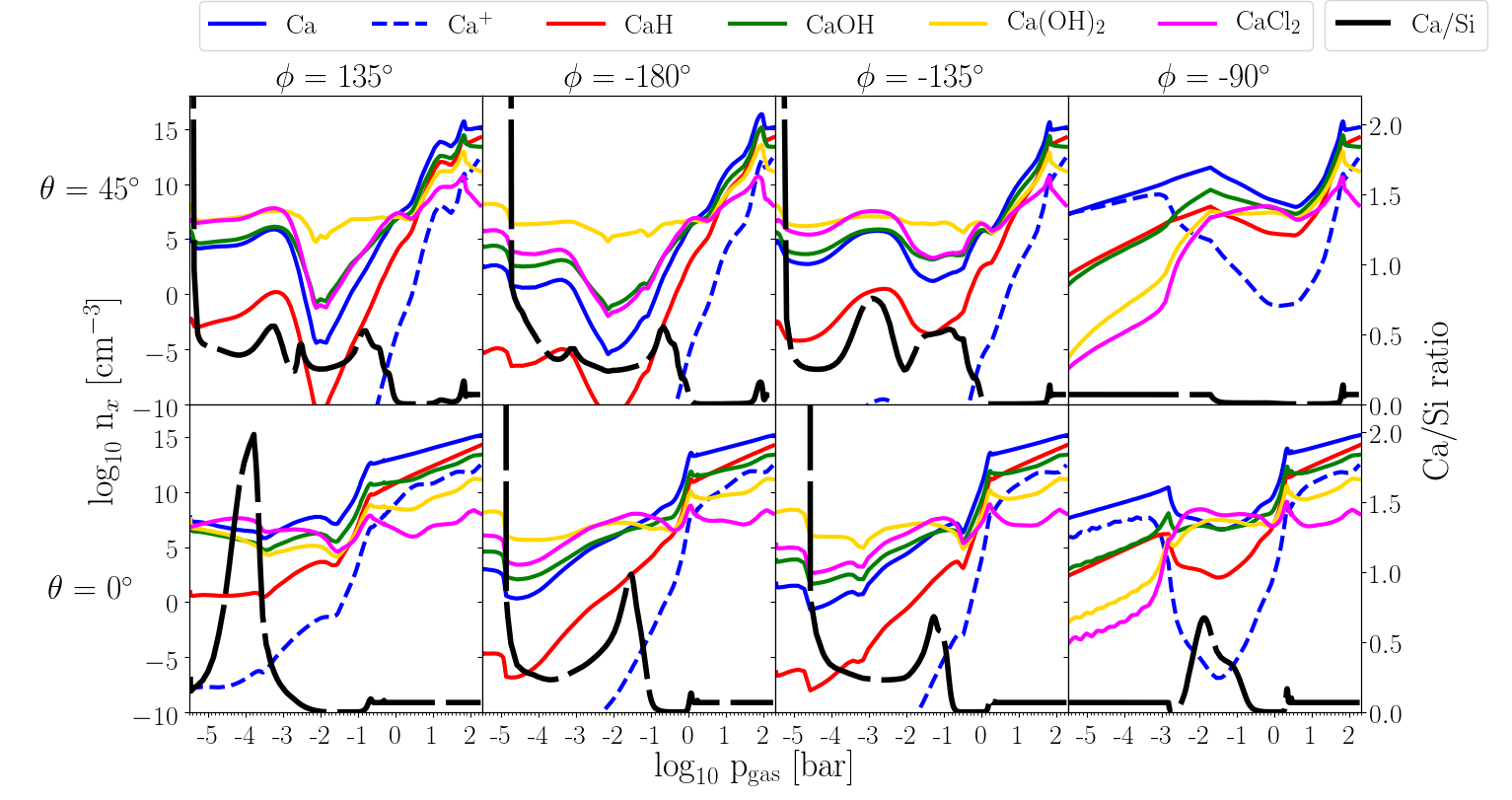}
   \caption{Number densities, log $n_{\rm x}$ [cm$^{-3}$],  of  Ca-binding gas-species (color coded, left axis). The Ca/Si ratio is overplotted and shows where cloud affects the atmosphere (black long-dashed line, right axis).}
   \label{Ca}
\end{figure*}

\begin{figure*}
   \centering
   \includegraphics[width=\textwidth]{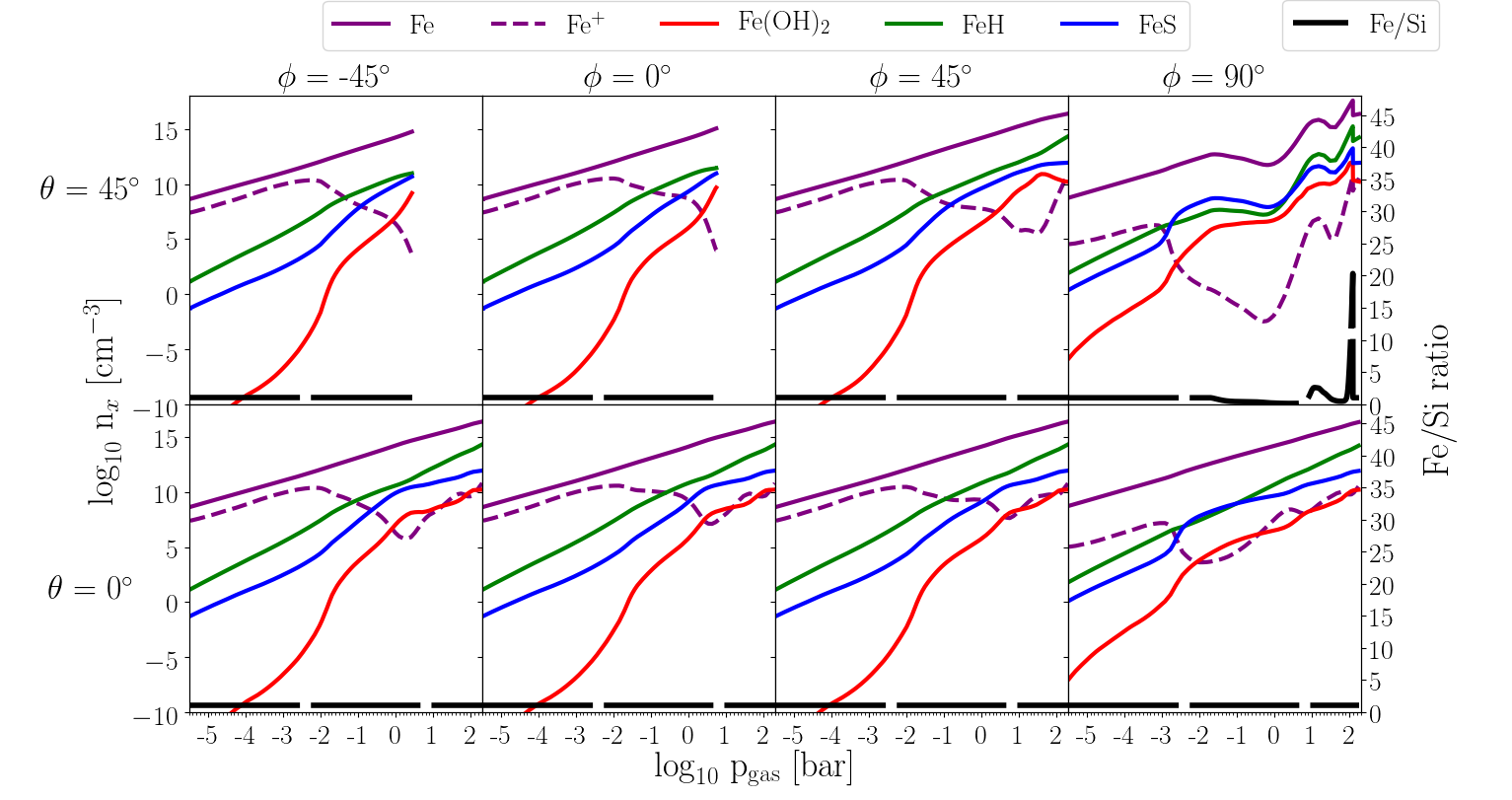}
   \includegraphics[width=\textwidth]{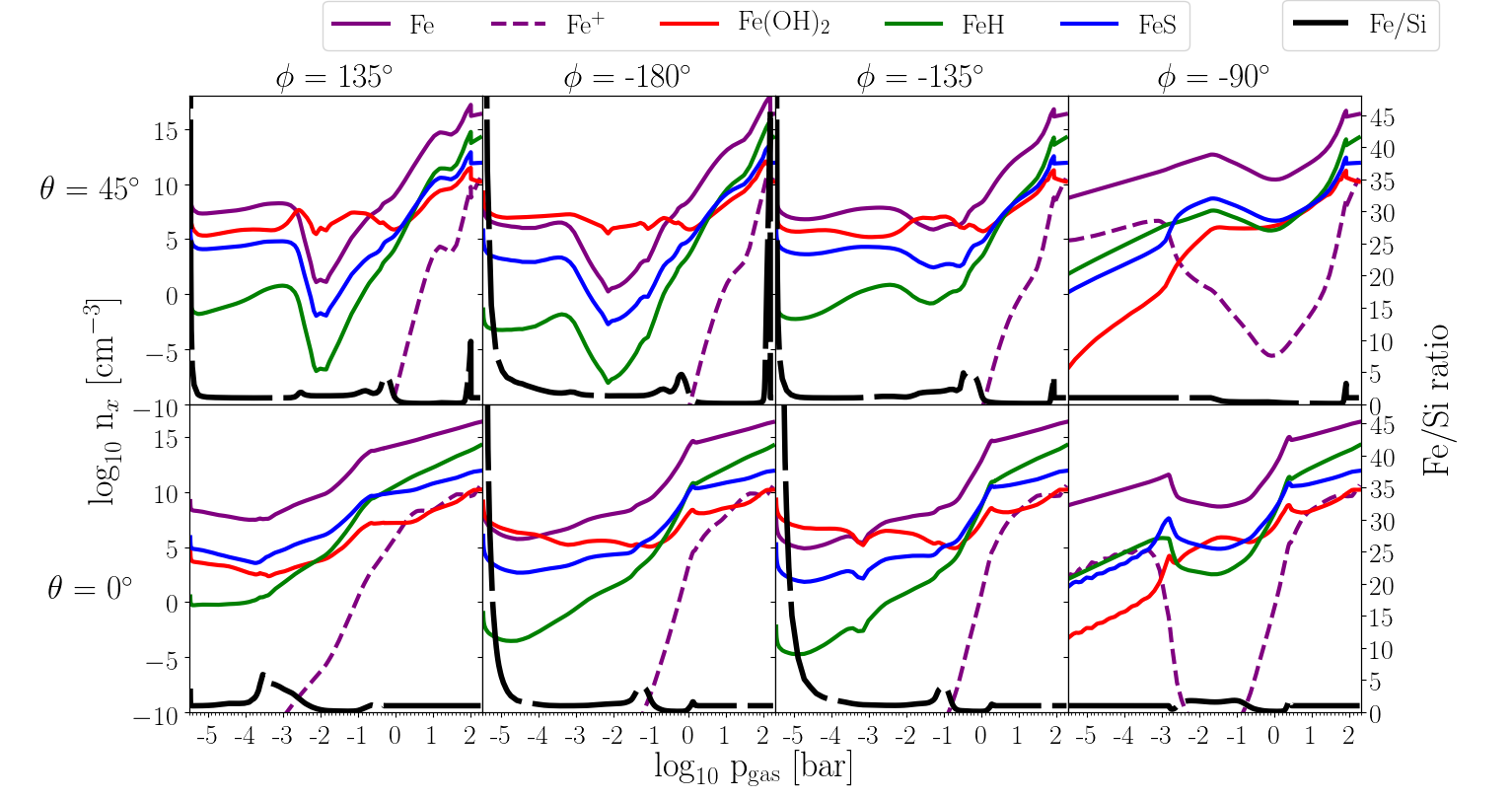}
   \caption{Number densities, log $n_{\rm x}$ [cm$^{-3}$],  of  Fe-binding gas-species (color coded, left axis). The Fe/Si ratio is overplotted and shows where cloud affects the atmosphere (black long-dashed line, right axis).}
   \label{Fe}
\end{figure*}

\begin{figure*}
   \centering
   \includegraphics[width=\textwidth]{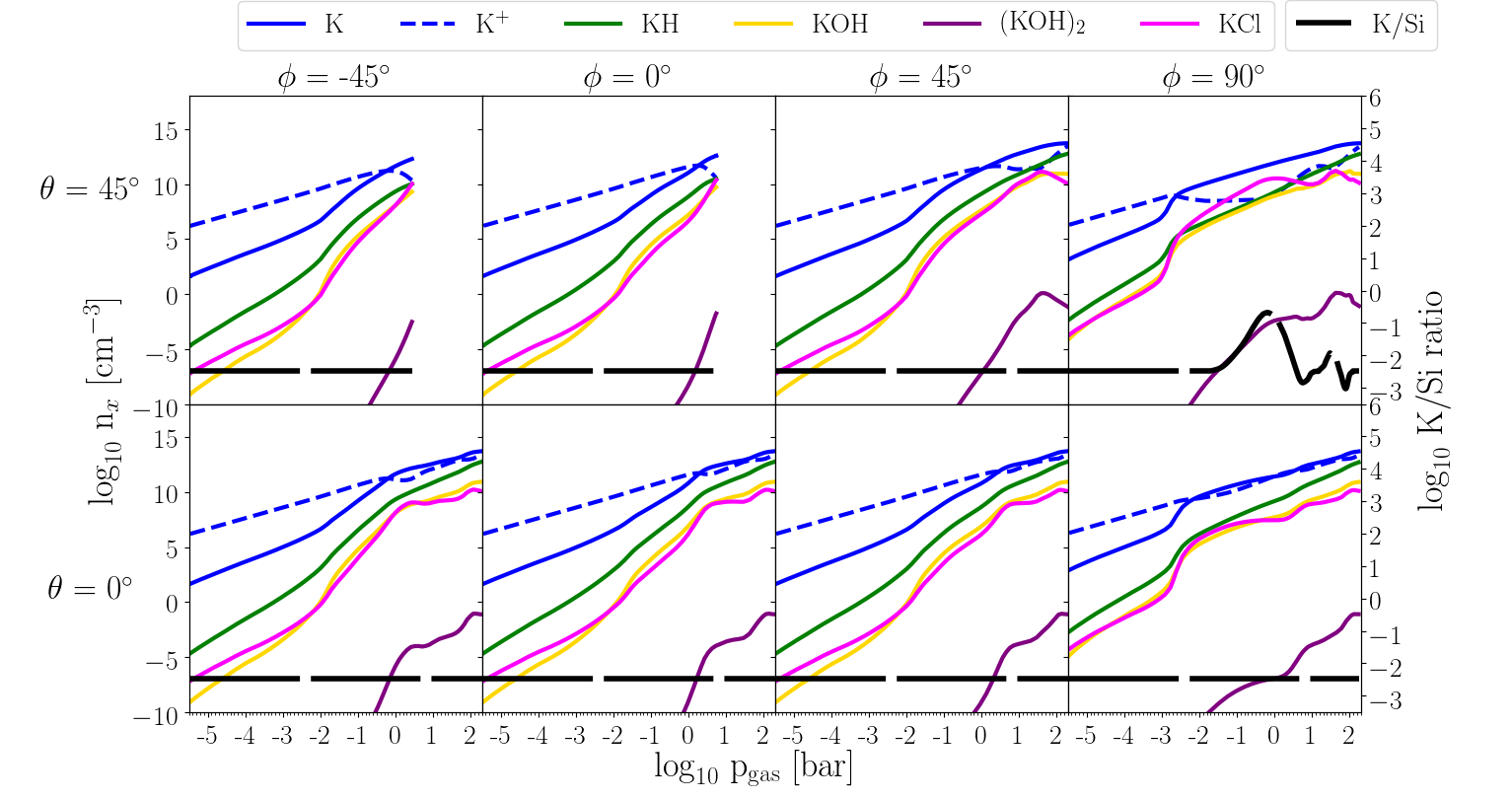}
   \includegraphics[width=\textwidth]{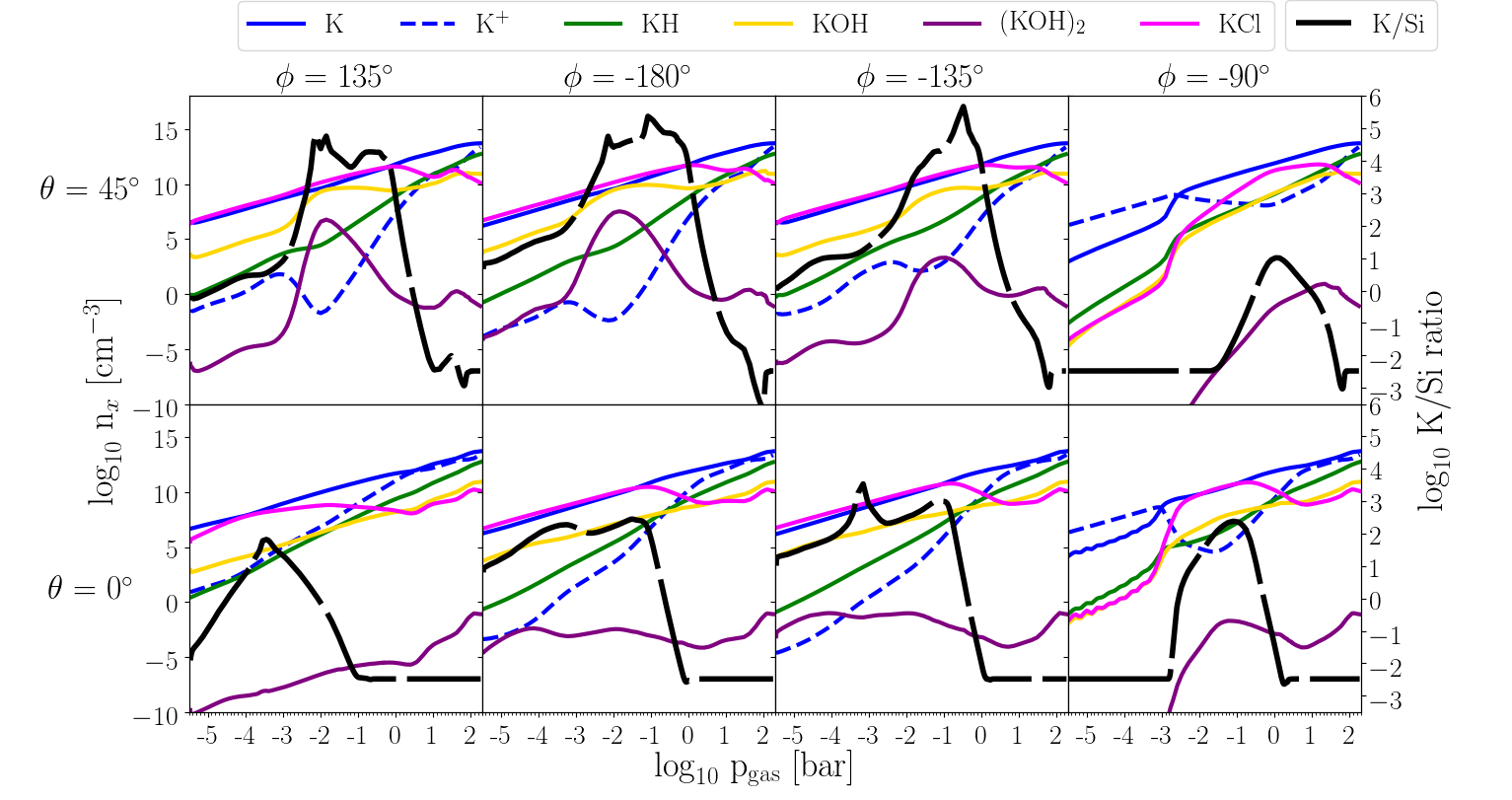}
   \caption{Number densities, log $n_{\rm x}$ [cm$^{-3}$],  of  K-binding gas-species (color coded, left axis). The log(K/Si) ratio is overplotted and shows where cloud affects the atmosphere (black long-dashed line, right axis).}
   \label{K}
\end{figure*}

\begin{figure*}
   \centering
   \includegraphics[width=\textwidth]{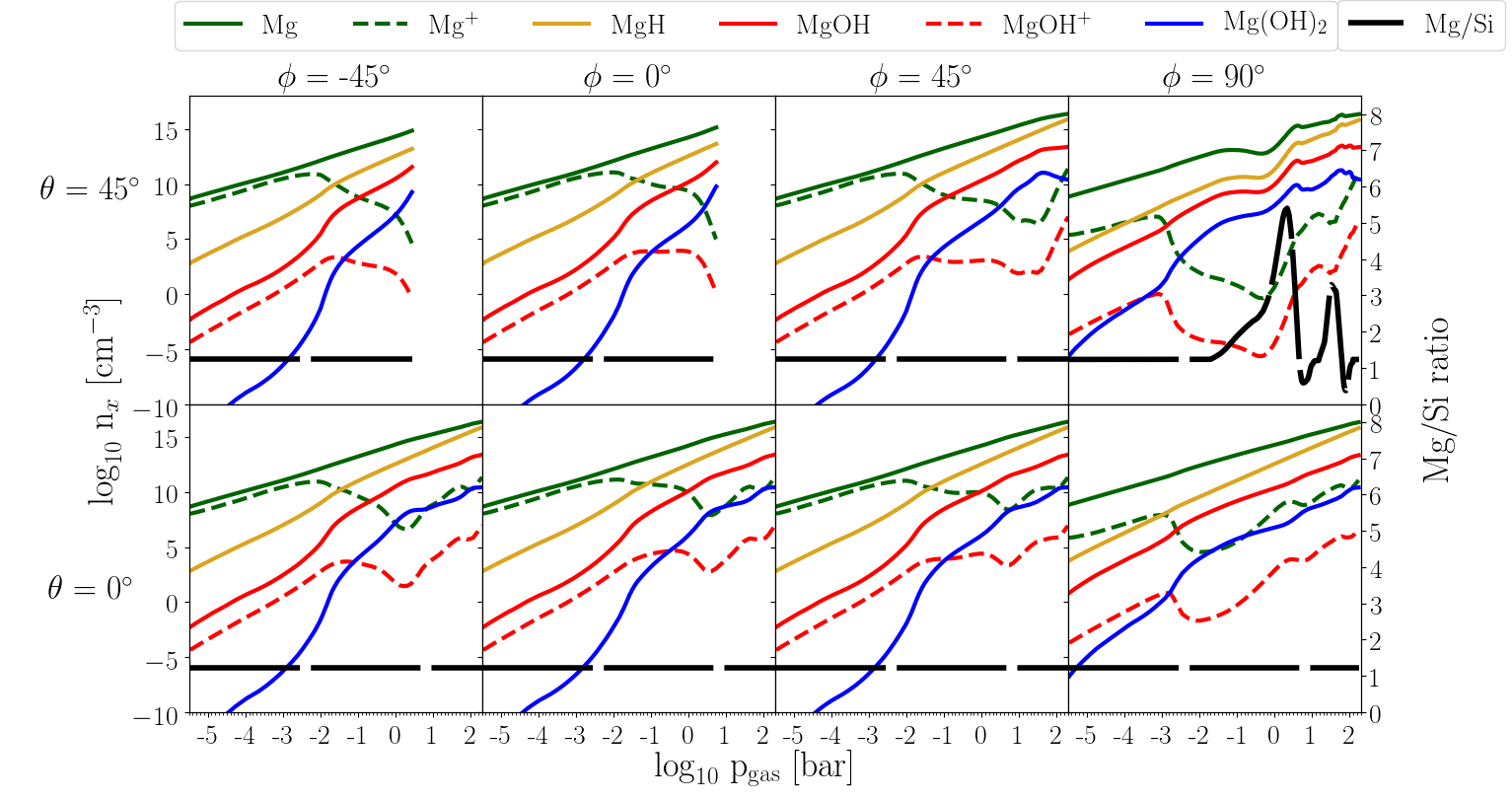}
   \includegraphics[width=\textwidth]{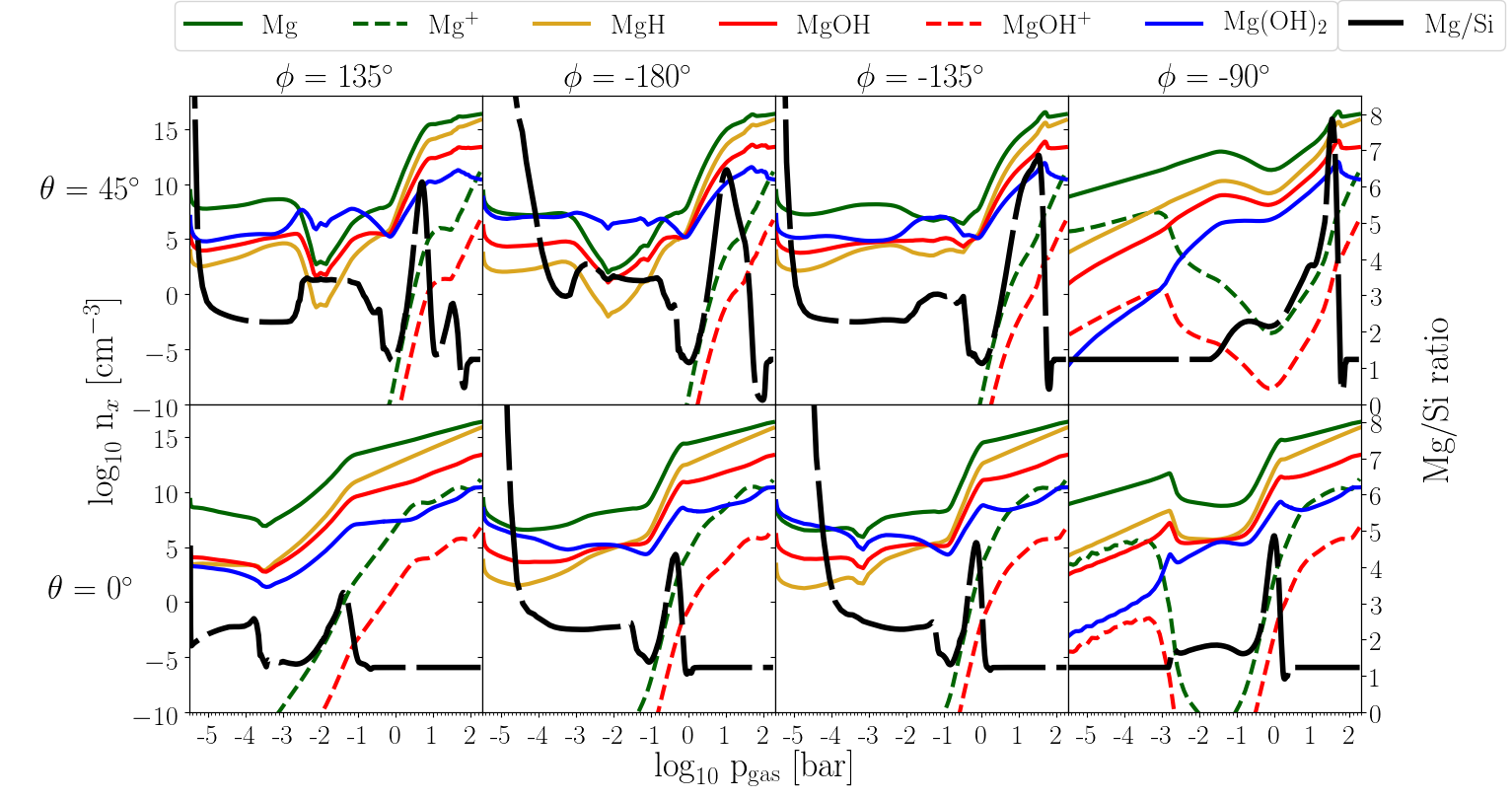}
   \caption{Number densities, log $n_{\rm x}$ [cm$^{-3}$],  of Mg-binding gas-species (color coded, left axis). The Mg/Si ratio is overplotted and shows where cloud affects the atmosphere (black long-dashed line, right axis).}
   \label{Mg}
\end{figure*}

\begin{figure*}
   \centering
   \includegraphics[width=\textwidth]{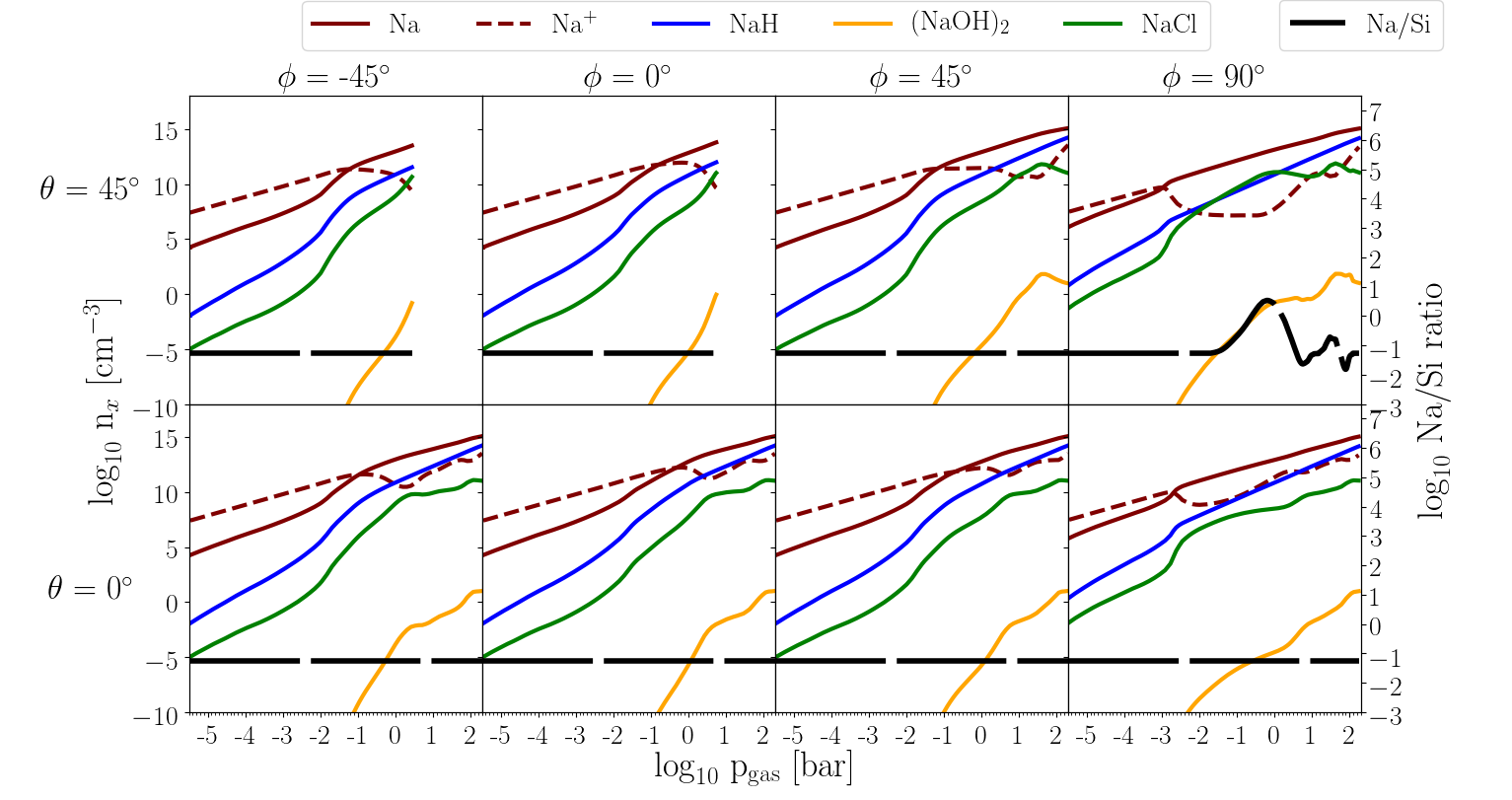}
   \includegraphics[width=\textwidth]{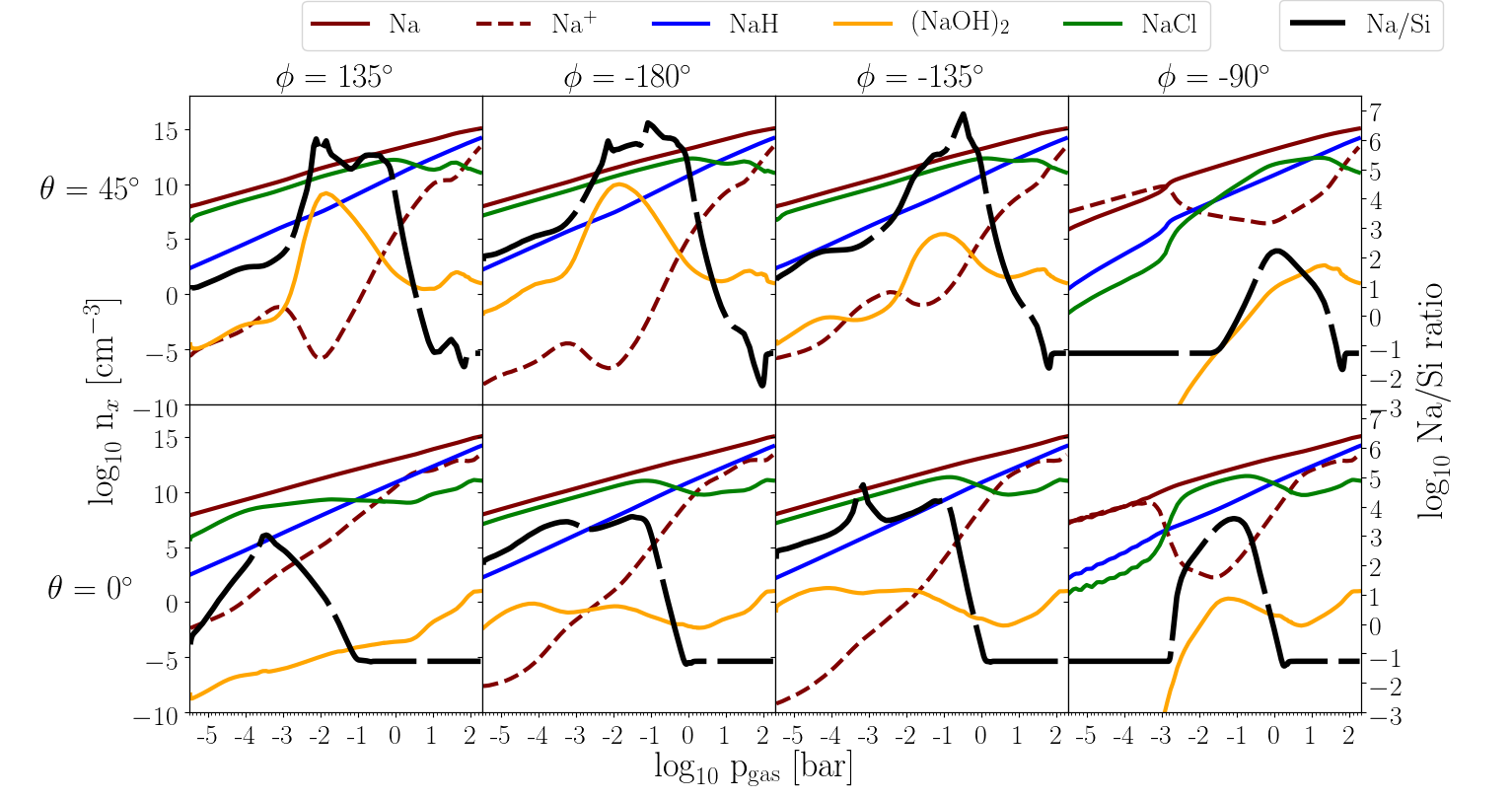}
   \caption{Number densities, log $n_{\rm x}$ [cm$^{-3}$],  of  Na-binding gas-species (color coded, left axis). The log(Na/Si) ratio is overplotted and shows where cloud affects the atmosphere (black long-dashed line, right axis).}
   \label{Na}
\end{figure*}

\begin{figure*}
   \centering
   \includegraphics[width=\textwidth]{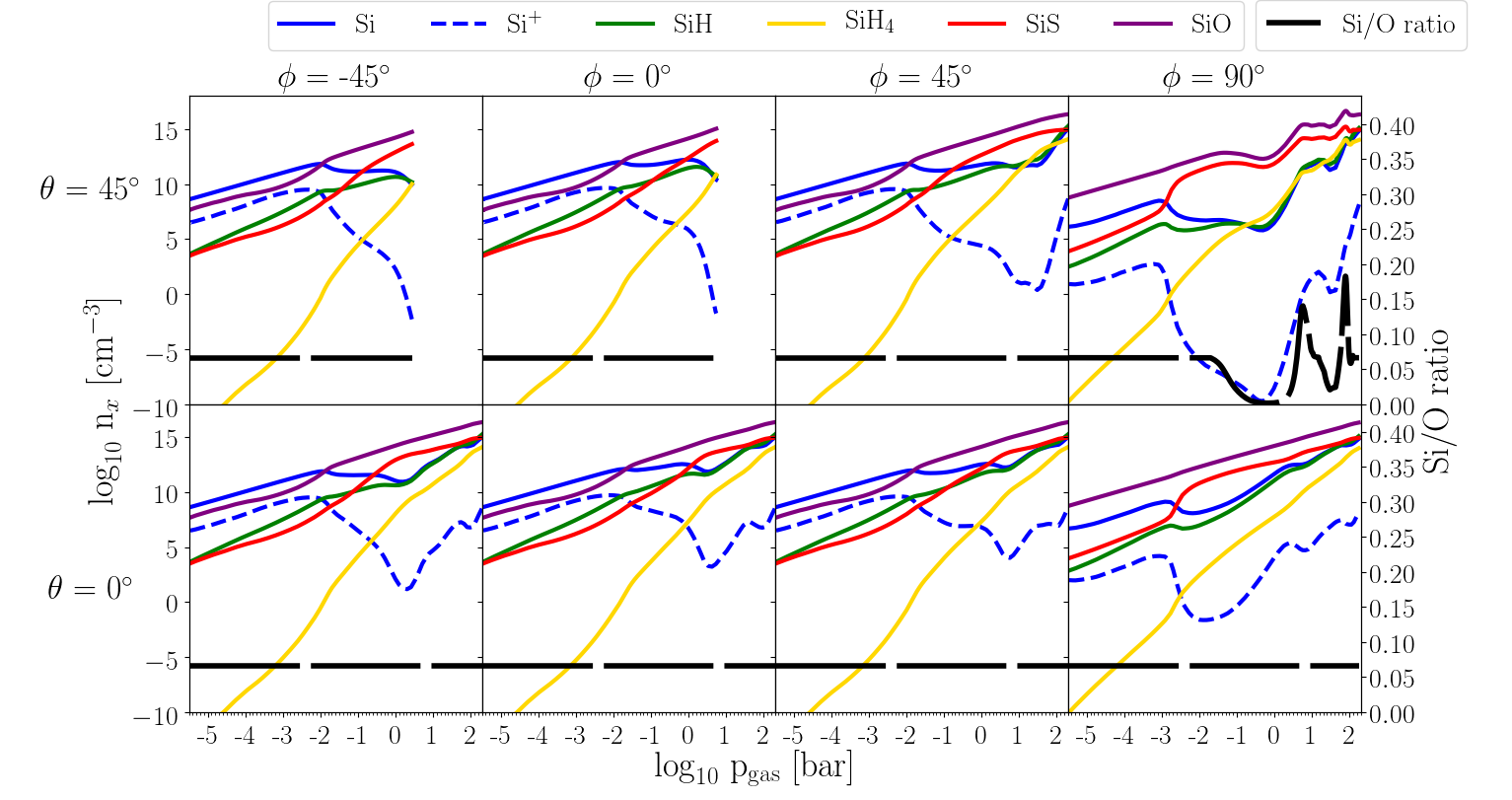}
   \includegraphics[width=\textwidth]{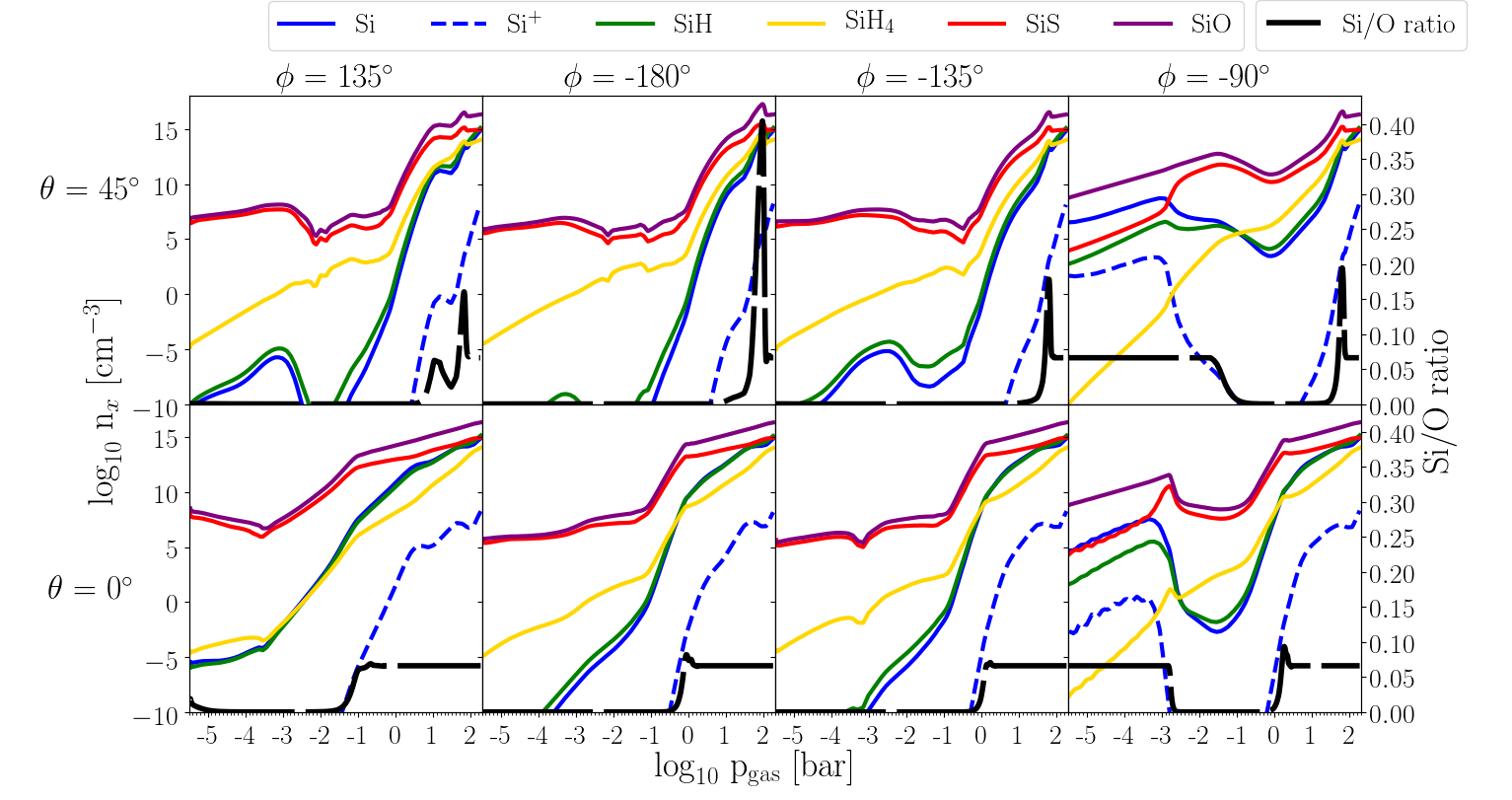}
   \caption{Number densities, log $n_{\rm x}$ [cm$^{-3}$], of Si-binding gas-species (color coded, left axis). The Si/O ratio is overplotted and shows where cloud affects the atmosphere (black long-dashed line, right axis).}
   \label{Si}
\end{figure*}

\begin{figure*}
   \centering
   \includegraphics[width=\textwidth]{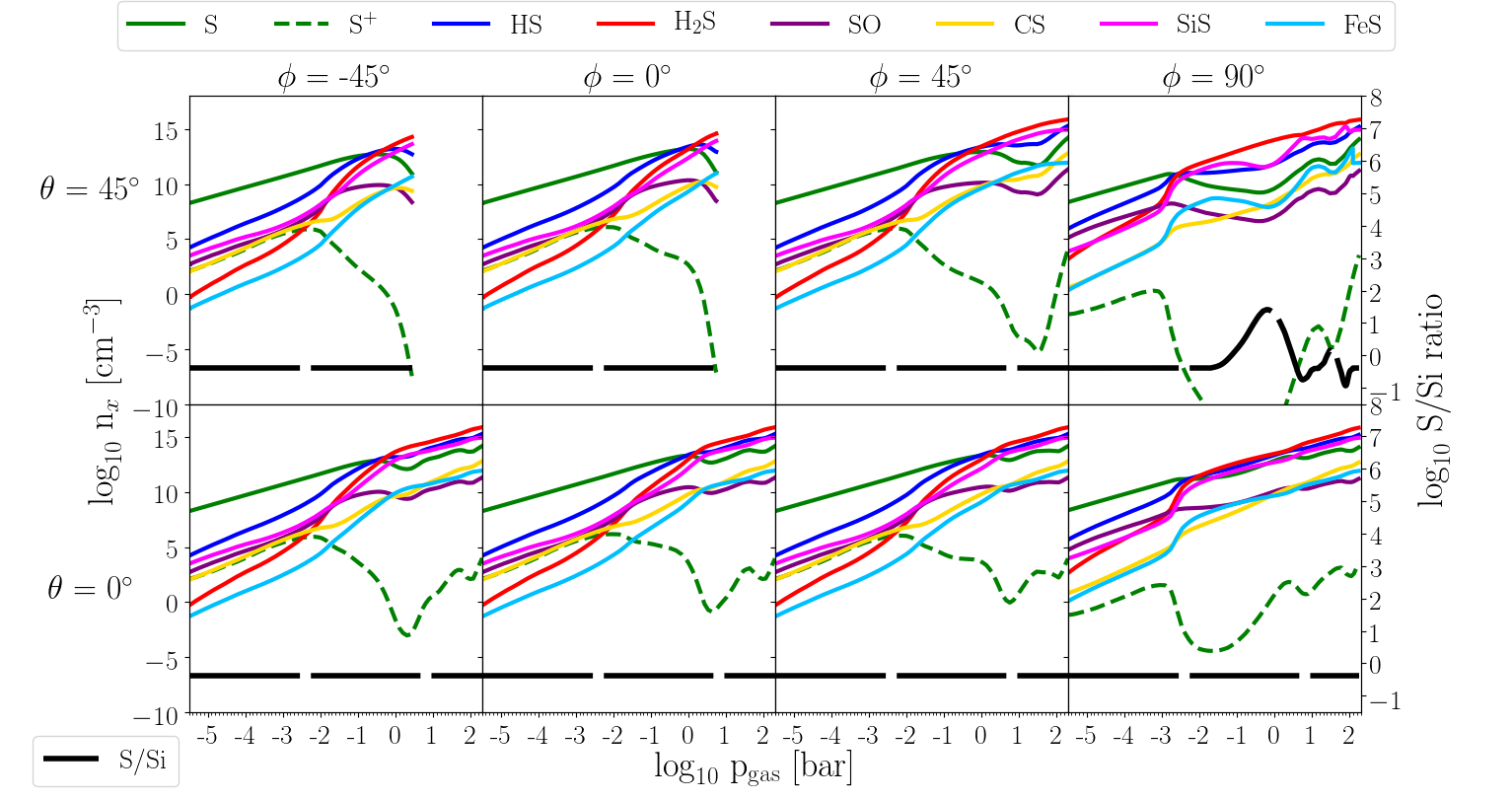}
   \includegraphics[width=\textwidth]{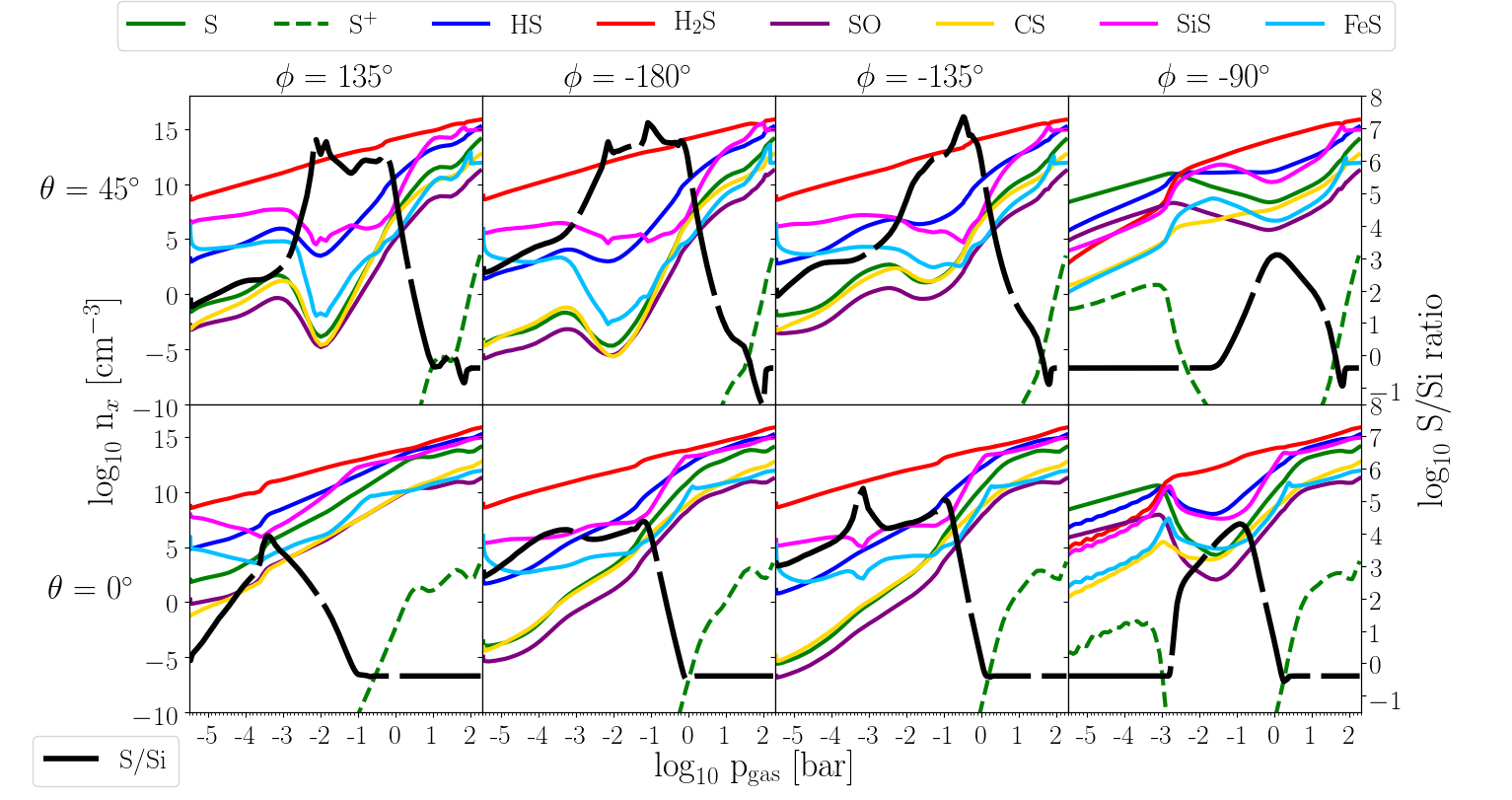}
   \caption{Number densities, log $n_{\rm x}$ [cm$^{-3}$],  of  S-binding gas-species (color coded, left axis). The log(S/Si) ratio is overplotted and shows where cloud affects the atmosphere (black long-dashed line, right axis).}
   \label{S}
\end{figure*}

\begin{figure*}
   \centering
   \includegraphics[width=\textwidth]{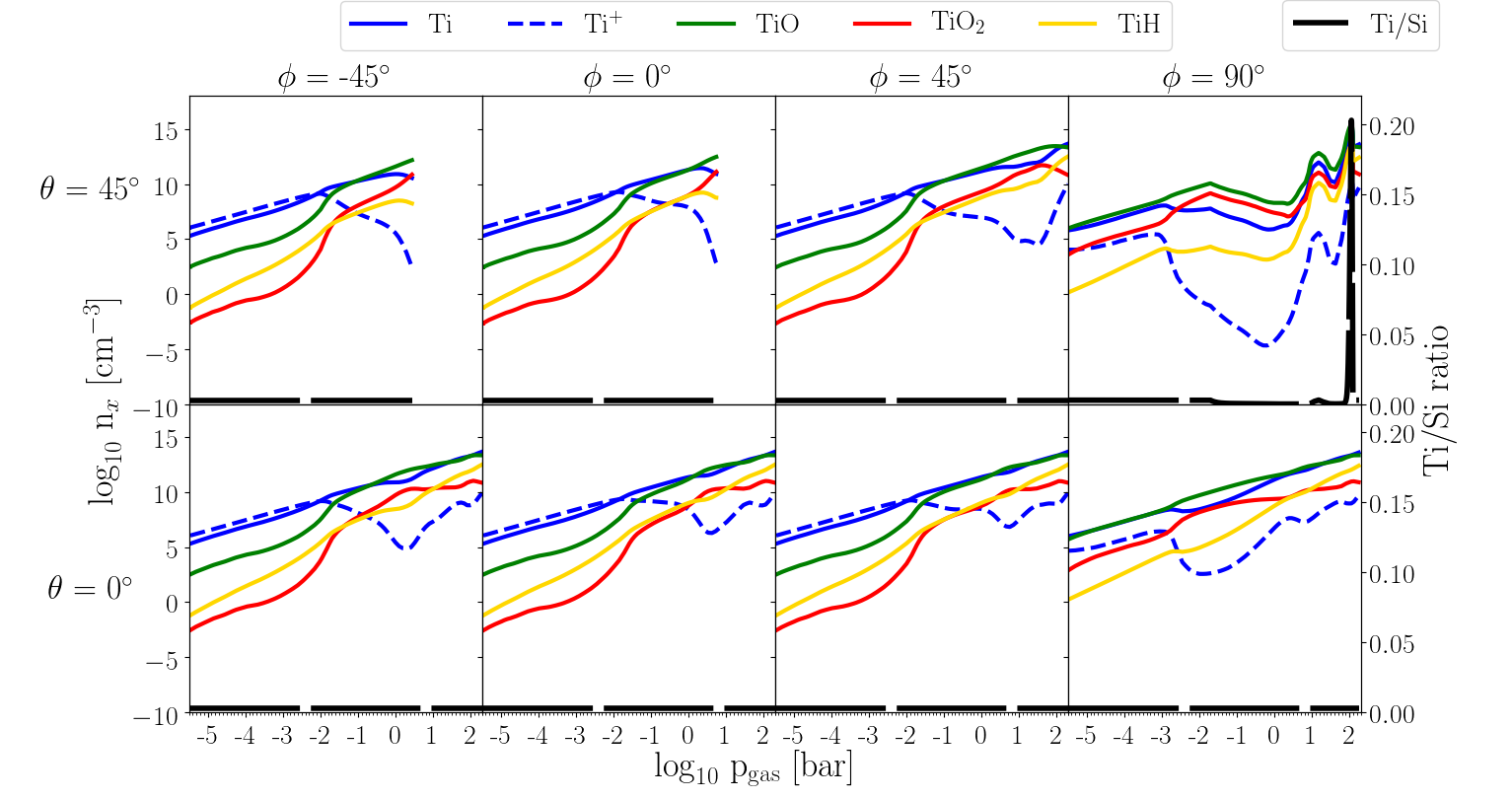}
   \includegraphics[width=\textwidth]{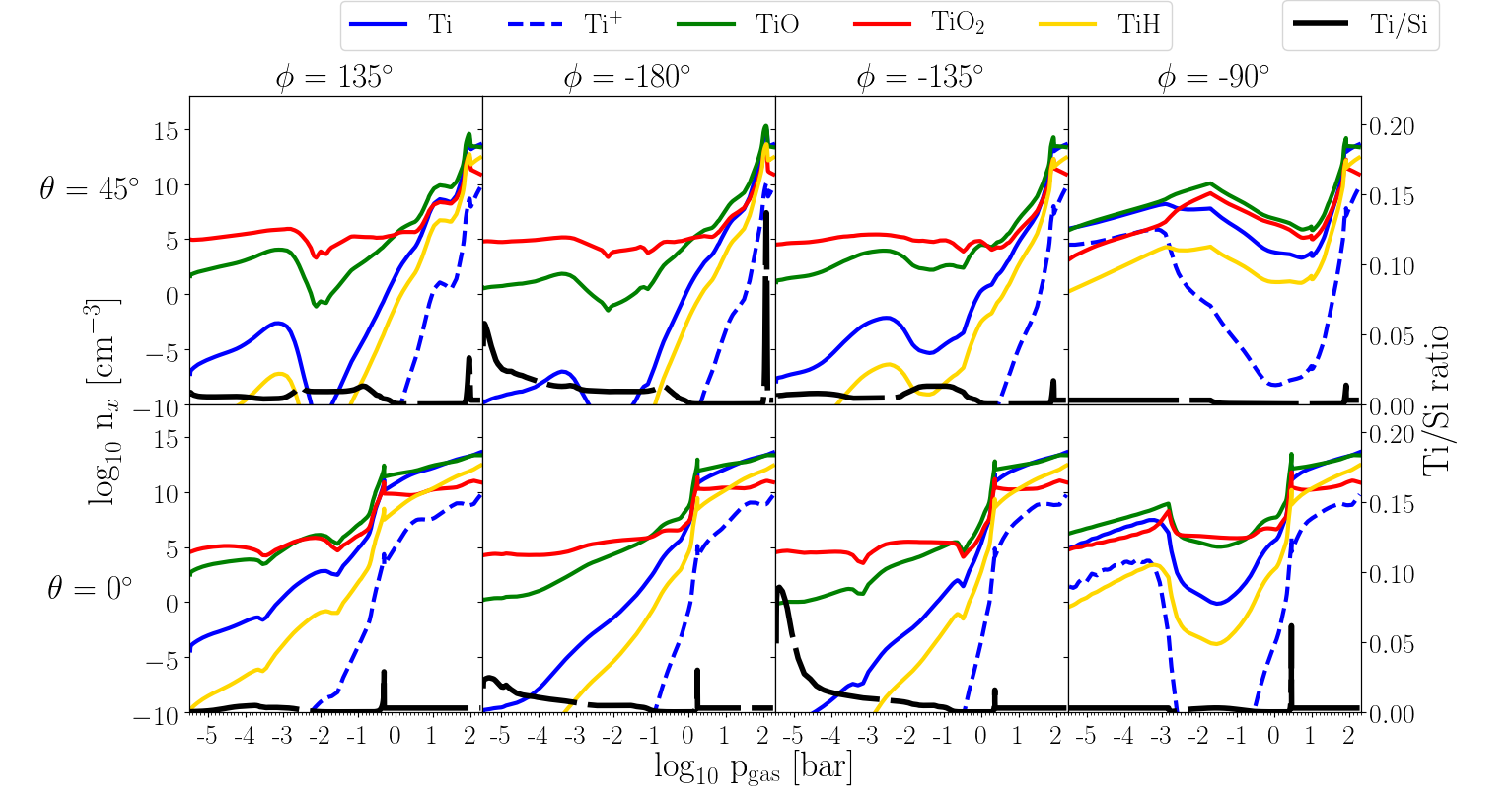}
   \caption{Number densities, log $n_{\rm x}$ [cm$^{-3}$],  of  Ti-binding gas-species (color coded, left axis). The Ca/Si ratio is overplotted and shows where cloud affects the atmosphere (black long-dashed line, right axis).}
   \label{Ti}
      \end{figure*}

\section{Input details}

\begin{table*}
\caption{Chemical surface reactions $r$ assumed to form the solid materials 
   s. The efficiency of the reaction is limited by
   the collision rate of the key species, which has the lowest
   abundance among the reactants. The notation $\half$ in the
   r.h.s.~column means that only every second collision (and sticking)
   event initiates one reaction. Data sources for the
   supersaturation ratios (and saturation vapor pressures): (1)
   Helling \& Woitke (2006); (2) Nuth \& Ferguson (2006); (3) Sharp \&
   Huebner (1990); (4) Woitke et al. (2017)}
\label{tab:chemreak}
\resizebox{15.2cm}{!}{
\begin{tabular}{c|c|l|l}
{\bf Index $r$} & {\bf Solid s} & {\bf Surface reaction} & {\bf Key species} \\
\hline 
1 & TiO$_2$[s]          & TiO$_2$ 
       $\longrightarrow$ TiO$_2$[s]                  & TiO$_2$ \\ 
2 & rutile              & Ti + 2 H$_2$O 
       $\longrightarrow$ TiO$_2$[s] + 2 H$_2$        & Ti     \\
3 & (1)                 & TiO + H$_2$O  
       $\longrightarrow$ TiO$_2$[s] + H$_2$          & TiO     \\ 
4 &                     & TiS + 2 H$_2$O
       $\longrightarrow$ TiO$_2$[s] + H$_2$S + H$_2$ & TiS     \\
\hline 
5 & SiO$_2$[s]          & SiH + 2 H$_2$O
       $\longrightarrow$ SiO$_2$[s] + 2 H$_2$ + H    & SiH \\ 
6 & silica              & SiO + H$_2$O 
       $\longrightarrow$ SiO$_2$[s] + H$_2$          & SiO     \\ 
7 &  (3)                & SiS + 2 H$_2$O 
       $\longrightarrow$ SiO$_2$[s] + H$_2$S + H$_2$ & SiS     \\
\hline 
8 & SiO[s]              & SiO 
       $\longrightarrow$ SiO[s]                  & SiO \\
9 & silicon mono-oxide  & 2 SiH + 2 H$_2$O 
       $\longrightarrow$ 2 SiO[s] + 3 H$_2$      & SiH   \\
10 & (2)                & SiS + H$_2$O 
       $\longrightarrow$ SiO[s] + H$_2$S         & SiS     \\
\hline   
11 & Fe[s]              & Fe 
       $\longrightarrow$ Fe[s]                  & Fe      \\ 
12 & solid iron         & FeO + H$_2$ 
       $\longrightarrow$ Fe[s] + H$_2$O         & FeO     \\
13 & (1)                & FeS + H$_2$ 
       $\longrightarrow$ Fe[s] + H$_2$S         & FeS     \\ 
14 &                    & Fe(OH)$_2$ + H$_2$ 
       $\longrightarrow$ Fe[s] + 2 H$_2$O       & Fe(OH)$_2$ \\ 
15 &                    & 2  FeH
       $\longrightarrow$ 2 Fe[s] + H$_2$         & FeH     \\ 
\hline 
16 & FeO[s]             & FeO 
       $\longrightarrow$ FeO[s]                  & FeO\\
17 & iron\,(II) oxide   & Fe + H$_2$O
       $\longrightarrow$ FeO[s] + H$_2$          & Fe\\
18 & (3)                & FeS + H$_2$O 
       $\longrightarrow$ FeO[s] + H$_2$S         & FeS\\
19 &                    & Fe(OH)$_2$
       $\longrightarrow$ FeO[s] + H$_2$          & Fe(OH)$_2$\\
20 &                    & 2 FeH + 2 H$_2$O
       $\longrightarrow$ 2 FeO[s] + 3 H$_2$      & FeH \\
\hline
21 & FeS[s]             & FeS
       $\longrightarrow$ FeS[s]                       & FeS\\
22 & iron sulphide      & Fe + H$_2$S
       $\longrightarrow$ FeS[s]     + H$_2$           & Fe\\
23 & (3)                & FeO + H$_2$S 
       $\longrightarrow$ FeS[s] + H$_2$O     & $\min\{$FeO, H$_2$S$\}$\\
24 &                    & Fe(OH)$_2$ + H$_2$S     
       $\longrightarrow$ FeS[s] + 2 H$_2$O   & $\min\{$Fe(OH)$_2$, H$_2$S$\}$\\
25 &                    & 2 FeH + 2 H$_2$S
       $\longrightarrow$ 2 FeS[s] + 3 H$_2$  & $\min\{$FeH, H$_2$S$\}$\\
\hline
26 & Fe$_2$O$_3$[s]     & 2 Fe + 3 H$_2$O 
       $\longrightarrow$ Fe$_2$O$_3$[s] + 3 H$_2$        & $\half$Fe\\
27 & iron\,(III) oxide  & 2 FeO + H$_2$O
       $\longrightarrow$ Fe$_2$O$_3$[s] + H$_2$          & $\half$FeO\\
28 & (3)                & 2 FeS + 3 H$_2$O
       $\longrightarrow$ Fe$_2$O$_3$[s] + 2 H$_2$S + H$_2$&$\half$FeS\\
29 &                    & 2 Fe(OH)$_2$ 
       $\longrightarrow$ Fe$_2$O$_3$[s] + H$_2$O + H$_2$ & $\half$Fe(OH)$_2$\\
30 &                    & 2 FeH + 3 H$_2$O
       $\longrightarrow$ Fe$_2$O$_3$[s] + 4 H$_2$ & $\half$FeH\\
\hline
31 & MgO[s]             & Mg + H$_2$O 
      $\longrightarrow$ MgO[s] + H$_2$                & Mg\\
32 & periclase          & 2 MgH + 2 H$_2$O
      $\longrightarrow$ 2 MgO[s] + 3 H$_2$            & $\half$MgH\\ 
33 & (3)                & 2 MgOH
      $\longrightarrow$ 2 MgO[s] + H$_2$              & $\half$MgOH\\
34 &                    & Mg(OH)$_2$
      $\longrightarrow$ MgO[s] + H$_2$O               & Mg(OH)$_2$\\
\hline
35 & MgSiO$_3$[s]     & Mg + SiO + 2 H$_2$O 
     $\longrightarrow$ MgSiO$_3$[s] + H$_2$
                                  & $\min\{$Mg, SiO$\}$\\ 
36 & enstatite        & Mg + SiS + 3 H$_2$O 
     $\longrightarrow$ MgSiO$_3$[s] + H$_2$S + 2 H$_2$ 
                                  & $\min\{$Mg, SiS$\}$\\ 
37 & (3)              & 2 Mg + 2 SiH + 6 H$_2$O 
     $\longrightarrow$ 2 MgSiO$_3$[s] + 7 H$_2$
                                  & $\min\{$Mg, SiH$\}$\\ 
38 &                  & 2 MgOH + 2 SiO + 2 H$_2$O
     $\longrightarrow$ 2 MgSiO$_3$[s] + 3 H$_2$    
                                  & $\min\{\half$MgOH, $\half$SiO$\}$ \\
39 &                  & 2 MgOH + 2 SiS + 4 H$_2$O
     $\longrightarrow$ 2 MgSiO$_3$[s] + 2 H$_2$S + 3 H$_2$ 
                                  & $\min\{\half$MgOH, $\half$SiS$\}$ \\
40 &                  & MgOH + SiH + 2 H$_2$O
     $\longrightarrow$ MgSiO$_3$[s] + 3 H$_2$
                                  & $\min\{\half$MgOH, $\half$SiH$\}$ \\
41 &                  & Mg(OH)$_2$ + SiO 
     $\longrightarrow$ 2 MgSiO$_3$[s] +  H$_2$
                                  & $\min\{$Mg(OH)$_2$, SiO$\}$ \\ 
42 &                  & Mg(OH)$_2$ + SiS + H$_2$O
     $\longrightarrow$ MgSiO$_3$[s] + H$_2$S+ H$_2$
                                  & $\min\{$Mg(OH)$_2$, SiS$\}$ \\
43 &                  & 2 Mg(OH)$_2$ + 2 SiH + 2 H$_2$O
     $\longrightarrow$ 2 MgSiO$_3$[s] + 5 H$_2$
                                  & $\min\{$Mg(OH)$_2$, SiH$\}$ \\
44 &                  & 2 MgH +  2 SiO + 4 H$_2$O
     $\longrightarrow$ 2 MgSiO$_3$[s]+ 5 H$_2$
                                  & $\min\{$MgH, SiO$\}$ \\
45 &                  & 2 MgH +  2 SiS + 6 H$_2$O
     $\longrightarrow$ 2 MgSiO$_3$[s]+ 2 H$_2$S + 5 H$_2$
                                  & $\min\{$MgH, SiS$\}$ \\
46 &                  & MgH + SiH + 3 H$_2$O
     $\longrightarrow$  MgSiO$_3$[s]+ 4 H$_2$
                                  & $\min\{$MgH, SiH$\}$ \\
\hline
47 & Mg$_2$SiO$_4$[s] & 2 Mg + SiO + 3 H$_2$O
     $\longrightarrow$ Mg$_2$SiO$_4$[s] + 3 H$_2$  
                                  & $\min\{\half$Mg, SiO$\}$\\
48 & forsterite       & 2 MgOH + SiO + H$_2$O
     $\longrightarrow$ Mg$_2$SiO$_4$[s] + 2 H$_2$
                                  & $\min\{\half$MgOH, SiO$\}$\\ 
49 & (3)              & 2 Mg(OH)$_2$ + SiO 
     $\longrightarrow$ Mg$_2$SiO$_4$[s] + H$_2$O + H$_2$
                                  & $\min\{\half$Mg(OH)$_2$, SiO$\}$ \\
50 &                  & 2 MgH + SiO + 3 H$_2$O
     $\longrightarrow$ Mg$_2$SiO$_4$[s] + 4 H$_2$
                                  & $\min\{\half$MgH, SiO$\}$ \\
51 &                  & 2 Mg + SiS + 4 H$_2$O            
     $\longrightarrow$ Mg$_2$SiO$_4$[s] + H$_2$S + 3 H$_2$
                                  & $\min\{\half$Mg, SiS\} \\
52 &                  & 2 MgOH + SiS + 2 H$_2$O 
     $\longrightarrow$ Mg$_2$SiO$_4$[s] + H$_2$S + 2 H$_2$
                                  & $\min\{\half$MgOH, SiS\}\\
53 &                  & 2 Mg(OH)$_2$ + SiS 
     $\longrightarrow$ Mg$_2$SiO$_4$[s] + H$_2$ + H$_2$S
                                  & $\min\{\half$Mg(OH)$_2$, SiS\} \\
54 &                  & 2 MgH + SiS + 4 H$_2$O
     $\longrightarrow$ Mg$_2$SiO$_4$[s] + H$_2$S + 4 H$_2$
                                  & $\min\{\half$MgH, SiS$\}$ \\
55 &                  & 4 Mg + 2 SiH + 8 H$_2$O            
     $\longrightarrow$ 2 Mg$_2$SiO$_4$[s] + 9 H$_2$
                                  & $\min\{\half$Mg, SiH\} \\
56 &                  & 4 MgOH + 2 SiH + 4 H$_2$O 
     $\longrightarrow$ 2 Mg$_2$SiO$_4$[s] + 7 H$_2$
                                  & $\min\{\half$MgOH, SiH\}\\
57 &                  & 4 Mg(OH)$_2$ + 2 SiH 
     $\longrightarrow$ 2 Mg$_2$SiO$_4$[s] + 5 H$_2$
                                  & $\min\{\half$Mg(OH)$_2$, SiH\} \\
58 &                  & 4 MgH + 2 SiH + 8 H$_2$O
     $\longrightarrow$ 2 Mg$_2$SiO$_4$[s] + 11 H$_2$
                                  & $\min\{\half$MgH, SiS$\}$ \\
\hline 
59 & Al$_2$O$_3$[s]   & 2 Al + 3 H$_2$O 
     $\longrightarrow$ Al$_2$O$_3$[s] + 3 H$_2$   & $\half$Al\\
60 & aluminia         & 2 AlOH + H$_2$O 
     $\longrightarrow$ Al$_2$O$_3$[s] + 2 H$_2$   & $\half$AlOH \\ 
61 & (3)              &  2 AlH + 3 H$_2$O 
     $\longrightarrow$ Al$_2$O$_3$[s] + 4 H$_2$   & $\half$AlH\\
62 &                  & Al$_2$O + 2 H$_2$O
     $\longrightarrow$ Al$_2$O$_3$[s] + 2 H$_2$   & Al$_2$O\\
63 &                  & 2 AlO$_2$H 
$\longrightarrow$ Al$_2$O$_3$[s] + H$_2$O    & $\half$AlO$_2$H\\
\end{tabular}}
\end{table*}

\begin{table*}
\caption{Table~\ref{tab:chemreak} continued}
\label{tab:chemreak2}
\resizebox{15.2cm}{!}{
\begin{tabular}{c|c|l|l}
{\bf Index $r$} & {\bf Solid s} & {\bf Surface reaction} & {\bf Key species} \\
\hline
64 & CaTiO$_3$[s]         & Ca + Ti + 3 H$_2$O     
      $\longrightarrow$ CaTiO$_3$[s] + 3 H$_2$       & $\min\{$Ca, Ti$\}$\\  
65 & perovskite           & Ca + TiO + 2 H$_2$O 
      $\longrightarrow$ CaTiO$_3$[s] + 2 H$_2$       & $\min\{$Ca, TiO$\}$\\
66 & (3)                  & Ca + TiO$_2$ + H$_2$O 
      $\longrightarrow$ CaTiO$_3$[s] + H$_2$         & $\min\{$Ca, TiO$_2\}$\\
67 &                      & Ca + TiS + 3 H$_2$O 
      $\longrightarrow$ CaTiO$_3$[s] + H$_2$S + 2 H$_2$  & $\min\{$Ca, TiS$\}$\\
68 &                      & CaO + Ti + 2 H$_2$O 
      $\longrightarrow$ CaTiO$_3$[s] + 2 H$_2$       & $\min\{$CaO, Ti$\}$\\
69 &                      & CaO + TiO + H$_2$O 
      $\longrightarrow$ CaTiO$_3$[s] + H$_2$         & $\min\{$CaO, TiO$\}$\\
70 &                      & CaO + TiO$_2$
      $\longrightarrow$ CaTiO$_3$[s]                 & $\min\{$CaO, TiO$_2\}$\\
71 &                      & CaO + TiS + 2 H$_2$O 
      $\longrightarrow$ CaTiO$_3$[s] + H$_2$S + H$_2$ & $\min\{$CaO, TiO$\}$\\
72 &                      & CaS + Ti + 3 H$_2$O 
      $\longrightarrow$ CaTiO$_3$[s] + H$_2$S + H$_2$ & $\min\{$CaS, Ti$\}$\\
73 &                      & CaS + TiO + 2 H$_2$O 
      $\longrightarrow$ CaTiO$_3$[s] + H$_2$S + 2 H$_2$ &$\min\{$CaS, TiO$\}$\\
74 &                      & CaS + TiO$_2$ + H$_2$O 
      $\longrightarrow$ CaTiO$_3$[s] + H$_2$S        & $\min\{$CaS, TiO$_2\}$\\
75 &                      & CaS + TiS + 3 H$_2$O 
      $\longrightarrow$ CaTiO$_3$[s] + 2 H$_2$S + H$_2$ &$\min\{$CaS, TiO$\}$\\
76 &                      & Ca(OH)$_2$ + Ti + H$_2$O 
      $\longrightarrow$ CaTiO$_3$[s] + 2 H$_2$  & $\min\{$Ca(OH)$_2$, Ti$\}$\\
77 &                      & Ca(OH)$_2$ + TiO 
      $\longrightarrow$ CaTiO$_3$[s] + H$_2$    & $\min\{$Ca(OH)$_2$, TiO$\}$\\
78 &                      & Ca(OH)$_2$ + TiO$_2$ 
      $\longrightarrow$ CaTiO$_3$[s] + H$_2$O   &$\min\rm\{Ca(OH)_2,TiO_2\}$\\
79 &                      & Ca(OH)$_2$ + TiS + H$_2$O
      $\longrightarrow$ CaTiO$_3$[s] + H$_2$S + H$_2$   & $\min\{$Ca(OH)$_2$, TiO$\}$\\
80 &                      & 2 CaH + 2 Ti + 6 H$_2$O
      $\longrightarrow$ 2 CaTiO$_3$[s] + 7 H$_2$   &   $\min\{$CaH, Ti$\}$\\
81 &                      & 2 CaH + 2 TiO + 4 H$_2$O
      $\longrightarrow$ 2 CaTiO$_3$[s] + 5 H$_2$   &   $\min\{$CaH, TiO$\}$\\
82 &                      & 2 CaH + 2 TiO$_2$ + 2 H$_2$O
      $\longrightarrow$ 2 CaTiO$_3$[s] + 3 H$_2$   &   $\min\{$CaH, TiO$_2$ $\}$\\
83 &                      & 2 CaH + 2 TiS + 6 H$_2$O
      $\longrightarrow$ 2 CaTiO$_3$[s] + 2 H$_2$S +5 H$_2$   &   $\min\{$CaH, TiS$\}$\\
84 &                      & 2 CaOH + 2 Ti + 4 H$_2$O
      $\longrightarrow$ 2 CaTiO$_3$[s] + 5 H$_2$   &   $\min\{$CaOH, Ti$\}$\\
85 &                      & 2 CaOH + 2 TiO + 2 H$_2$O
      $\longrightarrow$ 2 CaTiO$_3$[s] + 3 H$_2$   &   $\min\{$CaOH, TiO$\}$\\
86 &                      & 2 CaOH + 2 TiO$_2$
      $\longrightarrow$ 2 CaTiO$_3$[s] + H$_2$   &   $\min\{$CaOH, TiO$_2$ $\}$\\
87 &                      & 2 CaOH + 2 TiS + 4 H$_2$O
      $\longrightarrow$ 2 CaTiO$_3$[s] + 2 H$_2$S + 3 H$_2$   &   $\min\{$CaOH, TiS$\}$\\
\hline
88 & CaSiO$_3$[s]         & Ca + SiO + 2 H$_2$O
      $\longrightarrow$ CaSiO$_3$[s] + 2 H$_2$   &   $\min\{$Ca, SiO$\}$\\
89 & Wollastonite         & Ca + SiS + 3 H$_2$O
      $\longrightarrow$ CaSiO$_3$[s] + H$_2$S + 2 H$_2$   &   $\min\{$Ca, SiS$\}$\\
90 & (4)                  & 2 Ca + 2 SiH + 6 H$_2$O
      $\longrightarrow$ 2 CaSiO$_3$[s] + 7 H$_2$   &   $\min\{$Ca, SiH$\}$\\
91 &                      & CaO + SiO + 1 H$_2$O
      $\longrightarrow$ CaSiO$_3$[s] + H$_2$   &   $\min\{$CaO, SiO$\}$\\
92 &                      & CaO + SiS + 2 H$_2$O
      $\longrightarrow$ CaSiO$_3$[s] + H$_2$S + H$_2$   &   $\min\{$CaO, SiS$\}$\\
93 &                      & 2 CaO + 2 SiH + 4 H$_2$O
      $\longrightarrow$ 2 CaSiO$_3$[s] + 5 H$_2$   &   $\min\{$CaO, SiH$\}$\\
94 &                      & CaS + SiO + 2 H$_2$O
      $\longrightarrow$ CaSiO$_3$[s] + H$_2$S + H$_2$  &  $\min\{$CaS, SiO$\}$\\
95 &                      & CaS + SiS + 3 H$_2$O
      $\longrightarrow$ CaSiO$_3$[s] + 2 H$_2$S + H$_2$  &  $\min\{$CaS, SiS$\}$\\
96 &                      & 2 CaS + 2 SiH + 6 H$_2$O
      $\longrightarrow$ 2 CaSiO$_3$[s] + 2 H$_2$S + 5 H$_2$  &  $\min\{$CaS, SiH$\}$\\
97 &                      & 2 CaOH + 2 SiO + 2 H$_2$O
      $\longrightarrow$ 2 CaSiO$_3$[s] + 5 H$_2$   &  $\min\{$CaOH, SiO$\}$\\
98 &                      & 2 CaOH + 2 SiS + 4 H$_2$O
      $\longrightarrow$ 2 CaSiO$_3$[s] + 2 H$_2$S + 3 H$_2$   &  $\min\{$CaOH, SiS$\}$\\
99 &                      &  CaOH +  SiH + 2 H$_2$O
      $\longrightarrow$  CaSiO$_3$[s] + 3 H$_2$   &  $\min\{$CaOH, SiH$\}$\\
100 &                      &  Ca(OH)$_2$ +  SiO
      $\longrightarrow$  CaSiO$_3$[s] + H$_2$   &  $\min\{$Ca(OH)$_2$, SiO$\}$\\
101 &                      &  Ca(OH)$_2$ +  SiS + H$_2$O
      $\longrightarrow$  CaSiO$_3$[s] + H$_2$S + H$_2$   &  $\min\{$Ca(OH)$_2$, SiS$\}$\\
102 &                      & 2 Ca(OH)$_2$ + 2 SiH + 2 H$_2$O
      $\longrightarrow$  2 CaSiO$_3$[s] + 5 H$_2$   &  $\min\{$Ca(OH)$_2$, SiH$\}$\\
103 &                      & 2 CaH + 2 SiO + 4 H$_2$O
      $\longrightarrow$  2 CaSiO$_3$[s] + 5 H$_2$   &  $\min\{$CaH, SiO$\}$\\
104 &                      & 2 CaH + 2 SiS + 6 H$_2$O
      $\longrightarrow$  2 CaSiO$_3$[s] + 2 H$_2$S + 5 H$_2$   &  $\min\{$CaH, SiS$\}$\\
105 &                      & CaH + SiH + 3 H$_2$O
      $\longrightarrow$  CaSiO$_3$[s] + 4 H$_2$   &  $\min\{$CaH, SiH$\}$\\
\hline
106 & Fe$_2$SiO$_4$[s]      & 2 Fe + SiO + 3 H$_2$O
      $\longrightarrow$  Fe$_2$SiO$_4$[s] + 3 H$_2$  &  $\min\{$ $\half$Fe, SiO$\}$\\
107 & Fayalite              & 2 Fe + SiS + 4 H$_2$O
      $\longrightarrow$  Fe$_2$SiO$_4$[s] + H$_2$S + 3 H$_2$  &  $\min\{$ $\half$Fe, SiS$\}$\\
108 & (4)                   & 4 Fe + 2 SiH + 8 H$_2$O
      $\longrightarrow$  2 Fe$_2$SiO$_4$[s] + 9 H$_2$  &  $\min\{$ $\half$Fe, SiH$\}$\\
109 &                       & 2 FeO + SiO + H$_2$O
      $\longrightarrow$  Fe$_2$SiO$_4$[s] + H$_2$  &  $\min\{$ $\half$FeO, SiO$\}$\\
110 &                       & 2 FeO + SiS + 2 H$_2$O
      $\longrightarrow$  Fe$_2$SiO$_4$[s] + H$_2$S + H$_2$  &  $\min\{$ $\half$FeO, SiS$\}$\\
111 &                       & 4 FeO + 2 SiH + 4 H$_2$O
      $\longrightarrow$  2 Fe$_2$SiO$_4$[s] + 5 H$_2$  &  $\min\{$ $\half$FeO, SiH$\}$\\
112 &                       & 2 FeS + SiO + 3 H$_2$O
      $\longrightarrow$  Fe$_2$SiO$_4$[s] + 2 H$_2$S + H$_2$  &  $\min\{$ $\half$FeS, SiO$\}$\\
113 &                       & 2 FeS + SiS + 4 H$_2$O
      $\longrightarrow$  Fe$_2$SiO$_4$[s] + 3 H$_2$S + H$_2$  &  $\min\{$ $\half$FeS, SiS$\}$\\
114 &                       & 4 FeS + 2 SiH + 8 H$_2$O
      $\longrightarrow$  2 Fe$_2$SiO$_4$[s] + 4 H$_2$S + 5 H$_2$  &  $\min\{$ $\half$FeS, SiH$\}$\\
115 &                       & 2 Fe(OH)$_2$ + SiO
      $\longrightarrow$  Fe$_2$SiO$_4$[s] + H$_2$O + H$_2$  &  $\min\{$ $\half$Fe(OH)$_2$, SiO$\}$\\
116 &                       & 2 Fe(OH)$_2$ + SiS
      $\longrightarrow$  Fe$_2$SiO$_4$[s] + H$_2$S + H$_2$  &  $\min\{$ $\half$Fe(OH)$_2$, SiS$\}$\\
117 &                       & 4 Fe(OH)$_2$ + 2 SiH
      $\longrightarrow$  2 Fe$_2$SiO$_4$[s] + 5 H$_2$  &  $\min\{$ $\half$Fe(OH)$_2$, SiH$\}$\\
118 &                       & 2 FeH + SiO + 3 H$_2$O
      $\longrightarrow$  Fe$_2$SiO$_4$[s] + 4 H$_2$  &  $\min\{$ $\half$FeH, SiO$\}$\\
119 &                       & 2 FeH + SiS + 4 H$_2$O
      $\longrightarrow$  Fe$_2$SiO$_4$[s] + H$_2$S + 4 H$_2$  &  $\min\{$ $\half$Fe(OH)$_2$, SiS$\}$\\
120 &                       & 4 FeH + 2 SiH + 8 H$_2$O
      $\longrightarrow$  2 Fe$_2$SiO$_4$[s] + 11 H$_2$  &  $\min\{$ $\half$Fe(OH)$_2$, SiH$\}$ \\
\hline
121 & C[s]                  & C
      $\longrightarrow$  C[s]      & C \\
122 & Carbon                & C$_2$
      $\longrightarrow$  2 C[s]      & C$_2$ \\
123 & (4)                   & C$_3$
      $\longrightarrow$  3 C[s]      & C$_3$ \\
124 &                       & 2 C$_2$H
      $\longrightarrow$   4 C[s] + H$_2$      & $\half$C$_2$H \\
125 &                       & C$_2$H$_2$
      $\longrightarrow$   2 C[s] + H$_2$      & C$_2$H$_2$ \\
126 &                       & CH$_4$
      $\longrightarrow$   C[s] + 2 H$_2$      & $\half$CH$$
\end{tabular}}
\end{table*}

\end{document}